\documentclass[draft,numberedheadings]{aipproc}
\usepackage{amssymb}

\layoutstyle{6x9}

\begin{document}

\title{Phases and phase transitions in disordered quantum systems}


\author{Thomas Vojta}{
  address={Department of Physics, Missouri University of Science and Technology, Rolla, MO 65409, USA}
}

\begin{abstract}
These lecture notes give a pedagogical introduction to phase transitions in disordered quantum systems
and to the exotic Griffiths phases induced in their vicinity. We first review some fundamental concepts
in the physics of phase transitions. We then derive criteria governing under what conditions spatial
disorder or randomness can change the properties of a phase transition. After introducing the
strong-disorder renormalization group method, we discuss in detail some of the exotic phenomena arising
at phase transitions in disordered quantum systems. These include infinite-randomness criticality,
rare regions and quantum Griffiths singularities, as well as the smearing of phase transitions.
We also present a number of experimental examples.
\end{abstract}

\maketitle


\section*{Overview}

Phase transitions are fascinating phenomena that arise in a wide variety of physical, chemical,
and biological systems. These range from the never-ending transformations of water between
ice, liquid and vapor to transitions between different steady states in the dynamics of
bio-populations. A particularly interesting class of transitions, the quantum phase transitions,
occur when the quantum ground state of a material abruptly changes in response to a
change in external conditions. Many realistic systems are not perfectly clean but contain
impurities, defects, or other kinds of imperfections. The influence of such random disorder on
phase transitions is therefore of prime interest, both from a conceptual point of view and for
experimental applications.

This article gives an informal introduction into this topic at a level accessible for graduate students.
It is based on the notes of a series of lectures that the author delivered at the
XVII Training Course in the Physics of Strongly Correlated Systems in Vietri sul Mare, Italy
in October 2012. The emphasis of the paper is on a pedagogical exposition of the
fundamental ideas and methods rather than a comprehensive account of the literature.
More details can be found in several recent review articles \cite{Vojta06,Vojta10,IgloiMonthus05}.

To make these lecture notes self-contained, we start with a brief summary of the fundamental concepts of
classical and quantum phase transitions in Sec.\  \ref{sec:PT}. In Sec.\ \ref{sec:disorder}, we
consider various types of disorder or randomness, and we derive criteria that govern under what
conditions they can influence a phase transition. The strong-disorder renormalization group method
is introduced in Sec.\ \ref{sec:SDRG}. Sections \ref{sec:Griffiths} and \ref{sec:smeared} are devoted
to exploring some of the exotic phenomena arising at disordered quantum phase transitions,
viz., Griffiths singularities and smeared transitions. Applications to magnetic quantum phase transitions
in metals are discussed in Sec.\ \ref{sec:metals}. We conclude in Sec.\ \ref{sec:conclusions}.

\section{Phase transitions and quantum phase transitions}
\label{sec:PT}

In this section, we briefly collect some basic ideas about classical and quantum phase transitions
to the extent required for the rest of the article.
A reader unfamiliar with these topics will benefit from consulting text books such as
Refs.\ \cite{Goldenfeld_book92,Sachdev_book99}.

\subsection{Fundamental concepts}

\subsubsection*{What is a phase transition?}

\begin{figure}
\centerline{\includegraphics[width=10.5cm]{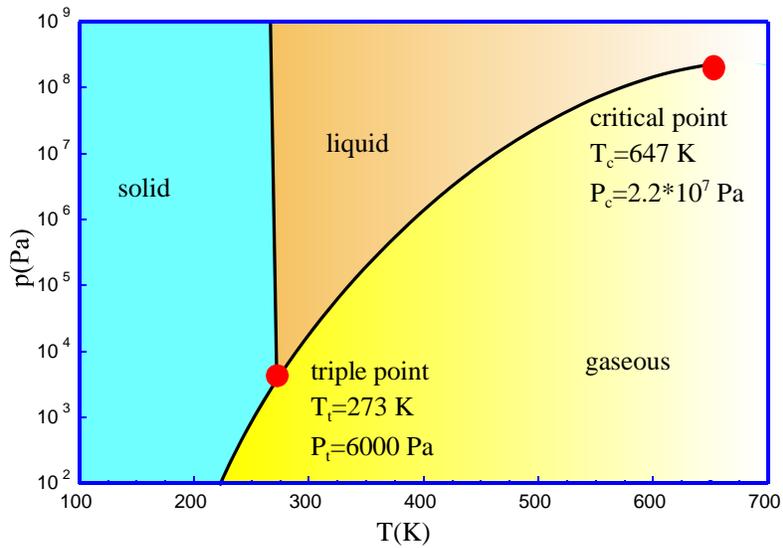}}
\caption{Phase diagram of water as a function of pressure $p$ and temperature $T$. The solid black lines
         mark first-order phase transitions. The liquid-gas phase boundary ends in
         a critical point at which the phase transition is continuous.}
\label{fig:h2o}
\end{figure}

Consider, for example, the phase diagram of water reproduced in Figure \ref{fig:h2o}.
It shows in which phase (solid, liquid or gas) water is for given values of
pressure $p$ and temperature $T$. If we cross one of the phase boundaries
(marked by solid black lines), the properties of water change \emph{abruptly} in response
to a \emph{smooth} change in pressure and/or temperature. This is the main characteristic of a phase transition.
Formally, we can define a phase transition as a singularity in the free energy as function
of the external parameters such as temperature, pressure, magnetic field, or chemical
composition.\footnote{A true singularity can only appear in the thermodynamic limit of infinite
          system size. For any finite system, the free energy is an analytic function because the partition
          function is a finite sum of exponentials.}
A singular free energy also implies singularities in other thermodynamic quantities including
internal energy, entropy, specific heat, and so on.

Phase transitions can be divided into two qualitatively different types or classes. To understand the difference,
consider again the example of water. Imagine we heat up a piece of ice, starting at a temperature  well below
the melting point of 0$^\circ$C. Initially, the ice' temperature will rise smoothly as it absorbs more and more
heat. However, after the temperature reaches 0$^\circ$C, the absorbed heat is used to melt the ice
while the temperature is stuck at the melting point. Only after all ice has melted, the temperature resumes its rise.
This means, the solid and liquid phases of water coexist at the phase transition point, and a nonzero amount of heat, the
{\emph{latent heat}, is required to turn the solid phase of water into the liquid phase.
Phase transitions following this scenario, namely phase coexistence and latent heat, are usually called first-order
phase transitions.\footnote{The name refers to Ehrenfest's classification \cite{Ehrenfest33} of phase transitions. At a first-order transition,
a \emph{first} derivative of the free energy is discontinuous.}

The phase transitions of water shown in Fig.\ \ref{fig:h2o} are all of first order, except for one single
isolated point. To understand this, focus on the liquid-gas phase boundary.
At ambient pressure (about $10^5$ Pa), the properties of liquid water and
water vapor are very different (for example, their densities differ by a factor of more than 1000). If we
follow the phase boundary to higher pressures and temperatures, liquid and vapor become more and more similar until
they become indistinguishable at $T_c=647$ K and $p_c=2.21\times10^7$ Pa (374$^\circ$C and 217 atm).
This point is the so-called critical point.
For $p>p_c$, water features only one fluid phase rather then two separate liquid and gas phases.
Phase transitions occurring at such a critical point are called continuous phase
transitions.\footnote{Ehrenfest further subdivided continuous transitions into second order transitions,
third order transitions and so on according to which derivative of the free energy is discontinuous.
Today, this subdivision is not used very much because there are few qualitative differences between
the sub-classes.}
 Continuous phase
transitions do not display phase coexistence because the two phases are indistinguishable at the transitions
point. This implies that there is no latent heat.

Further prototypical examples of phase transitions can be found in a ferromagnet such as
iron. Its magnetic phase diagram, shown in Fig.\ \ref{fig:iron},
\begin{figure}
\centerline{\includegraphics[width=7.5cm]{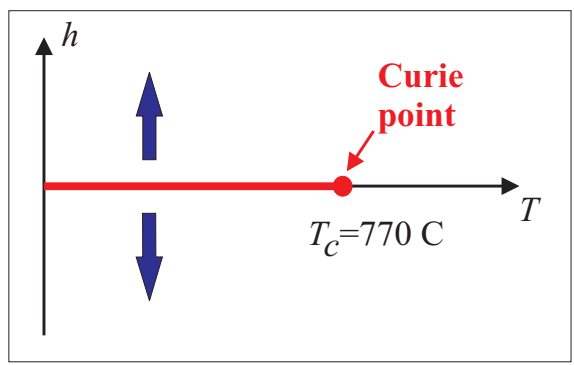}}
\caption{Schematic phase diagram of iron as a function of magnetic field $h$ and temperature $T$. The thick red line
         marks a first-order phase boundary between the spin-up and spin-down phases. It ends in
         a critical point (Curie point) at which the phase transition is continuous.}
\label{fig:iron}
\end{figure}
is qualitatively very similar to the liquid-gas phase diagram of water. There is a line
of first-order phase transitions at low temperatures and magnetic field $h=0$ that separates the spin-up and spin-down phases
which coexist on the phase boundary. If we follow the phase boundary to higher temperatures, the magnetization
decreases, thus the two phases become more similar. The phase boundary ends at the critical point (Curie point) at which
the magnetization vanishes and the two phases (spin-up and spin-down) thus become indistinguishable.

\subsubsection*{Critical behavior}

Critical points (continuous phase transitions) have many peculiar properties. One of the earliest and
most striking manifestations of criticality was discovered by Andrews in 1869 \cite{Andrews69}. He found that
a fluid (carbon dioxide in his case) becomes very milky, i.e. opalescent, close to its critical point.
Figure \ref{fig:opalescence} illustrates this so-called critical opalescence on the example of the mixing critical
point of hexane and methanol.
\begin{figure}
\centerline{\includegraphics[width=\textwidth]{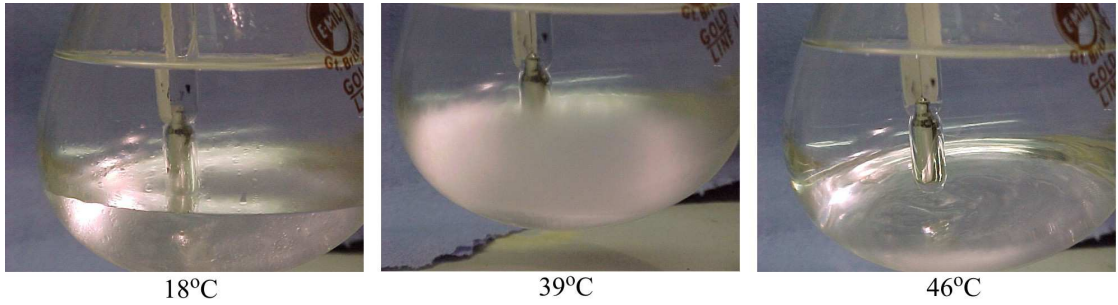}}
\caption{Critical opalescence near the mixing critical point of  hexane and methanol
         (taken with permission from Ref.\ \cite{Podesta}).}
\label{fig:opalescence}
\end{figure}
At high temperatures, hexane and methanol are miscible and form a homogeneous clear fluid.
The mixture phase-separates into a methanol-rich phase (below the meniscus)
and a hexane-rich phase (above the meniscus) at low temperatures. Close to the transition between the miscible and phase-separated
regimes, the mixture becomes very milky.

What is causing the critical opalescence? The fact that the fluid becomes milky means that it strongly
scatters visible light. This implies that it contains structures of a size comparable to the wavelength
of light (about 1 $\mu$m), much larger than typical microscopic lengths such as the distance
between molecules (a few {\AA}). Critical opalescence thus demonstrates that a critical point features
strong fluctuations on large length scales.

Formally, these fluctuations can be characterized in terms of correlation functions. For the mixing critical point
discussed above,
one could define a concentration correlation function $G(\mathbf{x}-\mathbf{x'}) = \langle \Delta c(\mathbf{x})\, \Delta c(\mathbf{x'})\rangle$.
Here $\Delta c(\mathbf{x})$ is the deviation of the hexane concentration $c$ at position $\mathbf{x}$ from its average value;
and $\langle \ldots \rangle$ denotes the thermodynamic average. The appropriate correlation function for a liquid-gas
critical point would be the density-density correlation function $\langle \Delta \rho(\mathbf{x})\, \Delta \rho(\mathbf{x'})\rangle$;
and for a magnetic critical point, it would be the spin-spin correlation function $\langle \Delta S(\mathbf{x})\, \Delta S(\mathbf{x'})\rangle$.
Under normal conditions, these correlation functions decay exponentially to zero for large distances $|\mathbf{x}-\mathbf{x'}|\to \infty$,
$G(\mathbf{x}-\mathbf{x'}) \sim \exp(-|\mathbf{x}-\mathbf{x'}|/\xi)$ which defines the correlation length
$\xi$. Upon approaching the critical point, the correlation length diverges as
\begin{equation}
\xi  \sim |r|^{-\nu} = |(T-T_c) / T_c|^{-\nu}
\label{eq:xi}
\end{equation}
where the reduced temperature $r=(T-T_c) / T_c$ is a dimensionless measure of the distance from criticality and
$\nu$ is the correlation length critical exponent.

Fluctuations close to criticality are not only large but also slow. One can define a time scale, the correlation time
$\xi_\tau$ in analogy to the correlation length $\xi$. The correlation time also diverges at the critical point,
\begin{equation}
\xi_\tau  \sim \xi^z \sim r^{-\nu z}
\label{eq:xit}
\end{equation}
which defines the dynamical critical exponent $z$. The fact that the fluctuations become arbitrarily slow
is also called the critical slowing down. It does occur not just in real experimental systems, it also greatly
hampers the efficiency of computer simulations of critical phenomena.\footnote{In some
cases, critical slowing down in computer simulations can be circumvented by clever ``cluster''
algorithms \cite{SwendsenWang87,Wolff89}, in other cases it is unavoidable.}

Power-law singularities in the length and time scales at a critical point generically lead to power-law
singularities in observable quantities. For example, at a ferromagnetic critical point, the magnetic
susceptibility diverges as $\chi \sim |T-T_c|^{-\gamma}$, and the magnetization vanishes as
$m \sim  |T-T_c|^{\beta}$. Here, $\beta$ is the order parameter critical exponent, and
$\gamma$ is called the susceptibility exponent. Analogously, at a liquid-gas critical point, the compressibility
diverges as $\kappa \sim |T-T_c|^{-\gamma}$, and the density difference between the liquid and gas phases vanishes
as $\Delta \rho \sim |T-T_c|^{\beta}$. Other observables show similar power laws. A summary of the
commonly used critical exponents is given in Table \ref{table:exponents}, using the example of a
ferromagnetic transition.
\begin{table}
\renewcommand{\arraystretch}{1.2}
\begin{tabular}{cccc}
\hline
&\textbf{exponent}& \textbf{definition} & \textbf{conditions} \\
\hline
specific heat &$\alpha$& $c \propto |r|^{-\alpha}$ & $r \to 0, ~h=0$\\
order parameter& $\beta$ & $m \propto (-r)^\beta$ & $r \to 0-$, ~$h=0$\\
susceptibility& $\gamma$ & $\chi \propto |r|^{-\gamma}$ & $r \to 0, ~h=0$\\
critical isotherm & $\delta$ & $h \propto |m|^\delta {\rm sign}(m)$ & $h \to 0, ~r=0$\\
\hline
correlation length& $\nu$ & $\xi \propto |r|^{-\nu}$ & $r \to 0, ~h=0$\\
correlation function& $\eta$ & $G(\mathbf{x}) \propto |\mathbf{x}|^{-d+2-\eta}$ & $r=0, ~h=0$\\
\hline
dynamical& $z$ & $\xi_\tau \propto \xi^{z}$ & $r \to 0, ~h=0$\\
activated dynamical  & $\psi$ & $\ln \xi_\tau \propto \xi^{\psi}$ & $r \to 0, ~h=0$\\
\hline
\end{tabular}
\caption{Definitions of the commonly used critical exponents using the example of a ferromagnetic
  transition.  $m$ is the magnetization and $h$ is an external magnetic field. $r$ denotes the
  distance from the critical point and $d$ is the space dimensionality.}
\label{table:exponents}
\end{table}
The collection of all these power laws characterizes the critical point and is usually
called the critical behavior.

The values of the critical exponents generally do not depend on the microscopic details
but only on the space dimensionality and the symmetries of the system under consideration.
For instance, all liquid-gas critical points have the same exponents. This phenomenon, called
universality, is very fortunate for theory, because it permits the determination of the
exact critical exponent values from simplified model problems as long as they have the
correct symmetries and dimensionalities. It also permits the definition of universality classes
comprising all critical points that share the same critical exponents.

Above, we have seen that fluctuations become very strong as one approaches a critical point.
Do they influence the critical behavior? The answer to this question depends on the space dimensionality
$d$.
In general, fluctuations become less important with increasing dimensionality.\footnote{Intuitively,
this happens because the number of neighbors of a given site increases with the dimensionality.
A large number of neighbors effectively averages out the fluctuations felt by a particular site.}
In sufficiently low dimensions, fluctuations are so strong that they completely destroy the
ordered phase at all (nonzero) temperatures. Consequently, there is no phase transition. This happens
for $d\le d_c^-$ where $d_c^-$ is the so-called lower critical dimension. Between $d_c^-$ and the
upper critical dimension $d_c^+$, an ordered phase and thus a phase transition  exist, but the
critical exponents are influenced by fluctuations and depend on $d$. Finally, for $d>d_c^+$,
fluctuations are unimportant for the critical behavior. The exponents become independent of $d$
and take their mean-field values. For example, for Ising ferromagnets, the critical dimensions are
$d_c^-=1$ and $d_c^+=4$, for Heisenberg ferromagnets they are  $d_c^-=2$ and $d_c^+=4$.

\subsubsection*{Scaling}

Scaling theory is a phenomenological description of critical points. It is an
extremely powerful tool for the analysis of experimental or numerical data.
Scaling theory was first put forward by Widom \cite{Widom65} as the scaling
hypothesis. Within modern renormalization group theory \cite{WilsonKogut74},
it can be derived from first principles.

The basic idea of scaling theory is that the correlation length $\xi$
is the only relevant length scale close to criticality. Right at the critical
point, $\xi$ diverges, and the system is scale-invariant. Rescaling all
lengths by a common factor should therefore leave the system unchanged. A little bit off
criticality, $\xi$ is large but finite. Thus, if we rescale all lengths, we need to
adjust the external parameters (temperature, field, etc.) such that $\xi$
has the same value. After that, the system should again be unchanged.

Let us now apply this idea to the free energy density  $f=F/V$ of a system
near a ferromagnetic critical point, written  as function of the reduced temperature $r$
and the external magnetic field $h$. Rescaling all lengths by an arbitrary
common factor $b$ leads to the homogeneity relation
\begin{equation}
f(r,h) = b^{-d} f(r\,b^{1/\nu}, h\,b^{y_h})
\label{eq:scaling}
\end{equation}
where the exponent $y_h$  is related to $\delta$  by $y_h=d \, \delta /(1+\delta)$.
The factor $b^{-d}$ arises from the change of volume upon the rescaling, the changes
in the arguments of $f$ reflect the adjustments necessary to keep $\xi$ at its old
value. Homogeneity relations for other thermodynamic variables can be obtained simply
by taking the appropriate derivatives of (\ref{eq:scaling}).  For example, the specific heat
is given by a second derivative of the free energy, $c \sim -\partial^2 f /\partial r^2$.
Its scaling form thus reads
\begin{equation}
c(r,h)  = b^{-d+2/\nu} c(r\,b^{1/\nu}, h\,b^{y_h})~.
\label{eq:c_scaling}
\end{equation}

How can homogeneity relations such as (\ref{eq:scaling}) and (\ref{eq:c_scaling}) be used to analyze experimental or
numerical data? The key is that the homogeneity relations hold for arbitrary $b$,
implying that we can set $b$ to whatever value we find useful. For example, if we were
interested in the dependence of $c$ on the reduced temperature at zero field, we could set
$b=r^{-\nu}$. This yields $c(r,0) = r^{d\nu-2} c(1,0)$ which could be used to determine
the exponent $\alpha=2-d\nu$ via a fit of the data.

The scaling form (\ref{eq:scaling}) of the free energy density $f$ depends on only two exponents. Since all thermodynamic
observables can be determined from $f$, this implies that the various exponents listed in Table
\ref{table:exponents} are not independent. Instead they are related by the so-called scaling relations
$2- \alpha =  2 \beta +\gamma$ and $2 - \alpha = \beta ( \delta + 1)$ as well as the
hyperscaling relations $2- \alpha =  d\,\nu$ and  $\gamma = (2-\eta) \nu$.

We note that scaling in the form discussed here holds only below the upper
critical dimension $d_c^+$. Above $d_c^+$, the situation is more complicated, and
we refer the reader to a textbook about critical phenomena such as Ref.\ \cite{Goldenfeld_book92}.

\subsubsection*{Landau theory}

Landau \cite{Landau37a,Landau37b,Landau37c,Landau37d}
provided a general framework for the description of phase transitions. According to
Landau, bulk phases can be distinguished according to their symmetries, and phase transitions
involve the  (spontaneous) breaking of these symmetries.\footnote{In recent years, there
has been an increasing interest in phase transitions that do \emph{not} follow Landau's
paradigm. These unconventional transitions often involve topological degrees of freedom
(see, e.g., Refs.\ \cite{AletWalczakFisher06,Sachdev09} for introductory discussions).
They are beyond the scope these lectures.}
As an example, consider the phase transition in an easy-axis (Ising) ferromagnet, i.e., a ferromagnet
in which the spins prefer to be either up or down (parallel or antiparallel to a particular
crystallographic direction). In the paramagnetic phase at high temperatures, the spins point
up or down at random, implying that the global up-down symmetry is not broken. In contrast,
in the ferromagnetic low-temperature phase, the spins spontaneously pick a preferred
direction (either up or down). Thus, the up-down symmetry is spontaneously broken.

A central concept in Landau's framework is the order parameter, a physical quantity that
characterizes the degree of symmetry breaking. It is zero in one phase (the disordered phase)
and nonzero and usually nonunique in the other phase (the ordered phase). In our example of
an Ising ferromagnet, the total magnetization $m$ is an order parameter. $m$ vanishes in the
paramagnetic phase while it is nonzero in the ferromagnetic phase. Moreover, $m$ is nonunique
(positive or negative) as the magnetization can be either up or down spontaneously.

In addition to providing the general framework for phase transitions discussed above,
Landau also developed an (approximate) quantitative description which is now known as
the Landau theory of phase transitions. It is based on the idea that the free energy density $f$
close to the transition can be expanded in powers of the order parameter $m$,
\begin{equation}
f = -h\,m + r\,m^2 + v\, m^3 + u\, m^4 + \ldots ~.
\label{eq:Landau}
\end{equation}
In the case of the ferromagnetic transition, the first term represents the external magnetic field.
Moreover, the cubic coefficient $v$ and all higher-order odd terms must vanish
because the free energy must be up-down symmetric in the absence of a field. For
other phase transitions (such as the nematic transition in liquid crystals), odd terms do occur in the
Landau expansion of the free energy.
The coefficients $r$, $v$, and $u$ depend on the material as well as the external conditions (temperature, pressure, etc.).
The correct physical state for given values of these coefficients is given by the minimum of $f$
with respect to the order parameter $m$.

The qualitative behavior of Landau theory can be easily discussed by plotting the
free energy density $f(m)$. Let us assume a vanishing external field, $h=0$, as well
as $v=0$ and $u>0$.\footnote{If $u$ were negative, the theory could not be truncated
 after the quartic term because it would be unstable against diverging order parameter $m$.}
Figure \ref{fig:landau} shows $f(m)$ for several values of
the quadratic coefficient $r$.
\begin{figure}
\centerline{\includegraphics[width=7.5cm]{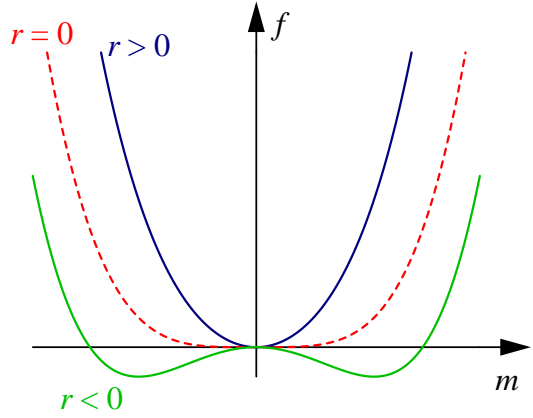}}
\caption{Sketch of the Landau free energy density $f$ as function of the order parameter $m$.}
\label{fig:landau}
\end{figure}
For $r>0$, the free energy has a single minimum at $m=0$. Thus, $r>0$ corresponds
to the paramagnetic phase. Conversely, the free energy has two degenerate
minima at nonzero $m$  for $r<0$, separated by a local maximum at $m=0$. This means,
the system is in the ferromagnetic phase for $r<0$. The transition between these two
phases occurs at $r=0$ implying that $r$ is indeed a measure of the distance from the transition,
$r \sim T-T_c$.

Quantitative results follow from solving $(\partial f / \partial m)=0$. In zero external field,
$h=0$, the magnetization behaves as $m \sim (-r)^{1/2}$ for $r<0$ while the magnetic susceptibility
diverges as $\chi = (\partial m / \partial h) \sim |r|^{-1}$. Thus, Landau theory predicts a
continuous phase transition, i.e., a critical point.\footnote{If the cubic coefficient in the
free energy (\ref{eq:Landau}) does not vanish, or if we have a free energy
with a negative quartic coefficient $u$ but a positive 6th-order coefficient,
the theory yields a first-order transition.}
On the critical isotherm, $r=0$,
the magnetization varies as $m \sim h^{1/3}$ with the external field. In summary, Landau theory gives
mean-field values for the critical exponents, viz., $\beta=1/2$, $\gamma=1$, and $\delta=3$.

How good is Landau theory? Landau theory does not contain any fluctuations because the order parameter
is treated as a constant. Landau theory therefore gives the correct critical behavior if the dimensionality
is above the upper critical dimension $d_c^+$ of the problem. In contrast, Landau theory fails
below $d_c^+$. In order to find the critical behavior for $d<d_c^+$, one needs to generalize
Landau theory by making the order parameter a fluctuating field $m({\mathbf x})$ that depends on position.
The Landau free energy gets replaced by a functional, the so-called Landau-Ginzburg-Wilson (LGW)
functional
\begin{equation}
F[m({\mathbf x})] = \int d^d x \left[-h\, m({\mathbf x}) + r\, m^2({\mathbf x})  + (\nabla\, m({\mathbf x}))^2 +u\, m^4({\mathbf x}) + \ldots \right]~.
\label{eq:LGW}
\end{equation}
The gradient term punishes rapid changes of the magnetization,
it thus encodes the ferromagnetic interaction between neighboring spins.
The partition function is now given by a functional integral
\begin{equation}
Z = \int D[m({\mathbf x})]\, \exp\left(- F[m({\mathbf x})] \right)~.
\label{eq:Z_LGW}
\end{equation}
In contrast to Landau theory, the problem defined by equations (\ref{eq:LGW}) and (\ref{eq:Z_LGW}) cannot be solved by
elementary means. However, it can be attacked successfully using modern renormalization group
techniques \cite{WilsonKogut74}.

\subsection{Introduction to quantum phase transitions}
\label{subsec:QPT_intro}
\subsubsection*{What is a quantum phase transition?}

The phase transitions we have considered so far all occur at nonzero temperatures,
and they are often tuned by changing the temperature. At these transitions, the ordered
phase is destroyed by thermal fluctuations. For instance, ice melts because
the thermal motion of the water molecules destroys the crystal lattice; and  a ferromagnetic
material becomes paramagnetic at the Curie point due to the thermal motion of the spins.

A different kind of phase transition occurs at the absolute zero of temperature
(in the quantum ground state) when a (nonthermal) parameter such as pressure,
magnetic field, or chemical composition is varied. Consider, for example, the magnetic
phase diagram of LiHoF$_4$ shown in Fig.\ \ref{fig:LiHoF4}.
\begin{figure}
\centerline{\includegraphics[width=8.5cm]{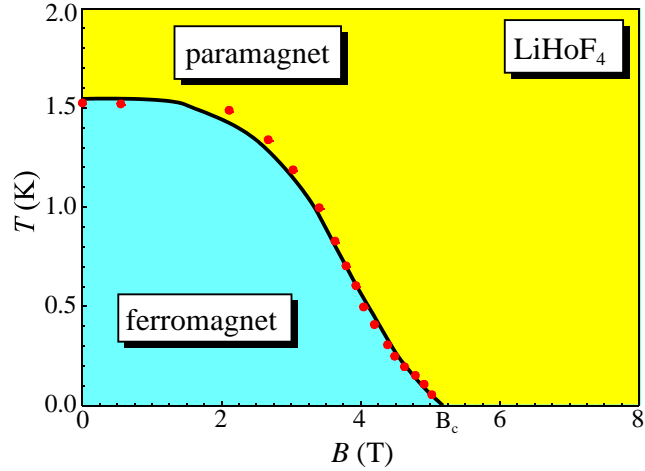}}
\caption{Magnetic phase diagram of LiHoF$_4$ as function of temperature $T$ and
   transverse magnetic field $B$. The red dots represent experimental data from Ref.\
   \cite{BitkoRosenbaumAeppli96}. }
\label{fig:LiHoF4}
\end{figure}
LiHoF$_4$ is an easy-axis (Ising) ferromagnet. In the absence of an external field,
the material is in a ferromagnetic phase for temperatures below the Curie point of
about 1.5 K. If the temperature is raised above this value, the thermal motion of the
holmium spins (which carry the magnetism) destroys ferromagnetic long-range order.
This phase transition is a thermal transition of the type discussed in the last section.
However, the figure also shows another way to destroy the ferromagnetism:
If one applies a magnetic field $B$ perpendicular (transverse) to the preferred direction
of the spins, the ferromagnetic order is suppressed. Beyond a critical field of about
5 T, the ferromagnetic phase vanishes completely, even at zero temperature. Thus
LiHoF$_4$ undergoes a zero-temperature ferromagnetic phase transition at this
critical magnetic field.

How does the magnetic field destroy the ferromagnetic phase? A qualitative understanding
can be gained from considering a toy Hamiltonian, the transverse-field Ising
model.\footnote{Note that the microscopic physics of LiHoF$_4$ is significantly more complicated
than the transverse-field Ising model. In particular, the Ho spins are not S=1/2 spins, their interaction is
of dipolar type, and the hyperfine coupling
between the electronic and nuclear magnetic moments is anomalously large \cite{BitkoRosenbaumAeppli96}.}
It reads
\begin{equation}
H = -J\sum_{\langle ij\rangle} \sigma_i^z \sigma_j^z - B\sum_i \sigma_i^x~,
\label{eq:TFIM}
\end{equation}
where the Pauli matrices $\sigma_i^z$ and $\sigma_i^x$ represent the $z$ and $x$ components of the spin operator at site $i$.
The first term implements an attractive interaction between nearest neighbor sites of a
cubic lattice. It prefers ferromagnetic order in the $z$-direction.
The second term represents the transverse magnetic field (in the $x$-direction).
To understand its effect, remember that $\sigma_i^x$ can be decomposed into spin-flip
operators, $\sigma_i^x=\sigma_i^+ + \sigma_i^-$. Thus, the transverse field term induces spin flips from
up to down and from down to up. If the field becomes sufficiently strong, these spin flips
destroy the ferromagnetic long-range order, even at zero temperature.

The zero-temperature phase transition between a ferromagnetic ground state for weak field $B$
and a paramagnetic ground state for strong $B$ is therefore driven by quantum fluctuations,
i.e., quantum zero-point motion, rather than thermal fluctuations. It can be viewed as a consequence
of Heisenberg's uncertainty principle. For this reason, this type of phase transition is
called a quantum phase transition. Quantum phase transitions can be divided into first-order and
continuous just like thermal phase transitions. First order quantum phase transitions correspond
to simple energy level crossings at which two different ground states are exactly degenerate.
Continuous quantum phase transitions, i.e., quantum critical points,
involve diverging length and time scales of the quantum fluctuations.

Quantum phase transitions have been identified in a wide variety of condensed matter systems.
Examples include ferromagnetic and antiferromagnetic transitions in strongly correlated electron
materials, metal-insulator transitions in doped semiconductors and metals, as well as superfluid-insulator
transitions in ultracold atomic gases (see Fig.\ \ref{fig:QPT_examples}).
\begin{figure}
\includegraphics[width=6.7cm]{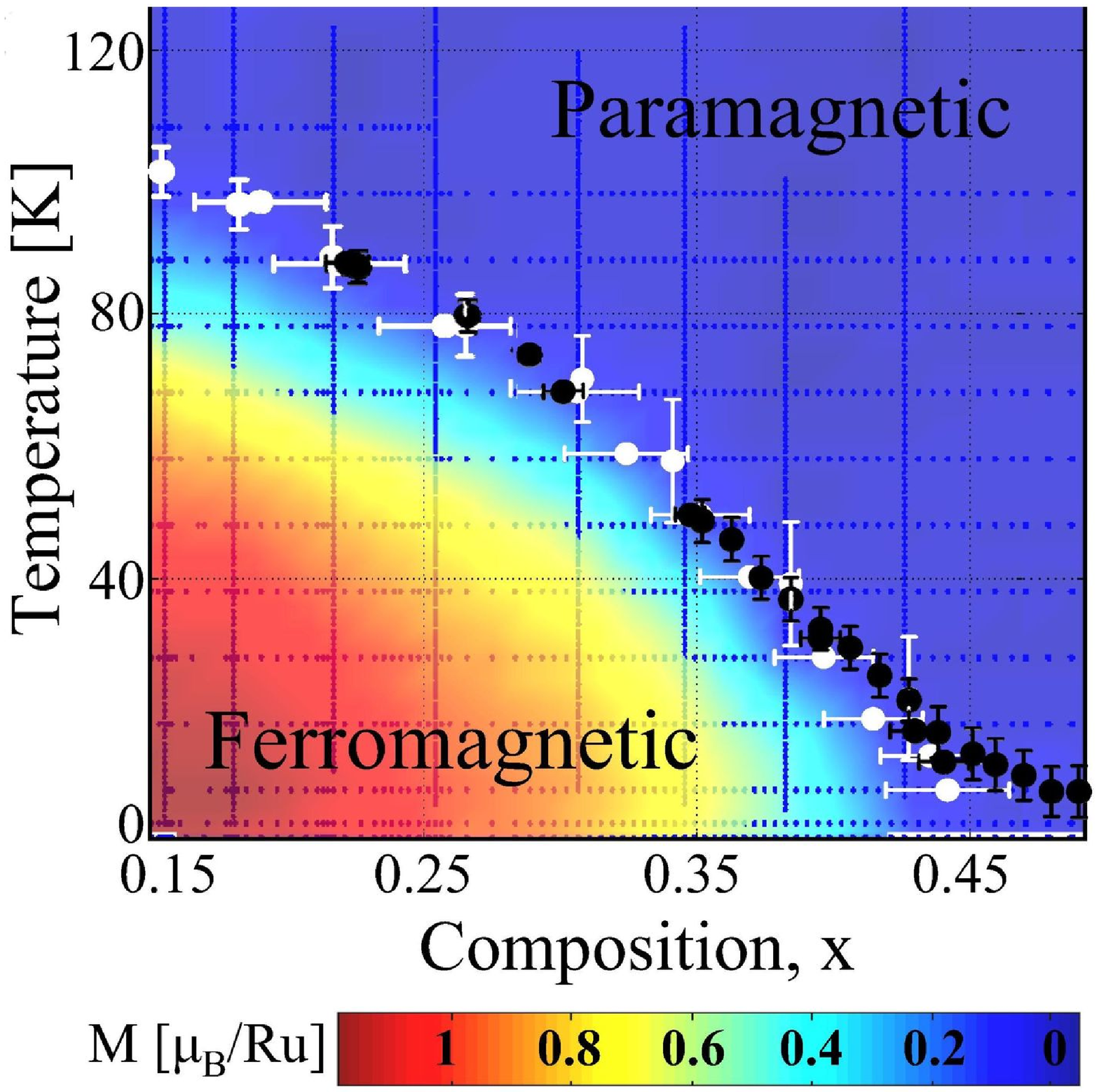}~\includegraphics[width=7.7cm]{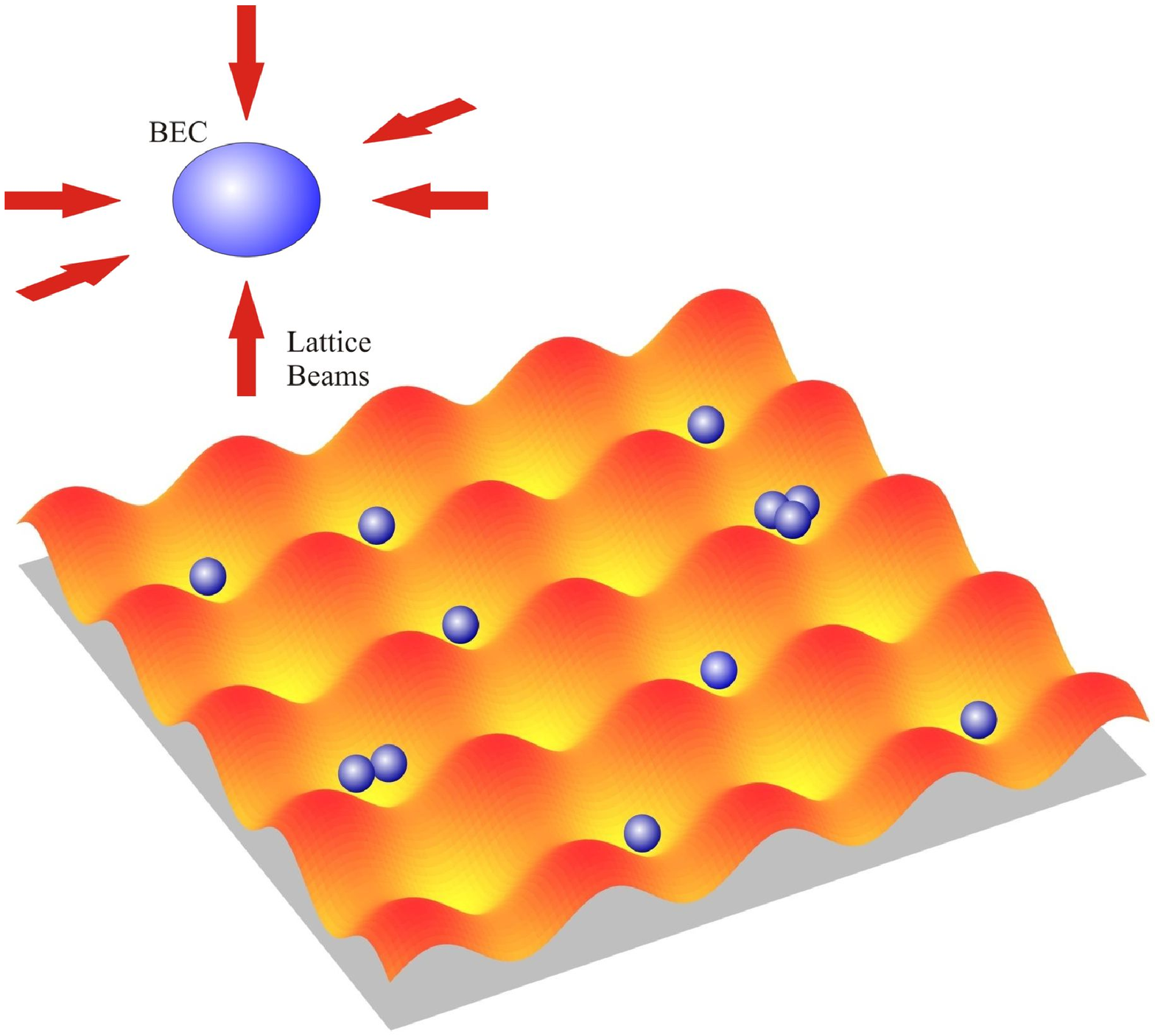}
\caption{Examples of quantum phase transitions. Left: Magnetic phase diagram of the alloy Sr$_{1-x}$Ca$_x$RuO$_3$.
         This material undergoes a quantum phase transition from a ferromagnetic metal to a paramagnetic metal
         as function of the Ca concentration $x$ (after Ref.\ \cite{Demkoetal12}).
         Right: Mott superfluid-insulator transition of an ultracold gas of bosons in an optical lattice produced
         by standing laser waves. If the optical lattice is weak, the ground state is a coherent superfluid. For strong lattice potential,
         the atoms localize in the wells, resulting in an insulating ground state (see Ref.\ \cite{GMEHB02} for an
         experimental realization of this idea).}
\label{fig:QPT_examples}
\end{figure}

\subsubsection*{Quantum-to-classical mapping}

Can we generalize the concepts for the description of phase transitions introduced in the last section
(such scaling and Landau theory) to the case of quantum phase transitions? An important idea that helps
addressing this question is the so-called quantum-to-classical mapping.

Consider the partition function of a classical many-particle system. It can be written as a (high-dimensional)
phase space integral. If the Hamiltonian is the sum of a kinetic part the depends only on momenta
and a potential part that depends only on positions, $H(\mathbf{p},\mathbf{q}) = T(\mathbf{p}) + V(\mathbf{q})$ (as is often the case),
the partition function factorizes according to
\begin{equation}
Z= \int d\mathbf{p} d\mathbf{q} ~\exp[-\beta H(\mathbf{p},\mathbf{q})] =
   \int d\mathbf{p}~ \exp[-\beta T(\mathbf{p})] \int d\mathbf{q}~ \exp[-\beta V(\mathbf{q})]
\label{eq:Z_cl}
\end{equation}
where $\beta=1/(k_B T)$ is the inverse temperature. The kinetic part is (usually) a product of independent Gaussian
integrals; it can not produce any singularities. The phase transition physics must therefore
originate in the configuration part of the partition function. Consequently, only fluctuations in space but not in time
need to be considered in the description of a classical phase transition.\footnote{This argument holds for the
thermodynamics which is governed by the partition function but not for the real-time dynamics at a classical transition.}
This explains why models without any internal dynamics such as the classical Ising model or our classical LGW
theory (\ref{eq:LGW}) correctly describe the thermodynamics of classical phase transitions.

In a quantum system, a factorization of the partition function into kinetic and potential parts
analogous to (\ref{eq:Z_cl}) is generally impossible because the operators of the kinetic energy,
$T(p)$, and the potential energy, $V(q)$, do not commute ($e^{A+B} \ne e^A \,e^B$ if $A$ and $B$ do not commute).
Thus, statics and dynamics are coupled, and we need to treat them on equal footing.
The problem can be addressed
by using the Trotter identity \cite{Trotter59} $e^{A+B} = \lim_{N\to\infty}[ e^{A/N} \,e^{B/N}]^N$, leading to
\begin{equation}
Z = \mathrm{Tr} \exp[-\beta H ] = \mathrm{Tr} \exp[-\beta (T+V) ] = \lim_{N\to \infty} \prod_{n=1}^N
\left( \exp[-\beta T/N] ^{\phantom 1}_{\phantom 1} \exp[-\beta V/N] \right)~.
\label{eq:Z_qu}
\end{equation}
The Trotter decomposition effectively cuts the inverse temperature $\beta$ into many ``slices''.
As $\beta$ plays the role of an imaginary time,\footnote{The Boltzmann factor $\exp(-\beta H)$
looks like a time evolution operator $\exp(-itH/\hbar)$ with imaginary $t$.}
this introduces an extra coordinate $\tau$, the imaginary time coordinate that goes
from 0 to $\beta$ in steps of $\beta/N$. At zero temperature, this direction becomes infinitely long.
We thus conclude that imaginary time plays the role of an additional dimension at a quantum
phase transition. By inserting appropriate resolutions of the unit operator between the factors
on the r.h.s. of (\ref{eq:Z_qu}), one can rewrite the partition function as a functional
integral\footnote{This is analogous to deriving Feynman's path integral for the
propagator in quantum mechanics \cite{FeynmanHibbs_book65}.}
\begin{equation}
Z = \int D[q(\tau)] \exp\left\{-S[q(\tau)]\right\}~.
\label{eq:Z_qu_pi}
\end{equation}
As a result, we arrive at the famous quantum-to-classical mapping which states:
\emph{A quantum phase transition in  $d$ space dimensions is equivalent to a classical
(thermal) phase transition in $d+1$ dimensions.}

This mapping is a very powerful tool
that we will use repeatedly to understand the properties of new quantum phase transitions.
However, it comes with a number of caveats. (i) The mapping works for thermodynamic quantities
only, as it is based on an analysis of the partition function. It is not applicable to
other properties
of quantum phase transitions such as real-time dynamics and transport properties.
(ii) The classical $(d+1)$-dimensional system arising from the mapping can be unusual and
anisotropic. (iii) The mapping only works if the resulting action $S[q(\tau)]$ is real
such that it can be interpreted as a classical free energy functional. As we will see later,
this is the case for some quantum phase transitions while others lead to complex actions, for
example due to Berry phases.\footnote{If the action $S[q(\tau)]$ is complex, the corresponding
Boltzmann factor $\exp\{[-S[q(\tau)]\}$, which acts as statistical weight in the path integral
(\ref{eq:Z_qu_pi}), is not positive definite. This is known as the notorious sign problem;
it strongly hampers quantum Monte-Carlo simulations of such systems.}
In some cases, it actually depends on the details of the
mapping procedure [for instance, on which sets of basis states are used in the decomposition of (\ref{eq:Z_qu})]
whether or not the resulting action is real. A summary of the relations between the quantum system and
its classical counterpart is given in Table \ref{tab:mapping}.
\begin{table}
\renewcommand{\arraystretch}{1.2}
\begin{tabular}{cc}
\hline
\textbf{Quantum System} & \textbf{Classical System}\\
\hline
$d$ space, 1 time dimensions & $d+1$ space dimensions\\
quantum coupling constant & classical temperature $T$\\
inverse physical temperature $1/(k_B T)$ & finite size in the imaginary time direction\\
spatial correlation length $\xi$ & spatial correlation length $\xi$\\
inverse energy gap $1/\Delta$ & correlation length $\xi_\tau$ in the imaginary time direction\\
\hline
\end{tabular}
\caption{Analogies between important quantities in the quantum-to-classical mapping (after Ref. \cite{SGCS97}).}
\label{tab:mapping}
\end{table}

\subsubsection*{Scaling at a quantum critical point}

With the quantum-to-classical mapping in mind, we can generalize Widom's scaling hypothesis
to quantum phase transitions. To do so, we need to include the imaginary time variable
$\tau$ in the scale transformation. In general, space and imaginary time do \emph{not} scale
in the same way. Thus, if we scale lengths according to $L \to b L$,
we need to scale imaginary time by a different factor, i.e., $\tau \to b^z \tau$. (Remember, the
dynamical exponent $z$ relates the divergencies of length and time scales at a critical point.)

Moreover, even though a quantum phase transition occurs at exactly zero temperature, experiments
are performed at nonzero temperatures. It is therefore useful to include the temperature as an
independent external parameter from the outset.\footnote{Under the quantum-to-classical mapping, the
inverse temperature corresponds to the system size in imaginary time direction. Temperature scaling
at a quantum critical point is thus equivalent to finite size scaling \cite{Barber_review83} in the corresponding (mapped)
classical system.}
 As the temperature is an energy, it scales like
an inverse time, i.e., $T \to b^{-z} T$.
Consequently, the scaling form of the free energy density at a quantum critical point reads
\begin{equation}
f(r,h,T) = b^{-(d+z)} f(r\,b^{1/\nu}, h\,b^{y_h},T\,b^z)~.
\label{eq:q-scaling}
\end{equation}
Note that $r$ measures the distance from the quantum critical point along the quantum
tuning parameter axis (for example the transverse magnetic field axis in the case of the transverse-field
Ising model), it is not related to the temperature.
Scaling forms of thermodynamic observables can again be obtained by taking the appropriate
derivatives; and those of other quantities (such as correlation functions) can be constructed
analogously.

Along the same lines, one can also generalize the Landau-Ginzburg-Wilson (LGW) order parameter
field theory (\ref{eq:LGW}) to quantum phase transitions. Since we need to include fluctuations
in space and (imaginary) time, the order parameter $m$ becomes a function of $\mathbf{x}$ and $\tau$.
Expanding in powers of $m$ as well as in gradients and time derivatives gives the LGW action or
free energy
\begin{equation}
S[m({\mathbf x},\tau)] = \int d^d x d\tau \left[-h\, m + r\, m^2
    + (\nabla\, m)^2 + (1/c^{2}) (\partial m/\partial \tau)^2 + u\, m^4 + \ldots \right]~.
\label{eq:q-LGW}
\end{equation}
Here, $c$ plays the role of a propagation speed for the order parameter fluctuations.
It must be emphasized that (\ref{eq:q-LGW}) is just the simplest example of a quantum
LGW theory. It applies. e.g., to the ferromagnetic transition in the transverse-field
Ising model. Many other quantum phase transitions have more complicated LGW theories.
In particular, if the system contains soft (gapless) excitations other than the order
parameter fluctuations, the expansion in powers of time derivatives generally breaks
down. The resulting time/frequency dependence then becomes nonanalytic
\cite{VBNK96,BelitzKirkpatrickVojta02,BelitzKirkpatrickVojta05}. We will see examples
of this behavior in later sections (when we discuss the effects of dissipation).

\subsubsection*{Phase diagram close to a quantum critical point}

The phase diagram of a system close to a quantum critical point is very rich;
it contains several qualitatively different regions depending on the presence
or absence of long-range order as well as the character of the fluctuations.
In particular, it is important to distinguish thermal and quantum fluctuations.
A fluctuation is of (predominantly) thermal character if its
frequency $\omega_c$ is below the thermal energy, $\hbar \omega_c \ll k_B T$.
In the opposite case, $\hbar \omega_c \gg k_B T$, the fluctuation is of quantum
character.

Figure \ref{fig:schema1} shows a schematic of such a phase diagram. The disordered
phase consists of three different regions.
\begin{figure}
\includegraphics[width=8.5cm]{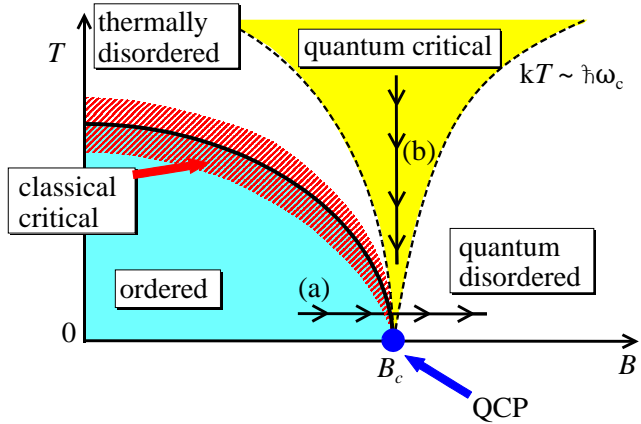}
\caption{Schematic phase diagram in the vicinity of a quantum critical point (QCP).
The thick solid line marks the boundary
     between the ordered and disordered phases. The dashed lines indicate the crossover
     between predominantly quantum or classical character of the fluctuations.
     In the hatched classical critical region, quantum mechanics is unimportant,
     and the leading classical critical singularities can be observed. Experiments
     following path (a) and (b) are discussed in the text.
}
\label{fig:schema1}
\end{figure}
In the thermally disordered region,
located on the ordered side of the quantum critical point but at high temperatures,
the long-range order is destroyed by thermal fluctuations. In the quantum disordered
region located at the disordered side of the quantum critical point at low temperatures, quantum fluctuations
destroy the long-range order while thermal effects are unimportant (even the ground state
is disordered). Between these regions is the so-called quantum critical region,
located at $B \approx B_c$ and, somewhat counter-intuitively, at comparatively high
temperatures where the character of the fluctuations is thermal. In this regime,
the system is critical with respect to $B$, and the critical singularities are cut-off
exclusively by the temperature.

It is important to note that the \emph{asymptotic} critical behavior at any nonzero temperature
is dominated by thermal fluctuations because the characteristic frequency $\omega_c \sim \xi_\tau^{-1}$ vanishes due to critical
slowing down. However, the classical critical region around the phase boundary
becomes very narrow at low temperatures. Another way to understand this result is to
compare the correlation time $\xi_\tau$ with the system size $\beta=1/(k_B T)$ in imaginary time
direction. Upon approaching the phase boundary, $\xi_\tau$ diverges. A any nonzero temperature,
it thus eventually becomes larger than the system size $\beta$. At this point, the imaginary
time dimension drops out, and the critical behavior becomes classical.

We thus conclude that the \emph{asymptotic} critical behavior is quantum only at precisely
zero temperature while it is classical for any nonzero temperature. This justifies
calling all finite-temperature phase transitions classical even if the underlying
microscopic physics is quantum. Nonetheless, the quantum
critical point controls large portions of the phase diagram at nonzero temperatures. An experiment
along path (a) in figure \ref{fig:schema1} will see the classical behavior only in an extremely narrow
range around the transition. Moreover, an experiment along path (b) in the quantum critical region
explores the temperature scaling of the quantum critical point.

\section{Phase transitions in disordered systems}
\label{sec:disorder}

In this section, we
consider various types of disorder or randomness and derive criteria that govern under what
conditions they can influence a phase transition.

\subsection{Types of disorder}
\label{subsec:types}

\subsubsection*{Quenched vs. annealed disorder}

Disorder\footnote{Unfortunately, the word ``disorder'' is used in two different meanings in the field of
phase transitions. First, it is used to characterize a phase without long-range order in
the Landau sense. In the case of the ferromagnetic transition, for example, we
call the paramagnetic phase a disordered phase. Second, the word ``disorder'' denotes
imperfections or randomness of the underlying system. This is the meaning here.}
 or randomness in a physical system can have many different origins including
vacancies or impurity atoms in a crystal as well as extended defects of the crystal lattice such as
dislocations or grain boundaries.
The solid could also be amorphous rather than crystalline.
In ultracold atom systems, disorder can be realized, for example, using speckle laser light.

To understand the physical consequences of such disorder, it is important to distinguish quenched and annealed
disorder. Quenched or frozen-in disorder is static; this means that the impurities or defects do not move or
change over the  relevant experimental time scales. In contrast, annealed disorder fluctuates during
the duration of the experiment. From a conceptually point of view, annealed disorder is  easier to deal with.
If the disorder degrees of freedom (for example, the impurity positions) are in thermal equilibrium over
the experimental time scales, they can simply be included in the usual statistical mechanics description of
the system. In other words, the thermodynamics of a system with annealed disorder is obtained by averaging the partition function
over the impurity degrees of freedom, $Z_{av} = [Z]_{dis}$ where $[\ldots]_{dis}$ denotes the disorder average.

In contrast, every sample, i.e., every disorder realization, is different in the case of quenched disorder.
To obtain average thermodynamic quantities one thus needs to average the free energy or, equivalently, the logarithm
of the partition function, $\ln Z_{av} = [\ln Z]_{dis}$.\footnote{A complete description of a large ensemble of
samples strictly requires working out the probability distributions of observables rather than just the average.
This will become an important point later in this article.}
Technically, averaging $\ln Z$ is much harder than averaging $Z$ itself; and this makes the physics of
quenched disordered systems a difficult research area. One way to overcome this difficulty is the so-called
replica trick \cite{EdwardsAnderson75}. It involves writing $\ln Z = \lim_{n\to 0} (Z^n -1)/n$ and averaging
$Z^n$ before taking the limit $n\to 0$. This method has met some success, for example, in the theory of spin
glasses. However, exchanging the average and the $n\to 0$ limit is mathematically problematic; and the replica
approach is known to fail in some cases. We will not be using this method.

In the remainder of these lectures, we will not consider annealed disorder but focus on the more interesting (and complicated) case of quenched
disorder.

\subsubsection*{Random mass, random fields, and all that}

In Section \ref{sec:PT}, we have seen that the qualitative properties of phase transitions
depend on the space dimensionality and on symmetries but not on microscopic details. This suggests
that we should classify the various kinds of quenched disorder according to their symmetries.

Consider, for example, a clean (three-dimensional) classical ferromagnet described an Ising Hamiltonian
\begin{equation}
H = -J \sum_{\langle ij\rangle} S_i S_j
\label{eq:clean_Ising}
\end{equation}
with spins $S_i = \pm 1$. The sum goes over pairs of nearest neighbors on a cubic lattice, and $J>0$ is the
exchange interaction. How could disorder change this Hamiltonian? One possibility is that the exchange interaction
becomes nonuniform and varies randomly from place to place. (In a real system this could be due to
nonmagnetic impurity atoms modulating the distances between the spins.) The disordered Hamiltonian reads
\begin{equation}
H = - \sum_{\langle ij\rangle} J_{ij}\, S_i S_j
\label{eq:random_mass_Ising}
\end{equation}
where the $J_{ij}$ are random variables. As long as all $J_{ij}$ remain positive, such randomness is a
weak and rather benign type of disorder. It does \emph{not} change the two bulk phases:
At sufficiently low temperatures, the system is still a ferromagnet, and at sufficiently high temperatures
it is still a paramagnet. Moreover the up-down spin symmetry is \emph{not} broken by the disorder.
The disorder just changes the local tendency towards ferromagnetism, in other words it changes the
``local critical temperature.'' Consequently, this type of disorder is often called random-$T_c$ disorder.
In a LGW description of the transition, such disorder would appear as a random variation
$\delta r(\mathbf{x})$ of the distance from criticality,
\begin{equation}
F[m({\mathbf x})] = \int d^d x \left\{-h\, m({\mathbf x}) + [r+\delta r(\mathbf{x})]\, m^2({\mathbf x})  + (\nabla\, m({\mathbf x}))^2 +u\, m^4({\mathbf x}) + \ldots \right\}~.
\label{eq:LGW_random_mass}
\end{equation}
The disorder couples to the $m^2$ term in the LGW free energy functional. In quantum field theory, this term
is usually called the mass term. Therefore, random-$T_c$ disorder is also called random-mass disorder.
(In addition to random exchange couplings, random-mass disorder can also be realized by random dilution
of the spins.)

Alternatively, we could imagine disorder that couples linearly to the order parameter $m$ rather than $m^2$.
In our example ferromagnet, this corresponds to a magnetic field that varies randomly from site to site.
The LGW theory of such a system reads
\begin{equation}
F[m({\mathbf x})] = \int d^d x \left\{-h(\mathbf{x})\, m({\mathbf x}) + r\, m^2({\mathbf x})  + (\nabla\, m({\mathbf x}))^2 +u\, m^4({\mathbf x}) + \ldots \right\}~
\label{eq:LGW_random_field}
\end{equation}
where $h(\mathbf{x})$ is the random variable. This type of disorder is called random-field disorder. It locally
breaks the up-down spin symmetry. Whether or not the symmetry is broken globally depends on the probability distribution
of the random fields. A particularly interesting situation arises if the distribution is even in $h$ such
that the up-down symmetry is globally preserved in the statistical sense. As we will see later in this section,
random-field disorder is generally stronger than random-mass disorder.

Many other types of disorder are possible. For example, in a Heisenberg magnet where the spin variables are
three-dimensional unit vectors, the disorder could break the rotational symmetry in spin space in a random
fashion. This  defines random-anisotropy disorder. In a superconductor or superfluid, the disorder could lead to
random phases of the complex order parameter. If the interactions in a spin system pick up random signs, the system
becomes frustrated which often results in spin-glass behavior \cite{FischerHertz_book91}.

In these lectures, we will mostly consider the cases of random-mass and random-field disorder.

\subsubsection*{Important questions}

If quenched disorder is added to a clean system undergoing a phase transition, a hierarchy of
important questions arises naturally:
\begin{itemize}
\itemsep2pt
\item Are the bulk phases (the phases separated by the transition) qualitatively changed by the disorder?
\item Is the phase transition still sharp, or is it smeared because different parts of the system undergo
      the transition independently?
\item If the phase transition is still sharp, does its order (first order vs. continuous) change?
\item If the phase transition remains continuous, does the critical behavior, i.e., the values of the critical exponents,
      change?
\end{itemize}

\subsection{Random-mass disorder and the Harris criterion}

In this section, we consider a clean system undergoing a continuous phase transition
(a critical point) and ask whether or not random-mass disorder changes the properties of the
phase transition. (We already know from the discussion in the last section that the bulk phases will
not change.)

Harris addressed this question in a famous paper \cite{Harris74} published in 1974.
Specifically, he found a powerful criterion for the stability of a clean critical point
against random-mass disorder. To derive this criterion, imagine that we are at a temperature
somewhat above the (global) critical temperature $T_c$ of the system. We can divide the system
into blocks whose linear size is the correlation length $\xi$ (see Fig.\ \ref{fig:Harris}).
\begin{figure}
\includegraphics[width=5.cm]{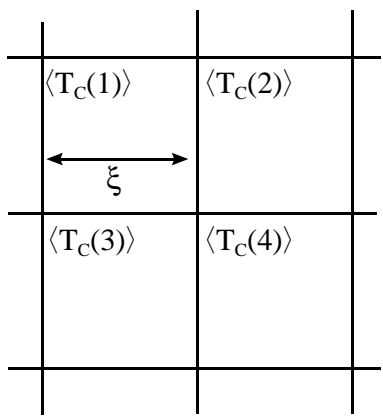}
\caption{Derivation of the Harris criterion. The system is divided into blocks of linear
         size $\xi$. Due to the disorder, each block has its own local ``critical temperature.''}
\label{fig:Harris}
\end{figure}
The spins within each block are effectively parallel and fluctuate together as a large
superspin. Due to the disorder, each block $i$ has its own ``local critical temperature''
$T_c(i)$ which is determined by the values of the interactions in this block.\footnote{It is
important to realize that, generally, finite-size blocks cannot undergo a true phase transition
at their respective local critical temperature $T_c(i)$. Instead, $T_c(i)$ marks the point
where the spins in the block lock together to form the superspin.}
Harris' idea consists in comparing the variations $\Delta T_c$ of the local critical temperature
from block to block with the distance $T-T_c$ from the global phase transition point. If
$\Delta T_c < T-T_c$, the blocks are all on the same side of the phase transition, and the system
is more or less uniform. In contrast, for $\Delta T_c > T-T_c$, some blocks are in the disordered
(paramagnetic) phase and some are in the ordered (ferromagnetic) phase, making a uniform transition
impossible.

For the clean critical behavior to be stable we thus must have $\Delta T_c < T-T_c$ as we approach
the transition, i.e., for $\xi \to \infty$. The dependence of $\Delta T_c$ on $\xi$ can be estimated
using the central limit theorem. As the local $T_c(i)$ is determined by an average of a large number
of random variables in the block [for example, the random $J_{ij}$ in the Hamiltonian (\ref{eq:random_mass_Ising})],
its variations decay as the square root of the block volume,
\begin{equation}
\Delta T_c \sim \xi^{-d/2}~.
\end{equation}
On the other hand, according to (\ref{eq:xi}), the global distance from criticality is related to $\xi$ via
\begin{equation}
T-T_c \sim \xi^{-1/\nu}~.
\end{equation}
The condition $\Delta T_c < T-T_c$ for $\xi \to \infty$ therefore
leads to the exponent inequality
\begin{equation}
d\nu >2 ~.
\label{eq:Harris}
\end{equation}
This is the famous Harris criterion for the stability of a clean critical point.

Let us discuss the interpretation of this inequality in more detail. If the
Harris criterion $d\nu >2$ is fulfilled, the ratio $\Delta T_c / (T-T_c)$ goes to zero
as the critical point is approached. This means, the system looks less and less
disordered on larger length scales; and the effective disorder strength vanishes
right at criticality. Consequently, the disordered system features the same critical
behavior as the clean one. This also implies that all thermodynamic observables
are self-averaging, i.e., their values are the same for all disorder realizations.
An example of a transition that fulfills the Harris criterion is the ferromagnetic transition
in a three-dimensional classical Heisenberg model. Its clean correlation
length exponent is $\nu \approx 0.69 > 2/d =2/3$.

In contrast, if $d\nu < 2$, the ratio $\Delta T_c / (T-T_c)$ increases upon
approaching the phase transition. This means, the blocks become effectively
more different on larger length scales. Eventually, some blocks will be on
one side of the transition while other blocks are on the other side. This makes
a uniform sharp phase transition impossible. The character of the transition must
therefore change, and the clean critical behavior is unstable. In the
marginal case $d\nu = 2$, more sophisticated methods are required to decide the
stability of the clean critical point.

What is the fate of the transition if the Harris criterion is violated? The Harris criterion itself
cannot answer this question, but research over the last four decades has established a number of
possible scenarios. In the simplest case, the phase transition remains sharp and continuous
but the critical behavior changes. The disordered system is in a new universality class
featuring a correlation length exponent that fulfills the inequality $d\nu >2$.
In this case, the disorder strength remains finite at large length scales, and
observables at the critical point are \emph{not} self-averaging. Many phase transitions
in classical disordered systems follow this scenario, for example the three-dimensional
classical Ising model. Its clean correlation length exponent is $\nu \approx 0.63$ which violates
the Harris criterion. In the presence of random-mass disorder, the critical behavior
therefore changes with the disordered correlation length exponent taking a value of
about 0.68.

Quantum phase transitions in disordered systems often display more exotic behavior
(more dramatic changes than just different critical exponent values). We will explore
several such phenomena in Sections \ref{sec:SDRG} to \ref{sec:smeared}.

Let us finish this section with a number of important remarks. (i) The Harris criterion holds in
the same form, $d\nu > 2$ for both classical (thermal) and quantum phase transitions. Specifically,
this means that the dimensionality $d$ should \emph{not} be replaced by $(d+1)$ or $(d+z)$
for a quantum phase transition (as one might have guessed from the quantum-to-classical mapping).
The reason is that the $d$ in the Harris criterion stems from averaging the disorder degrees of freedom
using the central limit theorem. As quenched disorder is time-independent, the number of random
variables in a block scales as $\xi^{d}$ for both classical and quantum phase transitions.
This insight also implies that the Harris criterion needs to be modified for disorder that is perfectly correlated
in one or more dimensions such as line or plane defects. In these cases, $d$ needs to be replaced by the number $d_r$ of
dimensions in which there actually is randomness, i.e., $d_r=d-1$ for line defects and $d_r=d-2$
for plane defects. This greatly enhances the relevance of such perfectly correlated disorder.

(ii) The Harris criterion $d\nu > 2$ applies to uncorrelated or short-range correlated disorder.
If the disorder displays long-range correlations in space, the inequality needs to be modified
because the central-limit theorem estimate of $\Delta T_c$ changes. Weinrib and Halperin
\cite{WeinribHalperin83} considered the case of power-law correlations that fall off as
$|\mathbf{x}-\mathbf{x}'|^{-a}$. They found that the Harris criterion does not change for $a>d$.
If $a<d$, the inequality needs to replaced by $a\nu > 2$. Thus, the criterion reads
\begin{equation}
\min(d,a) \nu >2 ~.
\label{eq:Winrib-Halperin}
\end{equation}
This implies that long-range correlated disorder is more dangerous for a clean critical point than
short-range correlated disorder. In other words, a given clean critical point may be stable against
uncorrelated disorder but unstable against disorder with sufficiently long-ranged
correlations.\footnote{At first glance, this result may appear counter-intuitive because long-range
correlated disorder looks ``less random''. The crucial point is, however, that the long-range
correlations favor the appearance of large fluctuations.}

(iii) The Harris inequality $d\nu > 2$ involves the \emph{clean} correlation length exponent. It does
not put a bound on the correlation length exponent of the disordered system. However, Chayes et al.
\cite{CCFS86} showed (under some mild assumptions) that the correlation length exponent of a
critical point in a disordered system has to fulfill the same inequality $d\nu \ge 2$. Counter examples
to this results are sometimes reported in the literature. In is not always clear whether they are
genuine violations of the theorem or whether they occur due to not fulfilling the underlying assumptions or maybe
because the reported exponents are not in the true asymptotic critical region.

\subsection{Random-field disorder and the Imry-Ma argument}

In this section, we investigate a system undergoing a phase transition in the presence of random-field disorder.
As an example, consider a (three-dimensional) Ising ferromagnet in a random magnetic field. Its Hamiltonian
reads
\begin{equation}
H= -J\sum_{\langle ij \rangle} S_i S_j - \sum_i h_i S_i~.
\label{eq:Ising_rf}
\end{equation}
Here, the $h_i$ are independent random variables with zero mean, $[h_i]_{dis}=0$ and variance
$[h_ih_j]_{dis}=W \delta_{ij}$.

As discussed in Sec.\ \ref{subsec:types}, the random field locally breaks the up-down
symmetry of the model. Spins on sites with positive random field, $h_i>0$,
prefer pointing up ($S_i =1$) while spins on sites with negative random field, $h_i<0$,
prefer pointing down ($S_i =-1$).
This raises the question: Does a ferromagnetic phase (in which all spins
align in the same direction) still exist in the presence
of the random field? For strong random fields, this question can be easily
answered in the negative. If $W \gg J^2$, almost all spins gain more energy from aligning
with their local random fields rather than with each other. Thus,
ferromagnetic order is impossible. What about weak random fields,
$W \ll J^2$? In this case almost all spins will prefer to align with their
neighbors rather than the field. One might be tempted to conclude that
ferromagnetism survives in this case. However, this would be premature because
long-range ferromagnetic order can be destroyed by the formation of domains.

Imry and Ma \cite{ImryMa75} developed a beautiful heuristic argument to test the
stability of the ferromagnetic state against domain formation. Consider a single
uniform spin-up domain of linear size $L$ (located in a region where the random field is mostly positive)
embedded in a spin-down bulk system (see Fig.\ \ref{fig:Imry-Ma}).
\begin{figure}
\includegraphics[width=6.5cm]{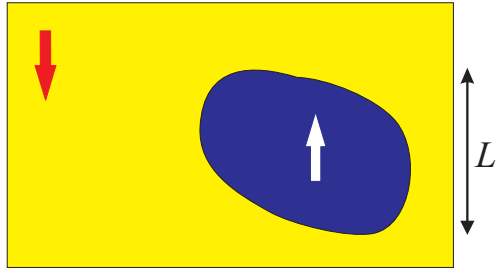}
\caption{To derive the Imry-Ma criterion, one needs to compare the energy gain of a domain (of size $L$)
         due to aligning with the average random field with the energy loss due to the domain wall.}
\label{fig:Imry-Ma}
\end{figure}
According to Imry and Ma, we need to compare the energy gain due to aligning the domain with the
average local random field with the energy loss due to the formation of the domain wall.
In $d$ space dimensions, the domain wall is a $(d-1)$-dimensional surface. Each bond crossing
this surface has antiparallel spins and thus energy $+J$ rather than $-J$; it therefore contributes $2J$
to the domain wall energy. As the number of such bonds is proportional to the domain wall area,
we can estimate the domain wall energy to be
\begin{equation}
\Delta E_{DW} \sim J \, L^{d-1}~.
\end{equation}
The energy gain due to aligning the uniform spin-up domain with the random field is simply given by the sum
over all random field values in the domain, $\Delta E_{RF} = -\sum_i' h_i$. As the $h_i$ are independent
random variables, the typical value of this sum can again be estimated by means of the central limit theorem,
giving
\begin{equation}
|\Delta E_{RF}| \sim W^{1/2} L^{d/2}~.
\end{equation}
Imry and Ma now observe that the uniform ferromagnetic state will be stable against the formation
of domains if $|\Delta E_{RF}| < \Delta E_{DW}$ for all possible domain sizes $L$.
Inserting the above estimates, this translates to $W^{1/2} L^{d/2} < J L^{d-1}$. In the case of weak
random field, $W \ll J^2$, this is certainly fulfilled for small domains containing just a few spins.
Whether or not it also holds for large domains depends on the dimensionality $d$. If $d>2$,
the domain wall energy increases faster with $L$ than the random field energy. Thus, domain formation
is always unfavorable, and the ferromagnetic state is stable against weak random-field disorder.

In contrast, for $d<2$, the random field energy increases faster with $L$ than the domain wall energy.
Even for weak random fields, there will be a critical $L$ beyond which forming domains that align
with the local random field becomes favorable. Consequently, the uniform ferromagnetic state is unstable
against domain formation for arbitrary random field strength. In other words, in dimensions $d<2$.
random-field disorder prevents spontaneous symmetry breaking. Analyzing the marginal case, $d=2$, again requires
more sophisticated methods.

Building on this heuristic argument,
Aizenman and Wehr \cite{AizenmanWehr89} proved rigorously that random fields destroy long-range order (and thus prevent spontaneous symmetry
breaking) in all dimensions $d \le 2$ for discrete (Ising) symmetry and in dimensions $d \le 4$ for
continuous (Heisenberg) symmetry. The difference between discrete and continuous symmetry can be easily
understood by comparing the domain walls (see Fig.\ \ref{fig:domain_wall}).
\begin{figure}
\includegraphics[width=9cm]{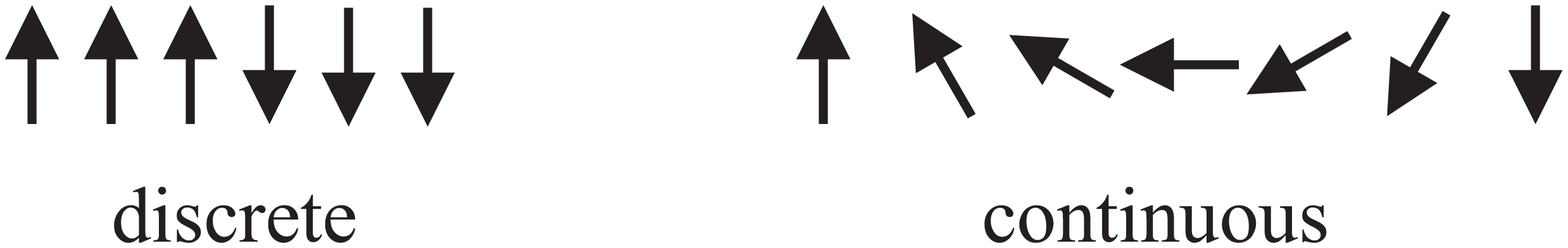}
\caption{For discrete (Ising) symmetry, a domain wall is a ``sharp'' up-down spin flip. For
        continuous symmetry, the domain wall consists of a smooth change of the spin orientation
        over some distance.}
\label{fig:domain_wall}
\end{figure}
In the discrete case, a domain wall consists of a sharp up-down spin flip at the surface of the
domain. As estimated above, its energy is proportional to the surface area. i.e., it scales as
$L^{d-1}$ with the domain size $L$. In the continuous case, the domain wall can be spread out
over the entire domain, i.e., over a length of order $L$. The domain wall energy can be estimated within
a LGW theory. The gradient term in (\ref{eq:LGW}) contributes $\Delta E_{DW} \sim L^d (\nabla m)^2 \sim
L^d (1/L)^2 \sim L^{d-2}$. The domain wall energy thus increases less rapidly with $L$ in the
continuous case, making the system more susceptible towards domain formation.
Inserting $\Delta E_{DW} \sim L^{d-2}$ into the Imry-Ma argument gives a critical dimensionality
of 4 rather than 2.

The above results hold for uncorrelated or short-range correlated random fields. Long-range correlated
random fields with correlations that fall off as $|\mathbf{x}-\mathbf{x}'|^{-a}$ modify the Imry-Ma argument
because the energy gain due to aligning the domain with the local random field changes, provided $a<d$ \cite{Nattermann83}.
In this case the uniform ferromagnetic state is stable for $a>2$ while domain formation is favored for $a<2$.
In the opposite case, $a>d$, the uncorrelated Imry-Ma argument is unmodified.

As in the case of the Harris criterion, the Imry-Ma argument gives identical results for classical and quantum phase
transitions because the dimensionality enters via the number of dimensions in which there
is randomness. This means only space dimensions count but not the imaginary time dimension.

\subsection{Destruction of first-order phase transitions by disorder}

Reasoning very similar to the Imry-Ma argument can be used to attack a different problem,
viz., the fate of a first-order phase transition in the presence of random-mass
disorder.

Remember, a first order phase transition is characterized by macroscopic phase coexistence
at the transition point.\footnote{Actually, the random-field problem considered in the last
subsection can be viewed in the same way. Imagine adding an extra uniform field $h$ to the random-field
Hamiltonian (\ref{eq:Ising_rf}). As $h$ is tuned through zero at low temperatures, the system undergoes
a first-order phase transition from the spin-down phase to the spin-up phase.}
 For example, at the liquid-gas phase transition of a fluid,
a macroscopic liquid phase coexists with a macroscopic vapor phase. Random-mass disorder
locally favors one phase over the other. We can thus pose the same question we asked in the
last section: Will the macroscopic phases survive in the presence of the disorder or
will the system form domains (droplets) that follow the local value of the random-mass
term (the ``local $T_c$'')?

This question can be answered by adapting the Imry-Ma argument to the problem
at hand \cite{ImryWortis79,HuiBerker89,AizenmanWehr89}. Consider a single domain or droplet (of linear size $L$) of one phase
embedded in the other phase. The free energy cost due to forming the surface is
\begin{equation}
\Delta F_{surf} \sim \sigma L^{d-1}
\end{equation}
where $\sigma$ is the surface energy between the two phases.\footnote{The surface term
scales as $L^{d-1}$ rather than $L^{d-2}$ because the two phases are generally \emph{not}
connected by a continuous transformation.}
 The energy gain from the
random-mass disorder can be estimated via the central limit theorem, resulting in a
typical magnitude of
\begin{equation}
|\Delta F_{dis}| \sim W^{1/2} L^{d/2}
\end{equation}
where $W$ is the variance of the random-mass disorder.

The macroscopic phases are stable if $|\Delta F_{dis}| < \Delta F_{surf}$. Using the same
reasoning as in the last subsection, this means that macroscopic phase coexistence is impossible
in dimensions $d\le 2$ no matter how weak the disorder is. In dimensions $d>2$,
phase coexistence is possible for weak disorder but will be destabilized for sufficiently strong
disorder.

We thus conclude that random-mass disorder destroys first-order phase transitions in
dimensions $d\le 2$. As before, this result holds for both classical (thermal) and quantum
phase transitions \cite{GreenblattAizenmanLebowitz09} as long as the disorder is uncorrelated
or short-range correlated in space. Long-range interactions can be taken into account as
in the random-field case in the last section.

What is the ultimate fate of the phase transition if the first-order character is destroyed by
the disorder? The Imry-Ma argument cannot answer this question. In many examples, the
first-order transition is replaced by (``rounded to'') a continuous one, but more complicated
scenarios cannot be excluded.

\section{Strong-disorder renormalization group}
\label{sec:SDRG}

After the introductory sections \ref{sec:PT} and \ref{sec:disorder}, we now
turn to the main topic of this article, quantum phase transitions in disordered
systems, taking the transverse-field Ising model as a paradigmatic
example. We introduce a very powerful technique to attack this problem,
the strong-disorder renormalization group (SDRG) which has become a standard
tool in this field.

\subsection{Random transverse-field Ising chain}
\label{subsec:RTIM}

We have already encountered the transverse-field Ising Hamiltonian (\ref{eq:TFIM}) in Sec.\ \ref{subsec:QPT_intro}
as a toy model for the quantum phase transition in LiHoF$_4$. Here, we will be considering a
one-dimensional disordered version of this model, the random transverse-field Ising chain. It is given
by
\begin{equation}
H = -\sum_{i} J_i \sigma_i^z \sigma_{i+1}^z - \sum_i B_i \sigma_i^x~,
\label{eq:RTIM}
\end{equation}
The interactions $J_i$ and the transverse fields $B_i$ are independent
random variables with probability distributions $P_I(J)$ and $R_I(B)$,
respectively. Both the $J_i$ and the $B_i$ are restricted to positive
values.\footnote{We denote the transverse field by $B$ to distinguish
it from a possible longitudinal  (conjugate to the order parameter)
field $h$. However, in the literature on the random
transverse-field Ising model, $h$ is often used for the transverse field.}

The qualitative features of the zero-temperature (ground state) phase diagram
of the transverse-field Ising chain are easily discussed. If the typical
interactions are much larger than the typical transverse fields, $J \gg B$,
the ground state displays ferromagnetic long-range order in the $z$ direction. In the opposite limit,
$J \ll B$, the ground state is a (field-polarized) quantum paramagnet. These two phases are
separated by a quantum phase transition at $J \sim B$. In fact, it has been
shown rigorously \cite{Pfeuty79} that the transition occurs when
$\prod_i J_i = \prod_i B_i$.\footnote{The location of the quantum phase transition point in the
random transverse-field Ising chain is fixed by the self-duality of the model:
The form of the Hamiltonian is invariant under the duality transformation
$\sigma_{i}^x = \mu_{i}^z \mu_{i+1}^z$
$\sigma_{i}^z = \prod_{(j\le i)} \mu_{j}^x$,
(where $\mu^x$ and $\mu^z$ are the dual Pauli matrices).
However, interactions and fields exchange their meaning,
 $J_i \rightleftarrows B_i$, under this transformation}

Traditional approaches to investigating the quantum phase transition in the
random transverse-field Ising chain would start by solving the clean problem,
$J_i \equiv J$, $B_i \equiv B$. The randomness would then be treated in a perturbative
fashion, maybe using the replica trick to carry out the disorder average.
Such approaches have been successful for classical (thermal) phase transitions
in disordered systems. However, they work very poorly for the quantum phase transition in
the random-transverse
field Ising model and many other quantum phase transitions.\footnote{Sometimes,
perturbative methods show the disorder strength diverging and thus signal their
own breakdown \cite{NVBK99a,NVBK99b}.}
Later, we will understand the deeper reasons for this; basically, disorder effects turn
out to be much stronger at quantum phase transitions than at classical ones.

In the next subsection we therefore introduce an alternative method that makes use
of the disorder from the outset rather than treating it as a small perturbation
of a clean system.

\subsection{Renormalizing the random transverse-field Ising chain}

\subsubsection*{Basic idea}

The strong-disorder renormalization group was proposed in 1979 by
Ma, Dasgupta, and Hu \cite{MaDasguptaHu79} to study random antiferromagnetic
spin chains. It was greatly developed by D.S.\ Fisher in the mid 1990's
\cite{Fisher92,Fisher94,Fisher95} who pioneered its application to
quantum phase transitions. Since then, the method has been employed in
a broad variety of problems ranging from disordered quantum systems to
classical nonequilibrium phase transitions (for a review, see, e.g.,
Ref.\ \cite{IgloiMonthus05}).

The basic idea of the strong-disorder renormalization group consists in identifying
the largest local energy scale in the entire system, i.e., the highest local
excited state. As this excited state does not contribute to the low-energy physics
important for the phase transition, it is integrated out, treating the neighboring
energies (couplings) as perturbations. This works well if the disorder is strong
[the distributions $P_I(J)$ and $R_I(B)$ are broad] because in this case, the largest
local energy will be much larger than the neighboring ones.
The method thus becomes controlled in the limit of strong disorder
which gives the strong-disorder renormalization group its name.

The process of eliminating the largest local energy is now iterated; this gradually reduces the maximum energy
in the system until the desired low-energy description of the phase transition
physics is achieved.

\subsubsection*{Renormalization group recursions}

Let us implement the strong-disorder renormalization group for the random transverse-field
Ising chain. The competing local energies in the Hamiltonian (\ref{eq:RTIM}) are the
interactions $J_i$ and the transverse fields $B_i$. We thus identify the largest local energy scale
as $\Omega = \max(J_i, B_i)$. To integrate out the corresponding excited states, we need to distinguish
$\Omega$ being an interaction and $\Omega$ being a field.

(i) If the largest local energy is a field, say $B_3$, the spin $\sigma_3$ is pinned in the
positive $x$ direction (in the $\sigma^x$ eigenstate $|\to\rangle$ with eigenvalue +1).
It does therefore not contribute to the $z$ magnetization and can be decimated (eliminated) from the system.
However, virtual excitations of $\sigma_3$ from $|\to\rangle$ to $|\leftarrow\rangle$
generate an effective coupling $\tilde J$ between the neighboring spins $\sigma_2$ and
$\sigma_4$.\footnote{If we simply eliminate $\sigma_3$ without taking the virtual excitations
into account, the chain would be cut into two independent pieces because spins the $\sigma_2$ and
$\sigma_4$ would not be coupled at all. Ferromagnetic long-range order would thus be impossible.
This is clearly too rough an approximation.}
To calculate $\tilde J$, we can consider the three-site system consisting of spins $\sigma_2$,
$\sigma_3$, and $\sigma_4$ with Hamiltonian
\begin{equation}
H=H_0 + H_1 \quad  \textrm{with} \quad  H_0 = -B_3 \sigma_3^x ~, \quad H_1 = -J_2 \sigma_2^z\sigma_3^z -J_3 \sigma_3^z\sigma_4^z~.
\end{equation}
Because $B_3 > J_2, J_3$, we can treat $H_1$ in perturbation theory. In second order, we obtain
\begin{equation}
\tilde J = J_2 J_3 / B_3~.
\label{eq:J_tilde}
\end{equation}
The new interaction $\tilde J$ is always smaller than either of the old ones. Thus, this
renormalization step  eliminates one degree of freedom (the spin $\sigma_3$) and reduces
the maximum energy scale $\Omega$.
\begin{figure}
\includegraphics[width=7.cm]{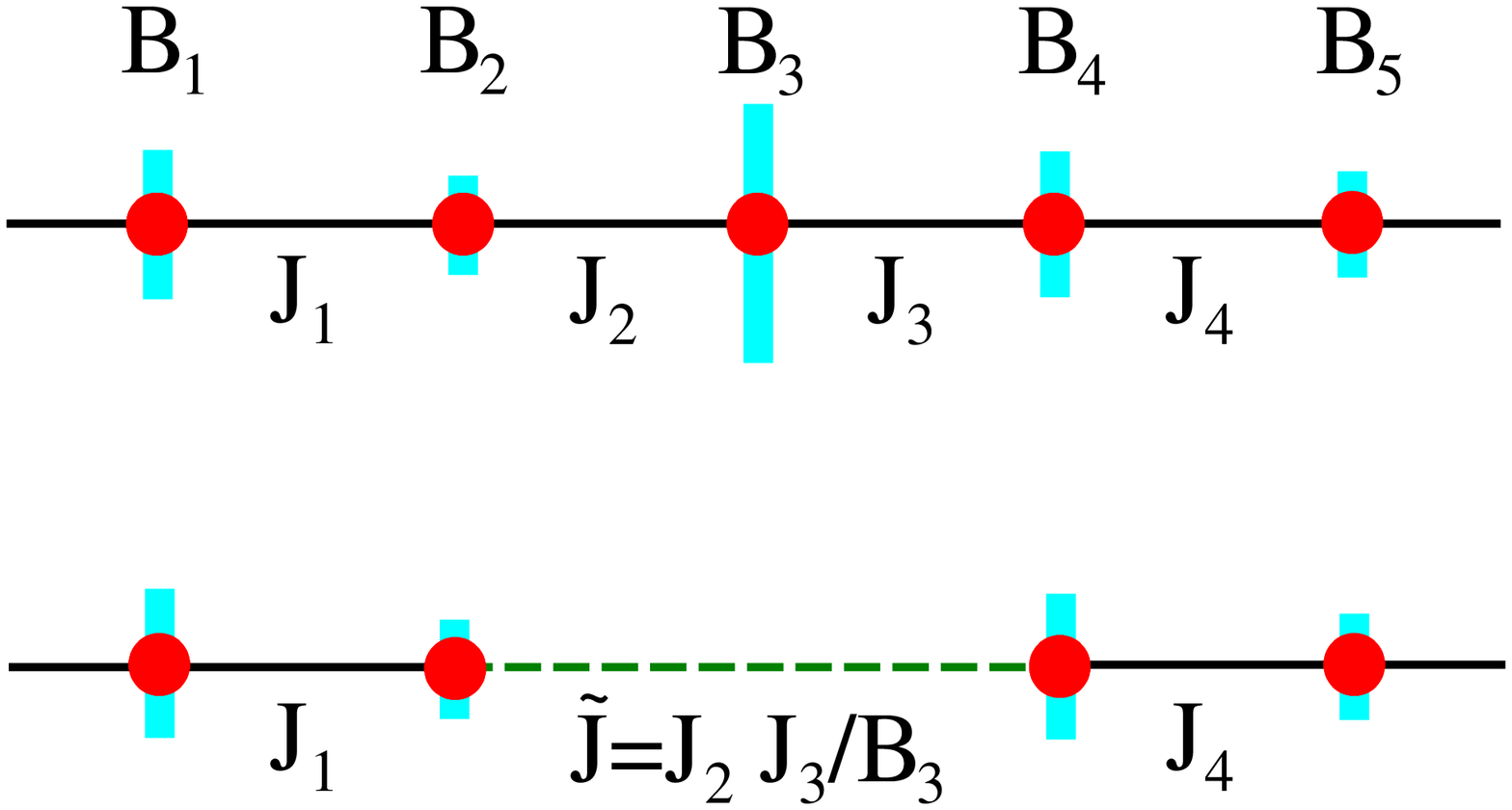}~~\includegraphics[width=7.cm]{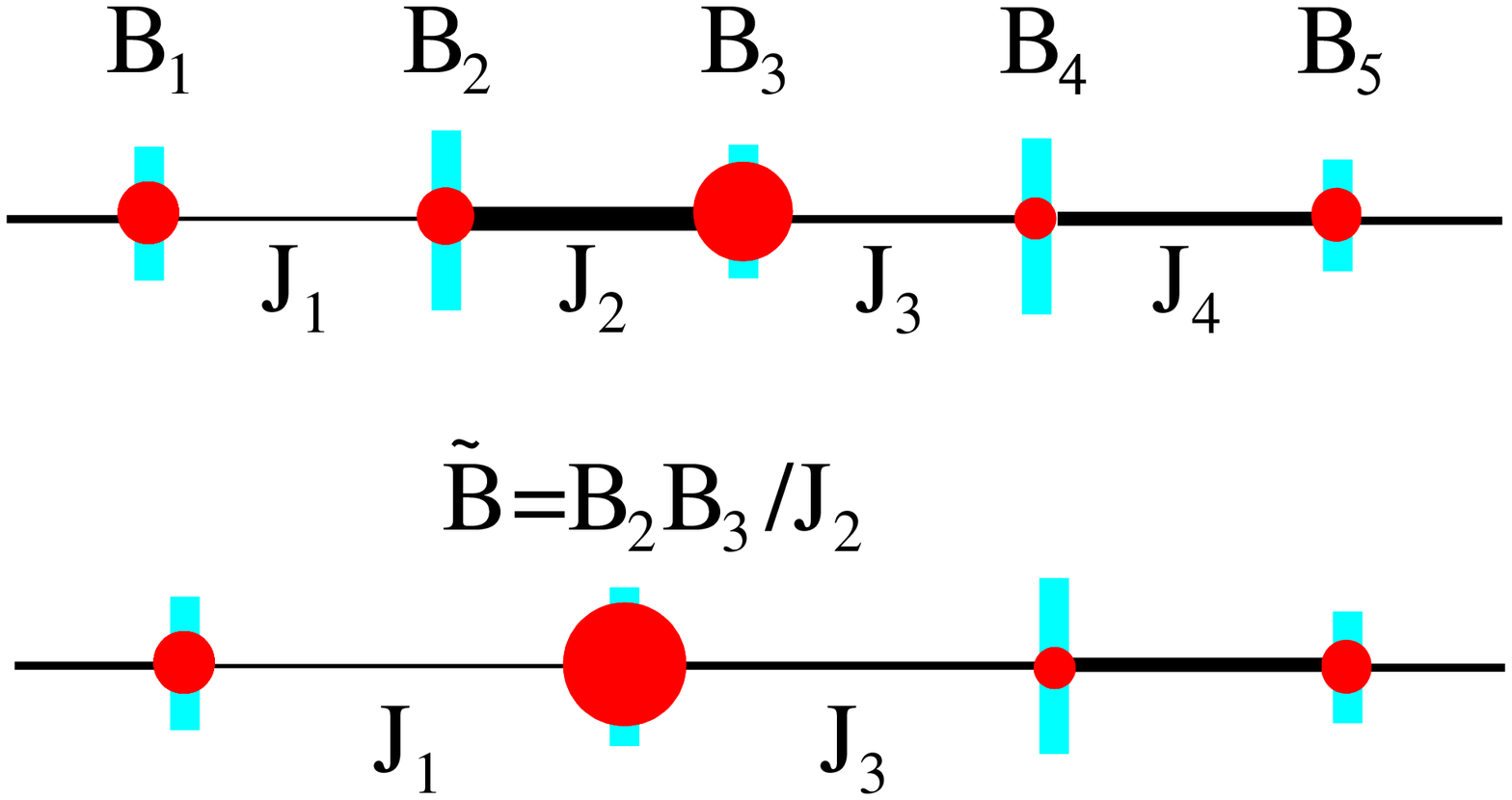}
\caption{Strong disorder renormalization group steps. Left: strong transverse field
  (here $B_3$). The spin $\sigma_3$ is integrated out in second-order perturbation theory,
  generating an effective interaction $\tilde J$ between the neighboring spins $\sigma_2$ and $\sigma_3$.
  Right: strong interaction (here $J_2$). The spins $\sigma_2$ and $\sigma_3$ are parallel to each other.
  They can be represented by a single effective spin $\tilde \sigma$ in the transverse field $\tilde B$.}
\label{fig:sdrg1a}
\end{figure}

(ii) If the largest local energy is an interaction, say $J_2$, the spins $\sigma_2$ and $\sigma_3$
prefer to be parallel. The cluster consisting of $\sigma_2$ and $\sigma_3$ can thus be treated as
a ``superspin'' $\tilde \sigma$ whose magnetic moment
\begin{equation}
\tilde \mu =\mu_2 + \mu_3
\label{eq:mu_tilde}
\end{equation}
is the sum of the moments
associated with $\sigma_2$ and $\sigma_3$. This means that out of the four basis states of the cluster
we keep the low-energy states $|\uparrow\uparrow \rangle$ and $|\downarrow\downarrow \rangle$ but not
the high-energy states $|\uparrow\downarrow \rangle$ and $|\downarrow\uparrow \rangle$.
Virtual excitations to these high-energy states need to be taken into account to evaluate the influence
of the transverse fields $B_2$ and $B_3$ on the effective spin $\tilde \sigma$.
The effective transverse field can be calculated in perturbation theory for the two-site cluster
consisting of $\sigma_2$ and $\sigma_3$ with
\begin{equation}
H=H_0 + H_1 \quad  \textrm{with} \quad  H_0 = -J_2 \sigma_2^z\sigma_3^z  ~, \quad H_1 = -B_2 \sigma_2^x  -B_3 \sigma_3^x ~.
\end{equation}
In second order in the transverse fields, we obtain
\begin{equation}
\tilde B = B_2 B_3 / J_2~.
\label{eq:h_tilde}
\end{equation}
The new transverse field $\tilde B$ is always smaller than either of the old ones.

The entire renormalization group step is summarized in Fig.\ \ref{fig:sdrg1a}. In both possible cases,
decimation of a site or decimation of an interaction, one degree of freedom is eliminated and
the maximum energy scale $\Omega$ is reduced. Otherwise, the structure
of the Hamiltonian (\ref{eq:RTIM}) is exactly preserved.
Notice the symmetry between
the recursion (\ref{eq:J_tilde}) for the interactions and the recursion (\ref{eq:h_tilde}) for the transverse fields.
It reflects the self-duality of the Hamiltonian. We also note that the recursion
(\ref{eq:mu_tilde}) for the moments is additive while the recursion (\ref{eq:h_tilde}) for the fields is
multiplicative, suggesting an exponential relation, $B \sim \exp(-c \mu)$ between the energy and the size
of a cluster. This is an important observation that we will return to repeatedly in the coming sections.

\subsubsection*{Flow equations}

The strong-disorder renormalization group method proceeds by iterating the
above renormalization group steps, thus gradually decreasing the maximum energy $\Omega$ to
zero. How do the interactions $J_i$ and transverse fields $B_i$ behave under this procedure?
As the $J_i$ and $B_i$ are random quantities, we need to analyze the evolution of their
probability distributions $P(J;\Omega)$ and $R(B;\Omega)$ with $\Omega$, starting from the initial (bare) distributions
$P_I(J)$ and $R_I(B)$.

To derive the renormalization group flow equations for $P(J;\Omega)$ and $R(B;\Omega)$, assume that we reduce
the maximum energy from $\Omega$ to $\Omega - d\Omega$ by decimating all interactions and fields
in the energy interval $[\Omega-d\Omega, \Omega]$. As a result, the distribution $P(J;\Omega)$
changes as
\begin{eqnarray}
-dP(J;\Omega) &=&
   d\Omega R(\Omega;\Omega) \left [ -2 P(J;\Omega) + \int dJ_1 dJ_2 ~ P(J_1;\Omega) P(J_2;\Omega) ~\delta(J-J_1 J_2/\Omega)  \right ] \nonumber \\
    &&+~  d\Omega [R(\Omega;\Omega) +  P(\Omega;\Omega)] P(J;\Omega)~.
\label{eq:dP(J)}
\end{eqnarray}
The terms in the first line are due to the decimation of strong transverse fields. The probability
for such a decimation is given by the probability for finding a field within $d\Omega$ of the
upper cutoff $\Omega$; it is identical to $d\Omega R(\Omega;\Omega)$.
Each such decimation removes two interactions from the system and introduces one new renormalized
interaction given by the recursion (\ref{eq:J_tilde}). This is encoded in the
terms inside the large bracket. As every decimation (of an interaction or field) reduces the number of remaining
interactions by one, the normalization of the distribution $P(J;\Omega)$ would change. The terms in the second line
of (\ref{eq:dP(J)}) compensate for that and keep $P(J;\Omega)$ normalized.\footnote{The overall
minus sign in front of $dP(J;\Omega)$ stems from the fact that $d\Omega$ changes the cutoff  $\Omega$
\emph{downwards}.}

The change of the field distribution $R(B;\Omega)$ can be found along the same lines. We thus arrive
at the renormalization group flow equations
\begin{eqnarray}
- \frac{\partial P}{\partial \Omega} &=& [P_\Omega - R_\Omega] P + R_\Omega \int dJ_1 dJ_2 ~~ P(J_1;\Omega) P(J_2;\Omega) ~ \delta(J-J_1 J_2/\Omega)~,
\label{eq:flow_eq_P}
\\
- \frac{\partial R}{\partial \Omega} &=& [R_\Omega - P_\Omega] R + P_\Omega \int dB_1 Bh_2 ~ R(B_1;\Omega) R(B_2;\Omega) ~ \delta(B-B_1 B_2/\Omega)~
\label{eq:flow_eq_R}
\end{eqnarray}
where $P$ and $R$ stand for $P(J;\Omega)$ and $R(B;\Omega)$ while $P_\Omega$ and $R_\Omega$ stand for
$P(\Omega;\Omega)$ and $R(\Omega;\Omega)$, respectively. The solutions of these integro-differential equations
in the limit $\Omega \to 0$ govern the low-energy physics of our system.

Before we start looking for solutions of the flow equations (\ref{eq:flow_eq_P}) and (\ref{eq:flow_eq_R}), let us change to
more suitable variables. The multiplicative structure of the recursions for $J$ and $B$ suggests using logarithmic variables.
We therefore introduce a logarithmic measure of the cutoff energy scale,
\begin{equation}
\Gamma = \ln(\Omega_I / \Omega)
\label{eq:Gamma}
\end{equation}
where $\Omega_I$ is the initial (bare) value of the cutoff. We also define logarithmic variables for the interactions
and fields by means of
\begin{equation}
\zeta = \ln (\Omega/J)~,  \qquad  \beta= \ln(\Omega / B)~.
\label{eq:log_vars}
\end{equation}
The probability distributions $\bar P(\zeta;\Gamma)$ and $\bar R(\beta;\Gamma)$  of these logarithmic variables
are related to the distributions of $J$ and $h$ via the transformations $P(J;\Omega) = \bar P (\zeta;\Gamma)/ J$ and
$R(B;\Omega) = \bar R (\beta;\Gamma)/ B$. In the following, we are going to drop the bar over the new distributions
as long as it is clear which distribution is meant.

By inserting the definitions (\ref{eq:Gamma}) and (\ref{eq:log_vars}) into (\ref{eq:flow_eq_P}) and (\ref{eq:flow_eq_R}),
we can rewrite the flow equations in terms of the logarithmic variables. This gives
\begin{eqnarray}
\frac{\partial P}{\partial \Gamma} &=& \frac{\partial P}{\partial \zeta} + [P_0 - R_0] P + R_0 \int_0^\zeta d\zeta_1 ~P(\zeta_1;\Gamma) ~P(\zeta-\zeta_1;\Gamma)~,
\label{eq:flow_eq_Plog}
\\
\frac{\partial R}{\partial \Gamma} &=& \frac{\partial R}{\partial \beta} + [R_0 - P_0] R + P_0 \int_0^\beta d\beta_1 ~ R(\beta_1;\Gamma)~ R(\beta-\beta_1;\Gamma)~.
\label{eq:flow_eq_Rlog}
\end{eqnarray}
Here, $P$ and $R$ stand for $P(\zeta;\Gamma)$ and $R(\beta;\Gamma)$ while $P_0$ and $R_0$ denote
$P(0;\Gamma)$ and $R(0;\Gamma)$.

How can one solve these flow equations? D.~S.\ Fisher provided an essentially complete analysis in two long papers
\cite{Fisher94,Fisher95}. It requires significant mathematical effort and is beyond the scope of these lectures.
Instead, we are going to use an ansatz for the distributions  $P$ and $R$ that will lead us to the
correct fixed point solutions.\footnote{This approach does \emph{not} guarantee, of course, that there are no other solutions
that are important for the physics of our problem. To address this question one needs to consult the complete solution.}
Specifically, let us assume that both distributions are simple exponentials,
\begin{equation}
P(\zeta;\Gamma) = P_0(\Gamma)\, \exp[-P_0(\Gamma)\, \zeta] ~, \qquad R(\beta;\Gamma) = R_0(\Gamma)\, \exp[-R_0(\Gamma)\, \beta]
\label{eq:PR_ansatz}
\end{equation}
with cutoff-dependent inverse widths $P_0$ and $R_0$. Inserting this ansatz into (\ref{eq:flow_eq_Plog})
and (\ref{eq:flow_eq_Rlog}), we find that the distributions indeed fulfill the flow equations provided that
the coefficients $P_0$ and $R_0$ are solutions of the coupled differential equations
\begin{eqnarray}
\frac {dP_0}{d\Gamma} &=& -R_0 \, P_0~,
\label{eq:flow_eq_P0}\\
\frac {dR_0}{d\Gamma} &=& -R_0 \, P_0~.
\label{eq:flow_eq_R0}
\end{eqnarray}
We thus have turned the flow equations for the probability distributions $P(\zeta;\Gamma)$ and $R(\beta;\Gamma)$
into flow equations for the coefficients $P_0$ and $R_0$ which can be solved much more easily.

In the following subsection, we will look for solutions of these flow equations and analyze their behavior in the low-energy
limit $\Gamma \to \infty$.

\subsection{Infinite-randomness scenario}

\subsubsection*{Critical point}

According to Pfeuty's exact result \cite{Pfeuty79}, the random transverse-field Ising model
is critical when $\prod_i J_i = \prod_i B_i$. At the critical point, the coefficients
$P_0(\Gamma)$ and $R_0(\Gamma)$ must therefore be identical. This also follows from the
self-duality of the Hamiltonian discussed in Sec.\ \ref{subsec:RTIM}.
The two flow equations (\ref{eq:flow_eq_P0})  and (\ref{eq:flow_eq_R0}) now coincide and read
\begin{equation}
\frac {dP_0}{d\Gamma} = - P_0^2
\label{eq:flow_eq_critical}
\end{equation}
This differential equation can be easily integrated, giving $P_0 = 1/(\Gamma-\Gamma_0)$.
The integration constant $\Gamma_0$ can be dropped as it just amounts to a redefinition
of the reference energy scale $\Omega_I$. We thus arrive at the ``fixed point''
solution\footnote{This solution is a fixed point in the sense that its functional form
does not change with $\Gamma \to \infty$. If we rescale $\zeta$ and $\beta$ by $\Gamma$,
the resulting distributions become true stationary points of the renormalization group flow.}
\begin{equation}
P(\zeta;\Gamma) = \frac 1 \Gamma e^{-\zeta / \Gamma}~, \qquad R(\beta;\Gamma) = \frac 1 \Gamma e^{-\beta / \Gamma}~.
\label{eq:critical FP}
\end{equation}
In the low-energy limit, $\Gamma  \to \infty$, these distributions become arbitrarily broad implying
that the randomness in the system becomes arbitrarily strong. For this reason, the resulting critical
point is called an infinite-randomness critical point. In terms of the original variables $J$ and $B$,
the distributions are highly singular,
\begin{equation}
P(J;\Omega) = \frac 1 {\Gamma J} \left(\frac J \Omega \right)^{1 / \Gamma}~, \qquad R(B;\Omega) = \frac 1 {\Gamma B} \left(\frac B \Omega \right)^{1 / \Gamma}~.
\label{eq:critical FP_Jh}
\end{equation}

Let us discuss some of the key properties of the critical fixed point solution (\ref{eq:critical FP})
in more detail. We start by analyzing the number $n_\Omega$ of sites (clusters) that survive in the renormalized
random transverse-field Ising chain at some energy scale $\Omega$. Every renormalization group step
reduces the number of clusters by one, either because a cluster is decimated (for a large transverse field)
or because two clusters are combined (for a large interaction). The number of clusters thus changes by
$dn_\Omega = - n_\Omega (P_\Omega+R_\Omega) d\Omega$ as the cutoff is reduced from $\Omega$ to $\Omega-d\Omega$.
Transforming to logarithmic variables leads to
\begin{equation}
\frac {dn_\Gamma}{d\Gamma} = -(P_0+R_0)\, n_\Gamma = - \frac 2 \Gamma n_\Gamma~.
\label{eq:n_Gamma_ODE}
\end{equation}
Integrating this differential equation shows that the number of clusters decreases as
\begin{equation}
n_\Gamma \sim \Gamma^{-2} \qquad \textrm{or} \qquad n_\Omega \sim \left[\ln ({\Omega_I}/{\Omega}) \right]^{-2}~.
\label{eq:n_Gamma}
\end{equation}
The typical distance $\ell(\Omega)$ between the surviving clusters is inversely proportional to their number,
yielding
\begin{equation}
\ln ({\Omega_I}/{\Omega})  \sim [\ell(\Omega)]^\psi  \qquad \textrm{with} \quad \psi=1/2~.
\label{eq:l_Omega}
\end{equation}
The self-duality of the critical solution implies that $\ell(\Omega)$ gives not just the
typical distance between clusters but also the typical length of a cluster.

We note that the  dependence of the energy scale on the length scale established by (\ref{eq:l_Omega})
is exponential and so is the dependence of the time scale on the length scale.
This differs from the power-law dependence (\ref{eq:xit}) found at conventional (clean) critical points.
An exponential relation between lengths and times is called activated or tunneling dynamical scaling (as opposed to the
usual power-law dynamical scaling) and $\psi$ is called the tunneling exponent. It also means
that the dynamical exponent $z$ is formally infinite.

In addition to the number of surviving clusters, one can also calculate their magnetic moment $\mu_\Omega$
(i.e., the number of original sites represented by an effective spin). The calculation is rather lengthy
\cite{Fisher95}; we therefore simply quote the result
\begin{equation}
\mu_\Omega \sim \left[\ln ({\Omega_I}/{\Omega}) \right]^{\phi}
\label{eq:mu_Omega}
\end{equation}
where the exponent $\phi=(\sqrt{5}+1)/2$ is given by the golden mean.\footnote{The fact that the exponent
$\phi$ which governs the magnetic moment is smaller than the exponent 2 which governs the length of a cluster
suggests that the surviving clusters have a fractal structure with lots of holes.}
These are highly unusual properties that are reflected in unusual behaviors of observables at the
infinite-randomness critical point.

The fact the probability distributions of the fields and interactions become infinitely broad as the critical
fixed point is approached implies that the method becomes asymptotically exact because the perturbative treatment
of the recursion steps becomes better and better. This also means that the above critical exponent values are
exact which is remarkable because exact exponent values are known for very few phase transitions only.
Moreover, the infinite-randomness character of the critical point explains why perturbative methods fail to capture the physics of this problem.
Even if the bare disorder strength is a small, the disorder becomes arbitrarily strong on large length scales.

\subsubsection*{Off-critical solutions}

Right at the critical point, the two decimation processes (eliminating clusters with strong
transverse fields and building larger clusters by decimating a strong interaction) balance each
other out such that $P_0 = R_0$ in the low-energy limit. In contrast, off criticality, one of the
processes wins.

On the paramagnetic side of the transition defined by $\prod_i B_i > \prod_i J_i$ or,
equivalently, $[\ln B]_{dis} > [\ln J]_{dis}$, the building
of clusters essentially stops at some value of the cutoff $\Omega$  because most $B$ are larger than most $J$.
From then on, only transverse fields are decimated, producing smaller and smaller interactions.
Thus, we expect the field distribution $R(\beta)$ to become stationary with $\Omega \to 0$ while $P(\zeta)$ should
rapidly shift to small interactions (large $\zeta$). On the ferromagnetic side, we expect the
opposite behavior, i.e., the elimination of clusters stops below some value of the cutoff $\Omega$.
Here, $P(\zeta)$ should become stationary and $R(\beta)$ should rapidly move to large $\beta$.

In the following, we focus on the paramagnetic side.
Off criticality, $P_0 \ne R_0$, thus the two flow equations (\ref{eq:flow_eq_P0})
and (\ref{eq:flow_eq_R0}) need to be solved together. By subtracting them from each other,
we obtain $(d/d\Gamma)[R_0 -P_0]=0$ which yields
\begin{equation}
R_0(\Gamma) = P_0(\Gamma) + 2\delta
\label{eq:R0P0delta}
\end{equation}
where $\delta$ is an energy-independent constant. What is the physical meaning of $\delta$?
Note that $[\ln (h/\Omega)]_{dis} =[-\beta]_{dis} = -1/R_0$ and  $[\ln (J/\Omega)]_{dis} = [-\zeta]_{dis} = -1/P_0$.
Thus $[\ln h]_{dis} - [\ln J]_{dis} = 1/P_0 - 1/R_0 \sim R_0 -P_0 = 2\delta$ for small $\delta$. We conclude that
$\delta$ is a measure of the distance from criticality.

After inserting  (\ref{eq:R0P0delta}), the flow equations (\ref{eq:flow_eq_P0})
and (\ref{eq:flow_eq_R0}) can be easily solved. We find
\begin{equation}
P_0=2\delta /[\exp(2\delta\Gamma)-1]~.
\label{eq:pm_P0}
\end{equation}
In the low-energy limit, $\Gamma \to \infty$, this results in the fixed-point distributions
\begin{eqnarray}
R(\beta) &=& 2\delta ~\exp[-2\delta \beta]
\label{eq:pm FP_R}\\
P(\zeta) &=& 2\delta e^{-2\delta \Gamma} ~\exp[-2\delta e^{-2\delta \Gamma}\beta]~.
\label{eq:pm FP_P}
\end{eqnarray}
As expected from our qualitative discussion above, the field-distribution $R(\beta)$ is indeed
independent of $\Gamma$ while the interaction distribution $P(\zeta)$ rapidly becomes
extremely broad. This means, almost all interactions $J$ are extremely small,
and the surviving clusters decouple. Transforming (\ref{eq:pm FP_R}) to the
original transverse fields $B$ yields a power-law distribution
\begin{equation}
 R(B;\Omega) =  \frac {2\delta}  B \left(\frac B \Omega \right)^{2\delta}~
 \label{eq:pm FP_R_h}
\end{equation}
characterized by a nonuniversal exponent that changes with distance $\delta$ from the critical point.

The number of clusters surviving at energy scale $\Gamma$ can be determined from the
differential equation $(dn_\Gamma)/(d\Gamma) = -(R_0+P_0) n_\Gamma$, as at the critical point.
This gives $n_\Gamma = n_0 \exp(-2\delta\Gamma)$ or, equivalently, $n_\Omega \sim (\Omega/\Omega_I)^{2\delta}$.
The typical distance $\ell(\Omega)$ between clusters surviving at energy scale $\Omega$
therefore behaves as
\begin{equation}
\ell(\Omega) \sim 1/n_\Omega \sim (\Omega/\Omega_I)^{-2\delta}~.
\label{eq:l_Omega_pm}
\end{equation}
In contrast to the activated dynamical scaling (\ref{eq:l_Omega}) at the critical point,
this is a power-law relation characterized by a nonuniversal dynamical exponent $z'=1/(2\delta)$.

To find the correlation length and time at a given distance $\delta$ from criticality
we can calculate the crossover energy scale $\Gamma_x$, beyond which the off-critical
flow (\ref{eq:pm_P0}) appreciably deviates from the critical flow $P_0 = 1/\Gamma$.
This gives $\Gamma_x \sim 1/(2\delta)$. According to (\ref{eq:n_Gamma}), the characteristic length
is $\ell(\Gamma_x) \sim \Gamma_x^2 \sim 1/(2\delta)^2$ at this energy scale. We conclude that the correlation
length exponent takes the value $\nu=2$. Interestingly, this means that it exactly
saturates the inequality $d\nu \ge 2$ derived by Chayes et al.\ \cite{CCFS86}.

\subsubsection*{Thermodynamic observables}

The general strategy for finding the temperature dependencies of observables within the strong-disorder
renormalization group approach consists in running the renormalization group until the cutoff reaches
the value $\Omega=k_B T$.
Close to criticality and at low energies $\Omega$, the probability distributions of $J$ and $B$ become very broad.
This means, all degrees of freedom eliminated during this process have energies much larger than $k_B T$
and do not contribute to the thermodynamics. In contrast, all the remaining interactions and transverse
fields are much smaller than $k_B T$. The remaining degrees of freedom can thus be considered as free.
Consequently, observables are simply the sums over independent contributions from the surviving clusters.

Let us apply this strategy to the thermodynamics right at criticality. According to
(\ref{eq:n_Gamma}) and (\ref{eq:mu_Omega}), the number of surviving clusters at energy scale
$\Omega$ behaves as $n_\Omega \sim [\ln(\Omega_I/\Omega]^{-1/\psi}$ with $\psi=1/2$
while the typical moment of such a cluster scales as $\mu_\Omega \sim [\ln(\Omega_I/\Omega]^{\phi}$
with $\phi = (\sqrt{5}-1)/2$. Each cluster (being effectively free) contributes
a Curie term $\mu^2/T$ to the magnetic susceptibility. The total susceptibility thus reads
\begin{equation}
\chi(T) = \frac 1 T ~ n_T \mu^2_T \sim \frac 1 T ~ [\ln(\Omega_I/k_B T)]^{2\phi-1/\psi}~.
\label{eq:RTIM_chi_critical}
\end{equation}
To estimate the entropy, we note that each surviving cluster is equivalent to a two-level system and
 contributes $k_B \ln(2)$ to the entropy. The total entropy is thus given by
\begin{equation}
S(T) =  k_B \ln(2)~ n_T  \sim k_B ~ [\ln(\Omega_I/k_B T)]^{-1/\psi}~.
\label{eq:RTIM_S_critical}
\end{equation}
The specific heat can be calculated by taking the appropriate temperature derivative of the
entropy. This yields
\begin{equation}
C_V(T) =  T (\partial S/\partial T) = (\partial S/ \partial \ln(T) )  \sim k_B ~ [\ln(\Omega_I/k_B T)]^{-1/\psi-1}~.
\label{eq:RTIM_CV_critical}
\end{equation}
All these quantities display unusual logarithmic temperature dependencies which stem from the activated
dynamical scaling (\ref{eq:l_Omega}).

The same strategy for determining observables can also be applied off criticality. According
to (\ref{eq:l_Omega_pm}), the number of surviving clusters on the paramagnetic side of the transition
scales as $n_\Omega \sim \Omega^{2\delta} = \Omega^{1/z'}$ with the energy cutoff $\Omega$. This
results in a total entropy of
\begin{equation}
S(T) =  k_B \ln(2)~ n_T  \sim T^{1/z'}~.
\label{eq:RTIM_S_pm}
\end{equation}
and a specific heat of
\begin{equation}
C_V(T) =  T (\partial S/\partial T)   \sim T^{1/z'}~.
\label{eq:RTIM_CV_pm}
\end{equation}
The magnetic susceptibility can be found along the same lines, yielding
\begin{equation}
\chi(T) = \frac 1 T ~ n_T \mu^2_T \sim T^{1/z' - 1} ~.
\label{eq:RTIM_chi_pm}
\end{equation}
Note that the magnetic moment of a surviving cluster on the paramagnetic side of the transition
increases only very slowly (logarithmically) with decreasing energy cutoff $\Omega$ because almost all decimations
are eliminations of sites \cite{Fisher95}. Therefore, $\mu_T$ does not contribute to the leading temperature dependence
of the susceptibility.

Equations (\ref{eq:RTIM_S_pm}) to (\ref{eq:RTIM_chi_pm}) show that the thermodynamic behavior of the
random transverse-field Ising chain is highly singular not just right at the critical point but also
in its vicinity. These off-critical singularities are examples of the celebrated Griffiths singularities
that we will study from a more general perspective in Sec.\ \ref{sec:Griffiths}.

\section{Griffiths singularities and the Griffiths phase}
\label{sec:Griffiths}

When a clean system undergoes a phase transition, thermodynamic quantities are singular right at the transition point.
Away from the transition point, they are generally nonsingular. However, at the end of the last section, we have seen that
the random transverse-field Ising model features singular thermodynamics not just at the critical point but also
in its vicinity. It turns out that such off-critical singularities generically occur at phase transitions  in disordered
systems. They are called the Griffiths singularities after R.~B. Griffiths who first proved their existence in
1969 \cite{Griffiths69}.

In this section, we will develop a general qualitative understanding of the physics leading to the Griffiths singularities;
and we will discuss their consequences for classical and quantum phase transitions.

\subsection{Rare regions}

Let us start from a clean classical ferromagnet given by an Ising model on a cubic
lattice, $H = -J \sum_{\langle ij\rangle} S_i S_j$. It undergoes a continuous phase transition
from a paramagnet to a ferromagnet at the clean critical temperature $T_c^0$. We now introduce
quenched disorder by randomly removing spins from the lattice. The Hamiltonian of the
resulting site-diluted Ising model  reads
\begin{equation}
H = -J\sum_{\langle ij \rangle} \epsilon_i\epsilon_j \,S_i S_j
\label{eq:diluted_Ising}
\end{equation}
where the quenched random variables $\epsilon_i$ can take the values 0 (representing a vacancy)
or 1 (representing a spin) with probabilities $p$ and $1-p$, respectively.

As the dilution reduces the overall tendency towards ferromagnetism, the critical temperature
$T_c(p)$ of the diluted system will decrease with increasing $p$
(see Fig.\ \ref{fig:PD_diluted}).
\begin{figure}
\includegraphics[width=7.7cm]{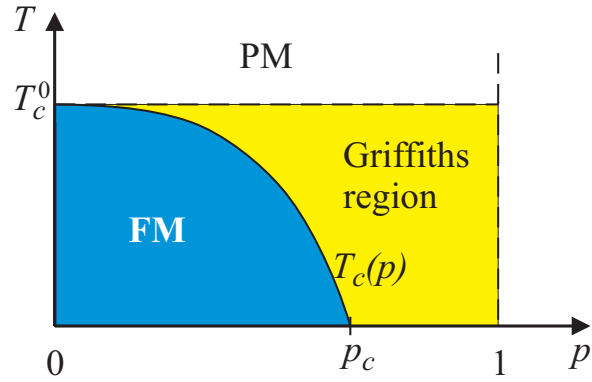}
\caption{Schematic temperature-dilution phase diagram of the site-diluted Ising model (\ref{eq:diluted_Ising}).
With increasing dilution, the critical temperature $T_c(p)$ is suppressed, it vanishes at the percolation
threshold $p_c$. The Griffiths region (or Griffiths phase) comprises the part of the paramagnetic phase
below $T_c^0$ where locally ordered ``clean'' clusters can exist.
}
\label{fig:PD_diluted}
\end{figure}
It vanishes at the percolation threshold
$p_c$ of the lattice because ferromagnetic long-range order is impossible if the lattice consists of
disconnected finite-size clusters only.

Due to statistical fluctuations in the vacancy distribution, large vacancy-free spatial regions can exist
even for high vacancy concentrations (with a very small but nonzero probability). As such rare regions are
finite-size pieces of the clean system, their spins align parallel to each other below the \emph{clean} critical
temperature $T_c^0$. Because they are of finite size, these regions cannot undergo a true phase transition
by themselves, but for temperatures between $T_c^0$ and the actual transition temperature $T_c(p)$,
they act as large superspins.

The parameter region where such locally ordered rare regions exist, but long-range ordered has not yet
developed is called the Griffiths phase. (More appropriately, it should be called the Griffiths region
as it is part of the paramagnetic phase.) In our example the Griffiths region comprises the area between
$T_c^0$ and $T_c(p)$ in the temperature-dilution plane.

So far, we have focused on the paramagnetic side of the transition. How can we generalize the idea of a
Griffiths region or Griffiths phase to the ferromagnetic side? The simplest idea seems to be to consider
rare ``holes'' in the magnetic order, i.e., rare vacancy-rich regions in a magnetically ordered bulk.
However, in contrast to the locally ordered rare regions considered above whose magnetization can fluctuate between
up and down, holes do not have an associated low-energy degree of freedom. Instead of simple ``holes''
one should thus consider locally ordered island \emph{inside} holes. Such islands can fluctuate between up
and down because they are only
very weakly coupled to the bulk ferromagnet outside the hole (see Fig.\ \ref{fig:rare_region}).
\begin{figure}
\includegraphics[width=10.7cm]{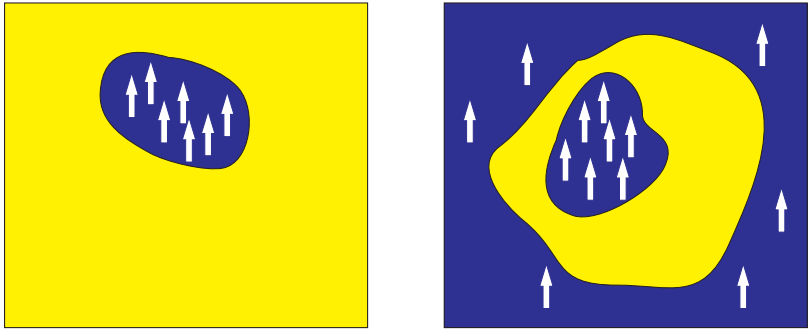}
\caption{Left: The paramagnetic Griffiths phase is due to rare locally ferromagnetic regions embedded
  in the paramagnetic bulk. Right: The ferromagnetic Griffiths phase is caused by locally ferromagnetic
  regions located inside paramagnetic ``holes'' in the bulk ferromagnet.}
\label{fig:rare_region}
\end{figure}
This conceptual difference will be responsible for slight differences in the resulting Griffiths singularities
on the two sides of the transition. In our example of a site-diluted Ising model, the ferromagnetic Griffiths
phase comprises all of the ferromagnetic phase for $p>0$.

Note that the precise location and extension of the Griffiths phases depend on the details of the system at hand.
For example, we could study an Ising model with bond (interaction) randomness instead of the site-diluted
Ising model (\ref{eq:diluted_Ising}).  It is given by the Hamiltonian
\begin{equation}
H=-\sum_{\langle ij\rangle} J_{ij} S_i S_j
\label{eq:H_JaJb}
\end{equation}
in which the nearest neighbor interactions $J_{ij}$ can take
the values $J_A$ and $J_B>J_A$ with probability $p$ and $1-p$, respectively.
In this model, the paramagnetic and ferromagnetic Griffiths phases are both located between
the transition temperature $T_c^B$ of a system in which all bonds have value $J_B$ and
the transition temperature $T_c^A$ of a system with all bonds having value $J_A$
(see Fig.\ \ref{fig:PD_JaJb}).
\begin{figure}
\includegraphics[width=10.cm]{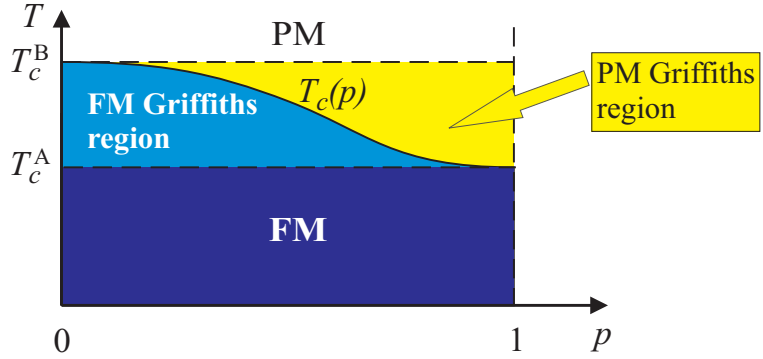}
\caption{Phase diagram of the random-bond Ising model (\ref{eq:H_JaJb}).
The Griffiths phases are located between the transition temperatures $T_c^A$ and $T_c^B$ of systems with uniform
interactions $J_A$ or $J_B$, respectively. The conventional
paramagnetic phase is located above $T_c^B$, and the conventional ferromagnet is below $T_c^A$}
\label{fig:PD_JaJb}
\end{figure}

Why are these rare large locally ordered regions interesting? In the next subsections, we will see that,
despite being rare, they are responsible for the peculiar Griffiths singularities in the vicinity of a phase transition
in a disordered system.

\subsection{Classical Griffiths singularities}
\label{sec:classical_Griffiths}

\subsubsection*{Thermodynamics of the Griffiths phase}

In 1969, Griffiths \cite{Griffiths69} showed that the free energy is a singular function of the external
control parameters everywhere in the Griffiths phase. The physical origin of the singular behavior was
later identified as being the contribution of the rare regions to the thermodynamic behavior.
To understand this mechanism, let us estimate the rare region contribution to the magnetization-field curve
$m(h)$ and to the magnetic susceptibility $\chi$.

If a weak magnetic field $h$ is applied to a disordered magnet in the (paramagnetic)
Griffiths phase, the locally ordered rare regions act as large superspins whose magnetic
moment is proportional to their volume, $\mu(L_{RR}) \sim \mu_B V \sim L_{RR}^d$, where
$L_{RR}$ is the linear size of the rare region. The energy gain due to aligning the moment
with the external field can be estimated as $\Delta E = -h \mu \sim h \mu_B V$.
If $|\Delta E| > k_B T$, the superspin is essentially fully polarized in field direction.
For $|\Delta E| < k_B T$, its magnetization is small and governed by linear response.

The singular contribution of the large rare regions to the magnetization-field curve can
therefore be estimated by summing over all rare regions with $|\Delta E| > k_B T$.
This gives
\begin{equation}
m_{RR}(h) \approx \sum_{|\Delta E| > k_B T} w(L_{RR}) ~ \mu(L_{RR})
\label{eq:m(H)_Griffiths_classical}
\end{equation}
where $w(L_{RR})$ is the probability for finding a locally ordered rare region  of linear size $L_{RR}$.
Basic combinatorics yields that the probability for finding a large spatial region devoid of
impurities decays exponentially with its volume,
\begin{equation}
w(L_{RR}) \sim \exp\left(-\tilde p\, L_{RR}^d \right)~.
\label{eq:w(L_RR)}
\end{equation}
For the diluted ferromagnet (\ref{eq:diluted_Ising}), the constant $\tilde p \sim -\ln(1-p)$.
To exponential accuracy, the rare region magnetization can now be estimated as
\begin{equation}
m_{RR}(h) \sim \int_{L_x}^\infty dL_{RR} ~\exp\left(-\tilde p\, L_{RR}^d \right)~\mu_B L_{RR}^d
\label{eq:m(H)_Griffiths_classical_result}
\end{equation}
where the integral is over all rare regions larger than a critical size $L_x$ defined by $L_x^d\sim k_B T / (\mu_B h)$.
This results in
\begin{equation}
m_{RR}(h) \sim \exp\left[-\tilde p k_B T / (\mu_B h)  \right]~.
\label{eq:m(H)_essential}
\end{equation}
The rare region magnetization is thus indeed a singular function of the applied field $h$,
as required by Griffiths' proof. However, the singularity is an essential singularity leading
to an extremely small contribution.

The magnetic susceptibility $\chi$ can be analyzed similarly. Each locally
ordered rare region makes a Curie contribution $\mu^2(L_{RR})/T$ to $\chi$.
The total rare region susceptibility can therefore be estimated as
\begin{equation}
\chi_{RR}(T) \sim \int dL_{RR}~ w(L_{RR}) ~\mu_B^2\, L_{RR}^{2d}/T~.
\label{eq:chi(T)_Griffiths_classical}
\end{equation}
This equation shows that the susceptibility of an individual rare regions does not
increase fast enough to overcome the exponential decay of the rare region probability $w$
with increasing size $L_{RR}$. Consequently, large rare regions only make an exponentially
small contribution to the susceptibility.

Analogous estimates can also be performed in
the ferromagnetic Griffiths phase on the ordered side of the phase transition.
The main difference between the paramagnetic and ferromagnetic Griffiths phases
is in the probability for finding a rare region. On the paramagnetic side, the rare
event is finding a \emph{vacancy-free} region of linear size $L_{RR}$ which leads to
(\ref{eq:w(L_RR)}). On the ferromagnetic side, the rare event is finding a large
enough \emph{vacancy-rich} region around a locally ordered island (see Fig.\ \ref{fig:rare_region}).
The required volume of this ``shell'' scales as $L_{RR}^{d-1} \xi_B$ where $\xi_B$ is the
bulk correlation length in this vacancy-rich region. Thus the probability of finding
such a ``island in a hole'' behaves as
\begin{equation}
w_{FM}(L_{RR}) \sim \exp\left(-c L_{RR}^{d-1} \right)~.
\label{eq:w_FM(L_RR)}
\end{equation}

The overall conclusion of these considerations is that classical
rare region phenomena are very weak because the singularity in the free energy is only an
essential one \cite{Wortis74,Harris75,Imry77,BrayHuifang89}. To the best of our knowledge,
classical Griffiths singularities in thermodynamic quantities have therefore not
been verified in experiment.\footnote{This conclusion holds for uncorrelated or short-range
correlated disorder only. We will see in Sec.\ \ref{subsec:quantum_Griffiths} that long-range
disorder correlations in space greatly enhance the rare region effects.}

\subsubsection*{Dynamical Griffiths singularities}

In contrast to the thermodynamics discussed above, the long-time dynamics in a classical
Griffiths phase is dominated by the rare regions.

The classical Ising Hamiltonians (\ref{eq:diluted_Ising}) and (\ref{eq:H_JaJb}) do not have any
internal dynamics. However, we can add a heuristic dynamics to these models. The simplest case is
a purely relaxational dynamics which corresponds to the so-called model A in the
famous classification of Hohenberg and Halperin \cite{HohenbergHalperin77}. Microscopically,
this type of dynamics is due to collisions with other degrees of freedom. In computer simulations, it can be realized
via the Glauber \cite{Glauber63} or Metropolis \cite{MRRT53} algorithms.

One of the simplest quantities for studying the real-time dynamics is the spin auto-correlation function
\begin{equation}
C(t) = \frac 1 N \sum_i \langle S_i(t) \, S_i(0) \rangle
\label{eq:autocorrelation_definition}
\end{equation}
where $S_i(t)$ is the value of the spin at site $i$ and time $t$, the sum is over
all $N$ sites of the lattice, and $\langle \ldots \rangle$ denotes the thermodynamic
average. It is related to the Fourier transform of the local dynamic susceptibility.

The fluctuations associated with the rare regions are much slower than the bulk fluctuations in the
paramagnetic Griffiths phase because they involve a coherent change of the magnetization in a
large volume. As different rare regions are independent of each other, their contribution to the
auto-correlation function can be estimated as
\begin{equation}
C_{RR}(t) \sim \int dL_{RR} ~w(L_{RR}) L_{RR}^d \exp\left[-t/\xi_t(L_{RR})\right]
\label{eq:autocorrelation_Griffiths}
\end{equation}
where $\xi_t(L_{RR})$ is the correlation time (or life time) of a single rare region, i.e.,
the typical time it takes to reverse the rare region magnetization. How can we estimate this
correlation time?
The most effective way of reversing the rare region magnetization is to create a domain wall
and then move it across the rare region. The energy cost of creating such a domain wall
is given by $\sigma A_{DW}$ where $\sigma$ is a temperature-dependent constant and $A_{DW}$ is
the domain wall area. Thus, the domain wall energy scales as $\sigma L_{RR}^{d-1}$ with the
rare region size. The creation of the domain wall is an activated process that happens at an exponentially
small rate, implying
\begin{equation}
\xi_t(L_{RR}) \sim \tau_0 \exp[C \sigma L_{RR}^{d-1} /(k_B T)]
\label{eq:RR_corr_time}
\end{equation}
where $C$ is some constant of order unity. After inserting the correlation time into the
autocorrelation function (\ref{eq:autocorrelation_Griffiths}), the integral  can be
estimated by means of the saddle point method. This yields
\begin{equation}
C_{RR}(t) \sim \exp\left[-\tilde C (\ln t)^{d/(d-1)} \right]
\label{eq:autocorrelation_Griffiths_result}
\end{equation}
in the long-time limit ($\tilde C$ is another constant). The long-time
decay of the rare region autocorrelation function is much slower than the exponential decay
of the bulk system off criticality. Thus, the long-time dynamics of the disordered Ising model
is dominated by the rare regions  \cite{Dhar83,RanderiaSethnaPalmer85,Bray88,Bray89}.

\subsection{Quantum Griffiths singularities}
\label{subsec:quantum_Griffiths}

\subsubsection*{Rare region density of states}

In the last subsection we have analyzed the Griffiths singularities close to a classical phase transition and
found them to be weak (at least in the case of thermodynamic quantities). Now, we are going to address the same
question for a quantum phase transition.

Our example is the three-dimensional transverse-field Ising model, and we introduce disorder via site-dilution.
The Hamiltonian reads
\begin{equation}
H = -J\sum_{\langle ij \rangle} \epsilon_i\epsilon_j \,\sigma_i^z \sigma_j^z - B \sum_i \epsilon_i \sigma_i^x~.
\label{eq:diluted_transverse_Ising}
\end{equation}
As in the last section, the $\epsilon_i$ are quenched random variables that take values 0 or 1 with probabilities
$p$ and $1-p$, respectively. Note that we again use $B$ for the transverse field to distinguish it
from a longitudinal field $h$ in $z$-direction.

As was discussed in Sec.\ \ref{subsec:QPT_intro}, the undiluted system undergoes a (zero-temperature)
quantum phase transition from a ferromagnet to a paramagnet at a critical value $B_c^0$ of the
transverse field. Site dilution reduces the tendency towards ferromagnetism and leads to a
phase diagram similar to the classical case (with the transverse field playing the role of
the temperature). It is shown in Fig.\ \ref{fig:PD_diluted_QPT}.
\begin{figure}
\includegraphics[width=8.cm]{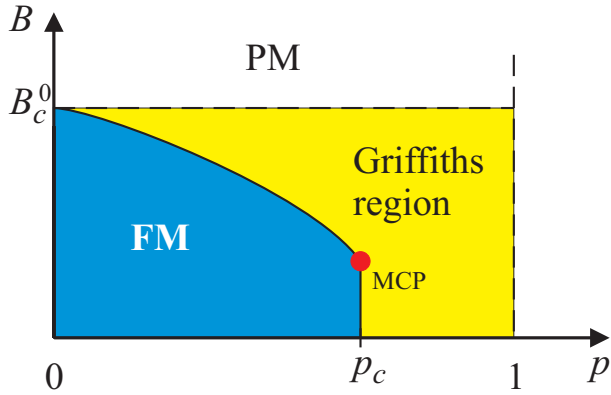}
\caption{Schematic zero-temperature phase diagram of the site-diluted transverse-field Ising model (\ref{eq:diluted_transverse_Ising}).
With increasing dilution, the critical field $B_c(p)$ is suppressed. A multicritical point (MCP) separates
the generic quantum phase transition for $p<p_c$ from the percolation transition at $p_c$. Ferromagnetic
long-range order is impossible for dilutions above $p_c$. The Griffiths region (or Griffiths phase) comprises the part of the paramagnetic phase
below $B_c^0$ where locally ordered ``clean'' clusters can exist.
}
\label{fig:PD_diluted_QPT}
\end{figure}
The paramagnetic Griffiths region consists of the area between $B_c^0$ and the ferromagnetic phase boundary
while the ferromagnetic Griffiths phase comprises the entire ordered phase.

In contrast to the classical case, the phase boundary $B_c(p)$ does not drop to
zero field at the percolation threshold $p_c$ of the lattice. It rather ends at a multicritical point
located at $p_c$ and some nonzero field $B_{\rm MCP}$. This implies that the site-diluted transverse
field Ising model features two different quantum phase transitions, a generic transition
at $p<p_c$, driven by quantum fluctuations and a quantum percolation transition at $p=p_c$
(across the vertical line in Fig.\ \ref{fig:PD_diluted_QPT}), driven by the lattice geometry.
By universality, we expect the generic transition to show the same critical behavior as
a three-dimensional version of the random transverse-field Ising model studied in Sec.\
\ref{sec:SDRG}. The percolation transition is in a different universality class which was
investigated by Senthil and Sachdev \cite{SenthilSachdev96}.

Why are the
classical and quantum phase diagrams (Figs.\ \ref{fig:PD_diluted} and \ref{fig:PD_diluted_QPT})
different? An intuitive picture of the difference follows from the quantum-to-classical mapping.
In the classical case, the critical temperature drops to zero at $p_c$ because the critical
infinite percolation cluster contains so-called red sites. A red site is a site that joins two otherwise
disconnected subclusters. Forming a domain wall at such a red site costs a \emph{finite}
energy of $2J$. The fluctuations of the relative magnetization orientations on the two subclusters
destroy the long-range order. In the quantum case, the red sites turn into red lines that are
infinitely long along the imaginary time direction. This suppresses the orientation
fluctuations between the two subclusters. Long-range order at $p_c$ is therefore possible.
A more detailed discussion is given, for example, in Ref.\ \cite{VojtaHoyos08b}.

We now focus on the paramagnetic Griffiths phase and study the contributions of the rare regions to
various observables. As in the classical case, we do this by combining the rare region probability
$w(L_{RR})$ with the properties of individual rare regions. The energy spectrum of a single locally
ordered region in the regime $B \ll B_c^0$ is sketched in Fig.\ \ref{fig:energy_spectrum}.
\begin{figure}
\includegraphics[width=5.cm]{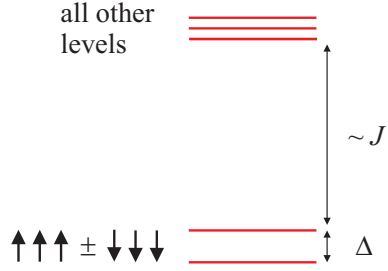}
\caption{Schematic energy spectrum of a single rare region in the regime $B\ll B_c^0$.
         The two low-lying states are the symmetric and antisymmetric combinations of
         fully polarized ``up'' and ``down'' states. They are separated by a large gap
         of order $J$ from the rest of the spectrum.}
\label{fig:energy_spectrum}
\end{figure}
For zero transverse field, $B=0$, the two fully polarized states $|\uparrow\uparrow\ldots\uparrow\rangle$
and $|\downarrow\downarrow\ldots\downarrow\rangle$ (all spins ``up'' or all ``down'')
are the degenerate ground states of the rare region. They are separated
from all other states by a large energy gap of order $J$. If a small field is switched on, the degeneracy
is lifted by forming the symmetric and antisymmetric superpositions
$|\uparrow\uparrow\ldots\uparrow\rangle \pm |\downarrow\downarrow\ldots\downarrow\rangle$ but
the large gap to the other energy levels remains.

Since we are interested in the low-energy physics of the system, we can neglect all these
high-energy states and treat the rare region as a two-level system. The value of the tunnel
splitting $\Delta$ between the two low-lying states $|\uparrow\uparrow\ldots\uparrow\rangle \pm |\downarrow\downarrow\ldots\downarrow\rangle$
can be estimated using perturbation theory in $B$. In $n$-th order in $B$, this produces terms
of the type $\langle \uparrow\uparrow\ldots\uparrow|\, (B\sum_i \sigma_i^x)^n\, | \downarrow\downarrow\ldots\downarrow\rangle$.
If the rare region contains $N \sim L_{RR}^d$ spins, the lowest order that gives a nonzero contribution is $n=N$
because each $\sigma_i^x$ operator in the  matrix element can only flip a single spin.
We thus arrive at the important conclusion that the tunnel splitting (or energy gap)  $\Delta$ must scale
like $B^N$. In other words, the gap is exponentially small in the volume of the rare region,
\begin{equation}
\Delta \sim B^N \sim \exp\left[  -a L_{RR}^d\right]
\label{eq:Delta_RR}
\end{equation}
with $a \sim \ln(J/B)$.

We can calculate a rare-region density of states by combining the size dependence (\ref{eq:Delta_RR})
of the energy gap with the probability (\ref{eq:w(L_RR)}) for finding a rare region. This gives
\begin{equation}
\rho(\epsilon) = \int dL_{RR} ~w(L_{RR}) \delta\left[\epsilon - \Delta(L_{RR})\right] \sim \epsilon^{p/a -1}= \epsilon^{\lambda -1}= \epsilon^{d/z' -1}~.
\label{eq:RR_DOS}
\end{equation}
We thus find the (quantum) Griffiths phase to be gapless, featuring a power-law density of states.
The exponent characterizing the singularity changes continuously throughout the Griffiths  phase.
It is often parameterized in terms of the Griffiths exponent $\lambda=p/a$ or the dynamical exponent
$z'=ad/p$.\footnote{To see that $z'$
plays the role of a dynamical exponent, consider the typical distance $r_{typ}$ between excitations
with energies below $\epsilon$. Using (\ref{eq:RR_DOS}), it follows that $r_{typ} \sim \epsilon^{-1/z'}$
or, equivalently, $\epsilon \sim r_{typ}^{-z'}$. Thus $z'$ indeed governs the relation between the
energy and length scales in the quantum Griffiths phase.}
In contrast to the classical case where the Griffiths singularities are weak essential singularities,
the (quantum) Griffiths singularity in the density of states of the diluted transverse-field Ising model is a
much stronger power-law singularity. If $z'>d$, the density of states even diverges for $\epsilon \to 0$.

In the ferromagnetic Griffiths phase, we can use the rare region probability (\ref{eq:w_FM(L_RR)})
instead of (\ref{eq:w(L_RR)}). The resulting density of states is still gapless and singular but takes the form
\begin{equation}
\rho(\epsilon)\sim \exp\{-\tilde c [\ln (\epsilon_0/\epsilon)]^{1-1/d}\}
\label{eq:RR_DOS_FM}
\end{equation}
rather than a power law.
Here, $\epsilon_0$ represents a microscopic energy scale, and the constant
$\tilde c = c / a^{1-1/d}$ changes continuously throughout the ferromagnetic Griffiths phase.

\subsubsection*{Observables in the quantum Griffiths phase}

Thermodynamic observables in the quantum Griffiths phase are easily calculated with the
help of the rare region densities of state (\ref{eq:RR_DOS}) or (\ref{eq:RR_DOS_FM}).
Rare regions (locally ordered clusters) having gaps $\epsilon < T$ are essentially free.
This means both low-lying states are accessible. In contrast, clusters having gaps
$\epsilon > T$ are in their quantum ground state. The total number $n(T)$ of free clusters
in the paramagnetic Griffiths phase can be obtained by integrating the density of states (\ref{eq:RR_DOS})
up to energy $T$,
\begin{equation}
n(T) = \int_0^T  d\epsilon \,\rho(\epsilon) \sim T^{d/z'}~.
\label{eq:n(T)_Griffiths}
\end{equation}
To determine the rare region entropy, we note that each free cluster contributes
$k_B \ln(2)$ while the clusters having $\epsilon > T$ do not contribute as they
are frozen in their quantum ground state. We thus find
\begin{equation}
S_{RR}(T) = k_B \ln(2) \, n(T) \sim T^{d/z'}~.
\label{eq:S(T)_Griffiths}
\end{equation}
The rare region contribution to the specific heat can be calculated simply by
taking the appropriate derivative
\begin{equation}
C_{RR}(T) = T(\partial S_{RR}/\partial T) \sim T^{d/z'}~.
\label{eq:CV(T)_Griffths}
\end{equation}
Each free cluster makes a Curie contribution $\mu^2/T$ to the uniform magnetic susceptibility.
According to (\ref{eq:Delta_RR}), the volume (and thus the moment $\mu$) of a rare region
depends logarithmically on its energy gap. The typical moment $\mu(T)$ of a free cluster at
temperature $T$ thus depends only logarithmically on $T$. This weak dependence can be neglected
if we are only interested in the leading power-law behavior. The rare region susceptibility
thus reads
\begin{equation}
\chi_{RR}(T) =  n(T) \, \mu^2 /T \sim T^{d/z' - 1}~.
\label{eq:chi(T)_Griffiths}
\end{equation}
To find the zero-temperature magnetization in a small \emph{longitudinal} magnetic field $h$,
we note that all clusters having $\epsilon < \mu h$ are almost fully polarized while clusters
having $\epsilon > \mu h$ have negligible magnetization. The magnetization-field curve can thus be
estimated as
\begin{equation}
m_{RR}(h) = \int_0^{\mu h} d\epsilon \, \rho(\epsilon) \, \mu \sim h^{d/z'}~.
\label{eq:m(H)_Griffiths}
\end{equation}
Here, we have again neglected the subleading logarithmic dependence of the moment $\mu$
on the energy $\epsilon$.

Many more observables can be worked out along the same lines including the nonlinear
susceptibility $\chi^{(3)}(T) \sim T^{d/z'-3}$, the zero-temperature dynamic susceptibility
${\rm Im} \chi(\omega) \sim \omega^{d/z'-1}$, and the NMR relaxation rate $1/T_1 \sim
\omega^{d/z' -2}$. All these power laws constitute the famous quantum Griffiths
singularities first discovered\footnote{Some of the unusual behavior was already
identified by McCoy and Wu \cite{McCoyWu68,McCoyWu68a,McCoyWu69} in a classical Ising model
with line defects. The transverse-field Ising chain maps onto this model under the
quantum-to-classical mapping. For this reason, quantum Griffiths singularities are
sometimes called Griffiths-McCoy singularities.}
} in the 1990's \cite{ThillHuse95,GuoBhattHuse96,RiegerYoung96,YoungRieger96}.
Recent overviews can be found, for example, in the review articles
Refs.\ \cite{Vojta06,Vojta10}. We emphasize that the power-law quantum Griffiths singularities are
much stronger than the
classical Griffiths singularities discussed in Sec.\ \ref{sec:classical_Griffiths}.
The magnetic susceptibility (\ref{eq:chi(T)_Griffiths}) even diverges (when $z'>d$)
while the system is still in the paramagnetic phase.

It is instructive the compare the quantum Griffiths singularities (\ref{eq:S(T)_Griffiths})
-- (\ref{eq:chi(T)_Griffiths})
with the off-critical behavior (\ref{eq:RTIM_S_pm}) -- (\ref{eq:RTIM_chi_pm})
of the random transverse-field Ising chain calculated within the strong-disorder renormalization group.
If we insert the correct dimensionality, $d=1$, the two sets of results agree.
We thus conclude that the unusual off-critical behavior of the random-transverse-field Ising chain
is a manifestation of the rare region physics discussed in the present section.

So far, we have focused on observables in the paramagnetic Griffiths phase.
By using the density of states (\ref{eq:RR_DOS_FM}) rather than (\ref{eq:RR_DOS}), one can
also derive the quantum Griffiths singularities on the ferromagnetic side of the transition.

\subsection{Classification of Griffiths (rare region) effects}
\label{subsec:Griffiths_classification}

In the last two subsections, we have compared the Griffiths singularities occurring at the thermal phase transition in a
classical diluted Ising model with those occurring at the quantum phase transition in the diluted transverse-field
Ising model. We have found the quantum Griffiths singularities to be much stronger as they are characterized by power laws
rather than weak essential singularities. In recent years, rare regions have been studied in many other disordered classical,
quantum, and nonequilibrium systems. The qualitative features of the resulting Griffiths phenomena
can be classified according to the effective dimensionality of the rare regions as was first suggested in Ref.\ \cite{VojtaSchmalian05}
and further developed in Ref.\ \cite{Vojta06}.

To understand this classification, let us return to our simple estimate of the rare region susceptibility
in the Griffiths phase. In general terms, it can be written as
\begin{equation}
\chi_{RR} \approx \int dL_{RR} \, w(L_{RR}) \, \chi_i(L_{RR})~
\label{eq:chi_estimate}
\end{equation}
where $w(L_{RR})$ is the probability for finding an individual rare region of size $L_{RR}$ and
$\chi_i(L_{RR})$ is its contribution to the susceptibility. The basic idea is to compare the decrease
of the probability $w(L_{RR})$ with the increase of $ \chi_i(L_{RR})$ as $L_{RR}$ is increased.

The functional form of $w(L_{RR})$ is governed
by combinatorics, it always decays exponentially with $L_{RR}$. More specifically,
$w(L_{RR}) \sim \exp(-\tilde p L_{RR}^{d_r})$ where $\tilde p$ is a nonuniversal constant and $d_r$
is the number of dimensions in which there is randomness. $d_r=d$ for point defects and other kinds of
uncorrelated or short-range correlated randomness, but for extended (line or plane) defects $d_r < d$.
The imaginary time dimension is never included in $d_r$ because quenched randomness is time-independent.

Depending on the behavior of $ \chi_i(L_{RR})$, three qualitatively different classes can be distinguished:
\begin{enumerate}
\item[(A)] $\chi_i(L_{RR})$ increases slower than exponential with $L_{RR}$. Therefore, the increase in
      $ \chi_i(L_{RR})$ cannot overcome the exponential decay of the probability. The contribution of large
      rare regions to the susceptibility integral (\ref{eq:chi_estimate}) is thus exponentially small.
      This case is realized, for instance, in the classical Ising model of Sec.\ \ref{sec:classical_Griffiths}
      where each rare region makes a Curie contribution $\mu^2(L_{RR})/T \sim L_{RR}^{2d}$ to $\chi$.

\item[(B)] $\chi_i(L_{RR})$ increases at least exponentially with $L_{RR}$ (but remains finite for any
      finite $L_{RR}$). In this case, the estimate (\ref{eq:chi_estimate}) diverges which means that
      the susceptibility is dominated by the largest rare regions. The quantum Griffiths singularities
      of the transverse-field Ising model of Sec.\ \ref{subsec:quantum_Griffiths} fall into this class
      because, according to (\ref{eq:Delta_RR}), the zero-temperature susceptibility of a single rare region
      behaves as $1/\Delta \sim \exp(a L_{RR}^d)$.

\item[(C)] In this class, $\chi_i(L_{RR})$ increases and diverges at some finite $L_{RR}$. This means that the
      cluster undergoes a true phase transition independently of the bulk system. As different rare regions generally
      undergo this transition at different values of the control parameter, this implies that the global
      phase transition is smeared. We will discuss this exotic possibility in Sec.\
      \ref{sec:smeared}.
\end{enumerate}

Which of the three classes a system falls into is determined by the relation between the effective
dimensionality $d_{RR}$ of the rare regions and the lower critical dimension $d_c^-$ of the problem.
In the case of a quantum phase transition, the imaginary time dimension needs to be included in $d_{RR}$.
If $d_{RR} > d_c^-$, the rare region can undergo a phase transition by itself, independently of the bulk.
The system is therefore in class C. In the case $d_{RR} < d_c^-$, an isolated rare region cannot undergo
the transition independently, and its contributions to observables grow like powers of $L_{RR}$.
This transition is therefore in class A. In the marginal case when the rare regions are right at the
lower critical dimension, $d_{RR} = d_c^-$, they still cannot undergo a transition by themselves,
but they ``almost can.'' Therefore, the rare region susceptibility (and other observables) increases
exponentially with its volume, putting the system into class B.

The result of these ideas is the classification shown in Table \ref{table:class}. It is expected to hold
for transitions between conventional phases in the presence of random mass disorder and short-range interactions
(such that interactions between the rare regions can be neglected).
\begin{table}
\renewcommand{\arraystretch}{1.1}
\begin{tabular}[c]{ccccc}
~\\[-1.3ex]
\hline
{\bf Class} & {\bf Rare region}  & {\bf Griffiths}
                   &{\bf   Global phase }   & {\bf Examples}~\\
~&{\bf dimension} &  {\bf singularities}   & {\bf transition}    &  {classical PT},
                                                                     ~{QPT}, {non-eq. PT}   \\
\hline
A & $d_{RR} < d_c^-$  & weak exponential    &   conventional         & {3D dilute Ising model \cite{BFMM98}}\\
  &                &                     &                         & {dilute bilayer Heisenberg model \cite{SknepnekVojtaVojta04}}\\[3ex]

B & $d_{RR} = d_c^-$  & strong power-law    &   infinite          & {Ising model with linear defects \cite{McCoyWu68}}\\
  &                &                     &      randomness         & {random transv.-field Ising model \cite{Fisher92}}\\
  &                &                     &                        & {disordered directed percolation \cite{HooyberghsIgloiVanderzande03}}\\[3ex]

C & $d_{RR} > d_c^-$  & RRs frozen &          transition   & {Ising model with planar defects \cite{Vojta03b}} \\
  &                & (undergo PT)&     smeared            & {itinerant quantum Ising magnet \cite{Vojta03a}}\\
  &                &                     &                        & {DP with extended defects \cite{Vojta04}}\\

\hline
\end{tabular}
\caption{Classification of rare region effects at classical, quantum, and non-equilibrium phase transitions
         (after Ref.\ \cite{VojtaSchmalian05}). The first example in each case is a classical transition,
         the second one is a quantum phase transition, and the third one, if any, is a nonequilibrium transition. A detailed
         discussion of the classification and these examples is given in the reviews, Refs.\ \cite{Vojta06,Vojta10}.}
\label{table:class}
\end{table}

\section{Smeared phase transitions}
\label{sec:smeared}

In all examples so far, the phase transition (if any) remained sharp
in the presence of disorder. This means that a nonzero order parameter appears
via a \emph{collective effect of the entire system}; and this onset is associated
with a singularity in the free energy. The reason for the transition remaining sharp is that a finite-size region
usually does not undergo a true phase transition by itself.

However, the arguments in Sec.\ \ref{subsec:Griffiths_classification} suggest,
that the global phase transition would be smeared rather than sharp if rare regions could
develop true static order independently of the bulk system. In this case, the
order parameter (e.g., the magnetization) would develop gradually when the external control
parameter is varied because one spatial region after
the other would undergo the phase transition.

How can an individual rare region undergo a true phase transition? One of the basic results
of statistical physics is that the partition function of a finite-size system is
analytic, and singularities develop only in the thermodynamic limit. This implies
that either (i) the rare regions themselves must be infinitely large or (ii) they must at
least be coupled to infinite baths. The first case is realized, for example, in a
three-dimensional classical Ising model with plane defects \cite{Vojta03b}. It will be discussed
in Sec.\ \ref{subsec:smeared_classical}. The second mechanism can be found
in dissipative quantum magnets \cite{Vojta03a} and will be analyzed in Sec.\ \ref{subsec:smeared_quantum}.

\subsection{Smearing of a classical phase transition}
\label{subsec:smeared_classical}

\subsubsection*{Randomly layered classical Ising model}

\begin{figure}
\includegraphics[width=8.cm]{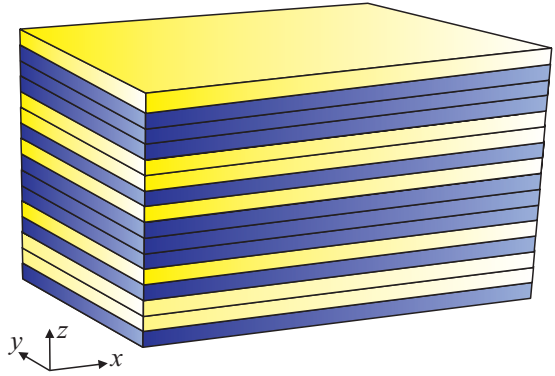}
\caption{Schematic of the layered magnet: layers of two different ferromagnetic materials
are arranged in a random sequence.}
\label{fig:layered}
\end{figure}

Imagine a material composed of a random sequence of layers of two different ferromagnetic
materials, as sketched in Fig.\ \ref{fig:layered}.
Nowadays, such structures can be created using modern nano-technology; magnetic multilayers
with systematic variations of the critical temperature from layer to layer have already been produced \cite{MPHW09}.

The randomly layered magnet can be modeled by a three-dimensional classical Ising model
\begin{equation}
H = - \sum_{\mathbf{r}} J^{\parallel}_z \, (S_{\mathbf{r}} {S}_{\mathbf{r}+\hat{\mathbf{x}}}
                                        +{S}_{\mathbf{r}} {S}_{\mathbf{r}+\hat{\mathbf{y}}} )
    - \sum_{\mathbf{r}} J^{\perp}_z \, {S}_{\mathbf{r}} {S}_{\mathbf{r}+\hat{\mathbf{z}}}
\label{eq:layered_Ising}
\end{equation}
where $S_\mathbf{r}$ is the spin at site $\mathbf{r}$, and $\hat{\mathbf{x}}$, $\hat{\mathbf{y}}$, $\hat{\mathbf{z}}$
are the unit vectors in the coordinate directions. The interactions $J^{\parallel}_z$ within the layers and $J^{\perp}_z$
between the layers are random but depend on the $z$ coordinate only. For definiteness,
let us assume all $J^{\perp}_z$ are identical to $J$ while the $J^{\parallel}_z$ are drawn from
a binary probability distribution
\begin{equation}
J_i^\parallel = \left \{ \begin{array}{cc}  J & (\textrm{with probability} ~1-p)\\
                                            aJ & (\textrm{with probability} ~p)\end{array} \right. .
\label{eq:binary}
\end{equation}
Here, $0<a\le 1$ parameterizes the values of the weak interactions. $p$ is the probability
for a weak layer while $1-p$ is the probability for a strong layer.

In the clean limit, $p=0$, the Hamiltonian (\ref{eq:layered_Ising}) is identical to the
usual three-dimensional Ising model. It undergoes a sharp continuous phase transition
at a critical temperature $T_c^0 \approx 4.511 J$. For nonzero $p$, the overall tendency
towards ferromagnetism is weakened. However, a large sample contains rare regions consisting
of thick slabs of consecutive strong layers.
In contrast to the rare regions considered in the earlier sections of this article, these
slabs are infinite in the $x$ and $y$ directions. Each such region is thus equivalent to a quasi
two-dimensional Ising model. As the two-dimensional Ising model is known to have a ferromagnetic phase,
each rare region in our layered magnet can undergo the phase transition independently of the
bulk system.

\begin{figure}
\includegraphics[width=12cm]{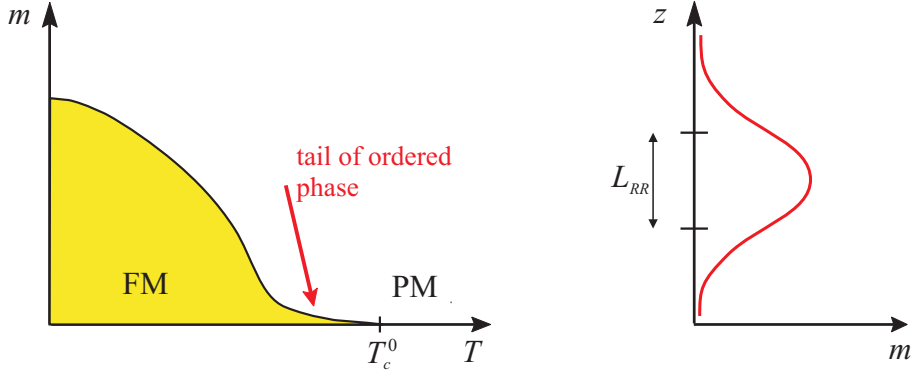}
\caption{Left: Schematic behavior of the magnetization $m$ as a function of the temperature $T$
        in the randomly layered Ising model (\ref{eq:layered_Ising}). As the temperature is lowered
        below the clean critical temperature $T_c^0$, the magnetization increases gradually because
        one slab after the other undergoes the phase transition. The global transition
        is thus smeared. Right: Magnetization profile $m(z)$ of a single ferromagnetic rare region
        embedded in a paramagnetic bulk.}
\label{fig:smeared_mag}
\end{figure}

The thickest slabs (rare regions) will undergo this transition very close to the
three-dimensional bulk critical temperature $T_c^0$. Thinner slabs order
at lower temperatures. The ferromagnetic slabs are weakly coupled via fluctuations
of the paramagnetic bulk, their magnetizations thus align.  As a result,
the global magnetization  develops gradually when the temperature
is lowered below $T_c^0$, as is sketched in Fig.\ \ref{fig:smeared_mag}.
This means that the global phase transition is smeared over a range of temperatures
below $T_c^0$. We emphasize that even though there is a singular onset of a nonzero magnetization
at $T_c^0$, this point is \emph{not} a critical point. Importantly, the initial onset of the global
magnetization is \emph{not} via a collective effect of the entire system but rather due to
the phase transition of one, particularly thick ``strong'' slab. This also implies that the correlation length in
$z$ direction remains finite for all $T$. The resulting magnetization in the tail of the
smeared transition is extremely inhomogeneous in space.

\subsubsection*{Optimal fluctuation theory}

We now use optimal fluctuation theory to determine the behavior in the tail of the smeared transition
\cite{Vojta03b}. The approach is similar to the calculation by Lifshitz and others \cite{Lifshitz64,Lifshitz64b,HalperinLax66}
of the electronic density of states in the band tails of doped semiconductors.

As pointed out above, a thick slab of $L_{RR}$ strong layers (i.e., layers with $J^\parallel=J$) undergoes a
ferromagnetic phase transition at some temperature $T_c(L_{RR})$ below the clean three-dimensional
bulk critical temperature $T_c^0$. We can use the LGW free energy (\ref{eq:LGW}) to determine
how $T_c(L_{RR})$ depends on the slab thickness $L_{RR}$. If we assume that the slab has undergone
the transition and is embedded in a nonmagnetic bulk, its local magnetization profile $m(z)$ must roughly
look like the right panel of Fig.\ \ref{fig:smeared_mag}. The (Gaussian part of the) free energy density due to this rare region
can be estimated as $\Delta f_{RR} \sim t m^2 + (\nabla m)^2 \approx t m^2 + m^2/L_{RR}^2$.
In mean-field theory, the transition of the slab occurs when the coefficient of $m^2$ vanishes, yielding
$t_c(L_{RR} )\sim T_c(L_{RR})-T_c^0 \sim -1/L_{RR}^2$.
The mean-field estimate can be refined using finite-size scaling \cite{Barber_review83} at
the three-dimensional bulk critical point. This results in
\begin{equation}
T_c^0 - T_c(L_{RR}) \sim L_{RR}^{-\phi}
\label{eq:T_c(L_RR)}
\end{equation}
where $\phi$ is the finite-size scaling shift exponent of the three-dimensional bulk critical point.
It takes the value $\phi=1/\nu$ because the three-dimensional Ising model is below its upper critical dimension
of $d_c^+=4$.

At a given temperature $T<T_c^0$, all slabs thicker than a critical thickness $L_c(T) \sim (T_c^0 -T)^{-\nu}$
are in the ferromagnetic phase while the rest of the system is still in the paramagnetic phase. Close to $T_c^0$,
the density of ferromagnetic slabs is very small, thus they can be considered independent. The total
magnetization in the tail of the smeared transition is therefore simply the sum over all ferromagnetic slabs,
\begin{equation}
m \sim \int_{L_c}^\infty dL_{RR} \, w(L_{RR}) \, m(L_{RR})~.
\label{eq:m_smeared}
\end{equation}
Here $m(L_{RR})$ is the magnetization of a single ferromagnetic slab, and $w(L_{RR})$ is the probability
of finding $L_{RR}$ consecutive strong slabs. This probability is given by the previous result (\ref{eq:w(L_RR)}),
but with the dimension $d$ replaced by $d_r=1$ because our layered Ising model has randomness in one
dimension (the $z$ direction) only. Thus, $w(L_{RR}) \sim \exp(-\tilde p L_{RR})$ with $\tilde p = -\ln(1-p)$.
While the probability $w(L_{RR})$ depends exponentially on the slab thickness, the dependence of the
slab magnetization $m(L_{RR})$ on $L_{RR}$ is at best a power law. It thus provides a subleading
correction to the integral (\ref{eq:m_smeared}). To exponential accuracy, we find
\begin{equation}
m \sim \exp(-\tilde p L_c) \sim \exp[-A(T_c^0-T)^{-\nu}]
\label{eq:m_smeared_result}
\end{equation}
with $A$ being a nonuniversal constant.

\begin{figure}
\includegraphics[width=7.5cm]{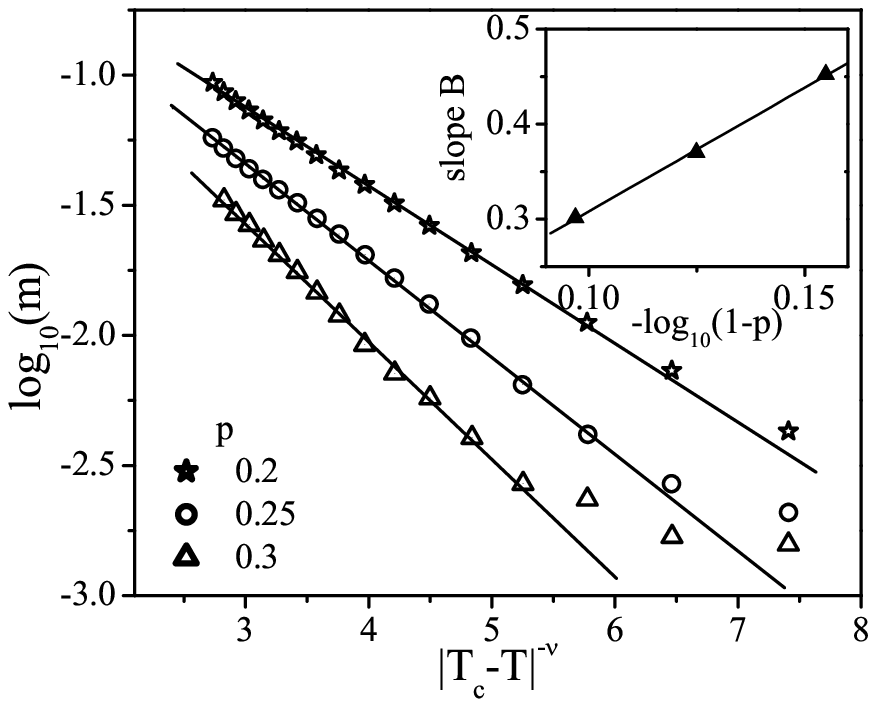}\includegraphics[width=7.5cm]{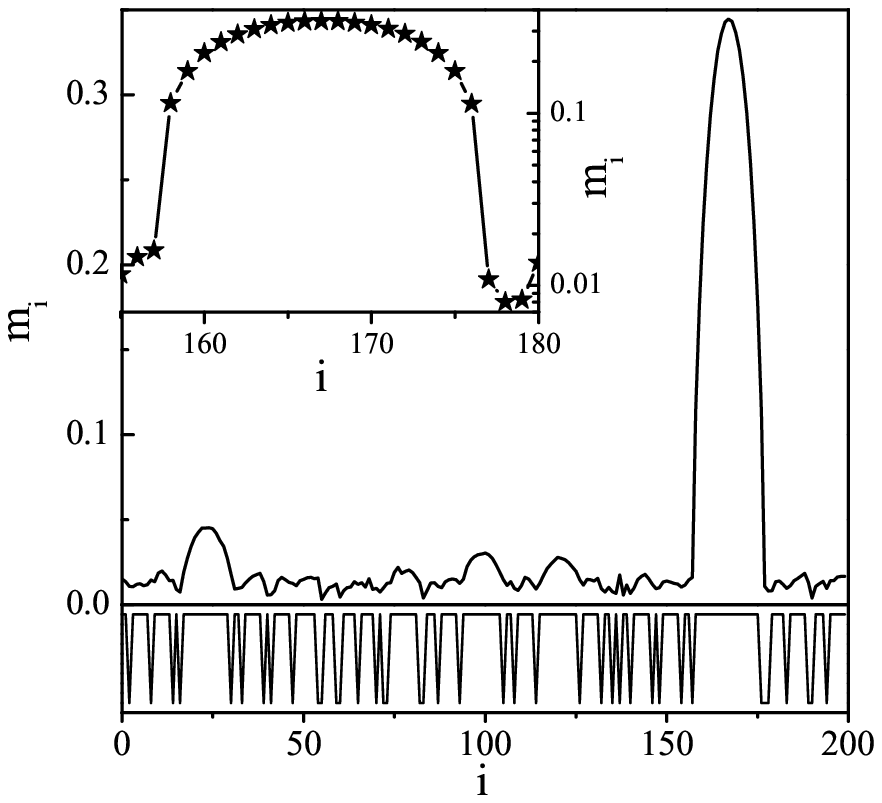}
\caption{Monte-Carlo simulations of a randomly layered Ising model. Left: Total magnetization $m$ in the tail
         of the smeared transition for several values of the weak bond probability $p$. Right:
         Local magnetization $m_i$ of layer $i$ for one disorder realization at $T=4.425$ very close to $T_c^0=4.511$.
         Only one slab has developed ferromagnetic order, and two or three more seem to be at the verge of ordering
         while the bulk magnetization vanishes within the Monte-Carlo noise.
         The interaction $J_i$ at the corresponding position
         is indicated in the lower panel (after Ref.\ \cite{SknepnekVojta04}).}
\label{fig:smeared_MC}
\end{figure}

The magnetization thus decays exponentially in the tail of the smeared transition and vanishes with
an essential singularity at the clean bulk critical temperature $T_c^0$. It is important to keep in mind
that this total (average) magnetization stems from a very inhomogeneous spatial distribution. This can be
seen in the results of large-scale Monte Carlo simulations \cite{SknepnekVojta04} of the layered Ising
model that are shown in Fig.\ \ref{fig:smeared_MC}.
Other observables can be worked out in a similar fashion \cite{Vojta03b,SknepnekVojta04}.

We emphasize that the functional form (\ref{eq:m_smeared_result}) of the magnetization as well as
those of other observables in the tail of the smeared transition are \emph{not} universal. In contrast to the
behavior at critical points, these functional forms do depend on the details of the disorder distributions
and on how the phase transition is tuned. If one uses, for example, a Gaussian interaction distribution
instead of the binary distribution (\ref{eq:binary}), the magnetization tail stretches all the way
to $T=\infty$ \cite{Vojta03b}. Moreover, if one tunes the phase transition by varying the concentration $p$
of the weak interactions at fixed $T$, the magnetization tail behaves exponentially at intermediate $p$
but vanishes as a power-law for $p\to 1$ \cite{HrahshehNozadzeVojta11,SNHV12}.
\footnote{See also the article by Nozadze {\it et al.} in this proceedings volume.}

\subsection{Smearing of a quantum phase transition}
\label{subsec:smeared_quantum}

\subsubsection*{Dissipative transverse-field Ising chain}

We now discuss the second possible route to smeared phase transitions in which individual rare regions
undergo their transitions independently because they are coupled to infinite (dissipative) baths.

We start from the random transverse-field Ising chain (\ref{eq:RTIM}) that we discussed in detail
in Sec.\ \ref{sec:SDRG}. Imagine that we now couple the $z$ component of each spin to an
independent (infinite) bath of quantum harmonic oscillators. The resulting Hamiltonian, the dissipative
random transverse-field Ising chain, can be
written as $H=H_I+H_B+H_C$. The Ising model part,
\begin{equation}
H_I = -\sum_{i} J_i \sigma_i^z \sigma_{i+1}^z - \sum_i B_i \sigma_i^x~,
\label{eq:H_I}
\end{equation}
is identical to (\ref{eq:RTIM}), the bath Hamiltonian takes the form
\begin{equation}
H_{B}=\sum_{k,i}\hbar \omega_{k,i}\left(a_{k,i}^{\dagger}a_{k,i}^{\phantom{]\dagger}}+{1}/{2}\right),
\label{eq:H_B}
\end{equation}
and the coupling between the spins and the dissipative baths is given by
\begin{equation}
H_{C}=\sum_{i}\sigma_{i}^{z}\sum_{k}\lambda_{k,i}\left(a_{k,i}^{\dagger}+a_{k,i}^{\phantom{\dagger}}\right)~.
\label{eq:H_C}
\end{equation}
Here, $a_{k,i}^{\dagger}$ and $a_{k,i}$ are the creation and destruction operators of oscillator
$k$ coupled to spin $i$, and $\omega_{k,i}$ is its frequency. The $\lambda_{k,i}$ parameterize
the strength of the coupling between spins and oscillator baths.
The character of the baths depends crucially on the low-frequency behavior of their spectral densities
\begin{equation}
{\cal E}_{i}(\omega)=\pi\sum_{k}\lambda_{k,i}^{2}\delta\left(\omega-\omega_{k,i}\right)~.
\label{eq:spectral-function}
\end{equation}
The most important case is, arguably, Ohmic
\footnote{An Ohmic dissipative bath leads to a friction force that is proportional to the velocity
and thus to Ohm's law when applied to the motion of charge carriers in a conductor.}  dissipation for which
the spectral density vanishes linearly with $\omega$ in the low frequency limit,
${\cal E}_{i} = (\pi/{2}) \alpha_{i} \,\omega$ for  frequencies below a cutoff $\omega_c$.

The dissipative random transverse-field Ising chain can be attacked by a generalization of the strong-disorder
renormalization group technique of Sec.\ \ref{sec:SDRG}. This was first done by implementing the recursion relations
(which now also include the renormalizations of the oscillators) numerically \cite{SchehrRieger06,SchehrRieger08}.
Later, a complete analytical solution was found \cite{HoyosVojta08,HoyosVojta12}. The renormalization group analysis
is mathematically rather involved and thus beyond the scope of these introductory lectures. Here, we instead develop
a heuristic picture of the rare region properties in the presence of Ohmic dissipation.

Consider again a rare locally ferromagnetic region of length $L_{RR}$ embedded in a paramagnetic bulk.
(This could be a region in which a number of consecutive interactions $J_i$ are atypically strong
or in which a few consecutive transverse fields are atypically weak.) As discussed in Sec.\
\ref{subsec:quantum_Griffiths}, the low-energy spectrum of such a region is identical to that of
a quantum two-level system. When coupled to Ohmic dissipative baths, each rare region is thus
equivalent to a dissipative two-level system which is a famous, paradigmatic model in
quantum physics \cite{CaldeiraLeggett83,LCDFGZ87,Weiss_book93}.

Importantly, the dissipative two-level system undergoes a (zero-temperature)
quantum phase transition from
a fluctuating ground state (a symmetric superposition of $|\uparrow\rangle$  and $|\downarrow\rangle$) to a localized ground
state (which prefers either $|\uparrow\rangle$ or $|\downarrow\rangle$) as the dissipation strength $\alpha$
is increased beyond a critical value $\alpha_c$.
In our case, the effective dissipation strength of a rare region is
simply the sum over the $\alpha_i$ of all baths that couple to the rare region, it is thus
proportional to its size, $\alpha_{\rm eff} = \sum_i \alpha_i \sim L_{RR}$.
Consequently, sufficiently large rare regions have effective dissipation strengths
$\alpha_{\rm eff} > \alpha_c$ even if the initial (bare) dissipation strengths $\alpha_i$ are weak.
Large rare regions therefore localize, i.e., they undergo the ferromagnetic quantum phase transition
independently of the bulk system. This implies that the global quantum phase transition is smeared.
This conclusion agrees with the explicit strong-disorder renormalization group calculations
\cite{SchehrRieger06,SchehrRieger08,HoyosVojta08,HoyosVojta12}.
The behavior of observables can be determined either by means of the renormalization group or by employing
an optimal fluctuation approach similar to that of Sec.\ \ref{subsec:smeared_classical}.

\subsubsection*{Higher dimensions}

Similar phenomena also arise if we
couple Ohmic dissipative oscillator baths to the two-dimensional or
three-dimensional site-diluted transverse-field Ising model
(\ref{eq:diluted_transverse_Ising}). This problem was studied in Ref.\ \cite{HoyosVojta06}. Sufficiently
large percolation clusters are on the localized side of the dissipative two-level system quantum
phase transition because their effective dissipation strength $\alpha_{\rm eff}$ is larger than
$\alpha_c$. This implies that their quantum dynamics is frozen.

\begin{figure}
\includegraphics[width=11cm]{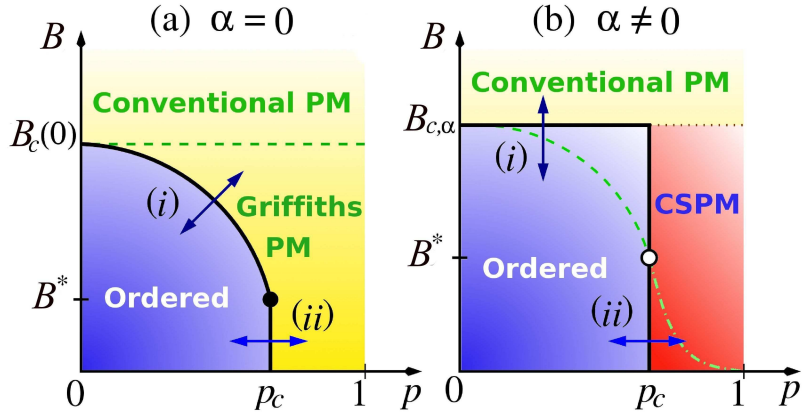}
\caption{Schematic zero-temperature phase diagrams
of diluted transverse-field Ising magnets without (a) and with (b) Ohmic dissipation.
In the classical super-paramagnetic phase (CSPM), large rare regions are frozen and
act as classical superspins. The dashed
line in (b) marks the crossover between homogeneous and inhomogeneous
order in the smeared transition scenario (after Ref.\ \cite{HoyosVojta06}).}
\label{fig:percdis}
\end{figure}

Consequently, the phase diagram of the model changes qualitatively compared to the
dissipationless case, as is demonstrated in Fig.\ \ref{fig:percdis}.
For dilutions $p<p_c$, the lattice remains connected. The large ferromagnetic rare regions
thus align with each other. This leads to a smeared phase transition as $B$ is varied at fixed $p$ [transition
(i) in Fig.\ \ref{fig:percdis}(b)]. The tail of the ferromagnetic phase stretches all the way to the clean critical field $B_{c,\alpha}$.
For dilutions $p>p_c$, in contrast, the lattice consists of disconnected clusters. Long-range
ferromagnetic order is thus impossible. The large ferromagnetic rare regions therefore act as
independent superspins. Because their quantum dynamics is frozen, their behavior is purely classical.

We conclude that the Ohmic dissipation completely destroys the quantum Griffiths phase.
For $p<p_c$, it is replaced by the tail of the smeared ferromagnetic phase transition, and for $p>p_c$,
it is supplanted by a classical super-paramagnetic phase. In contrast to the field-driven transition (i),
the percolation transition (ii) between the ferromagnetic phase and the classical super-paramagnet
remains sharp because it is driven by the geometry of the underlying lattice.

We emphasize that all these conclusions hold at zero temperature. At nonzero temperatures, the static order
of the rare regions is destroyed, and Griffiths singularities may be present in a transient temperature
and energy range. We will come back to this point in the next subsection.

\section{Magnetic quantum phase transitions in metals}
\label{sec:metals}
\subsection{Landau-Ginzburg-Wilson theory}
\label{sec:LGW_metals}

A particularly important application of the ideas developed in the last
sections are magnetic quantum phase transitions in metallic materials.
The standard theory of quantum phase transitions in (clean) Fermi liquids is the
Hertz-Millis theory \cite{Hertz76,Millis93}.
It can be derived by starting
from a microscopic Hamiltonian of interacting electrons and integrating out
all fermionic degrees of freedom; only the order parameter (magnetization)
fluctuations $m(\mathbf{x},\tau)$ are kept in the partition function.
The result is a quantum LGW free energy (or action) of the form
\begin{eqnarray}
S[m(\mathbf{x},\tau)] &=& \int d^dx_1 d\tau_1 \int d^dx_2 d\tau_2 ~ m(\mathbf{x_1},\tau_1)
               \Gamma(\mathbf{x_1},\tau_1,\mathbf{x_2},\tau_2) m(\mathbf{x_1},\tau_1) \nonumber \\
 &&+ u \int d^dxd\tau ~m^4(\mathbf{x},\tau) + O(m^6)~.
\label{eq:Hertz_action}
\end{eqnarray}
The order-parameter $m$ is a scalar for easy-axis (Ising) magnets but a two-component or
three-component vector for the cases of XY or Heisenberg symmetries, respectively.
The two-point vertex $\Gamma(\mathbf{x_1},\tau_1,\mathbf{x_2},\tau_2)$ is given by its
Fourier transform
\begin{equation}
\Gamma(\mathbf{q},\omega_{n})=r+\mathbf{q}^{2}+\gamma(\mathbf{q}) \left|\omega_{n}\right|~.
\label{eq:bare_Gamma}
\end{equation}
Here, $\mathbf{q}$ is the wave number of the order parameter fluctuations, and
$\omega_{n}$ is a Matsubara frequency.

In contrast to the example LGW theory (\ref{eq:q-LGW}), the dynamic term
$\gamma(\mathbf{q}) \left|\omega_{n}\right|$ is nonanalytic in the frequency.
This is an example of the nonanalyticities that generally occur whenever gapless (soft)
modes are integrated out in the derivation of the LGW theory \cite{BelitzKirkpatrickVojta05}.
Physically, this term accounts for the so-called Landau damping of the order parameter fluctuations
by the excitation of fermionic particle-hole pairs. The wave number dependence
of the prefactor $\gamma(\mathbf{q})$ depends on the type of
quantum phase transition. For a ferromagnetic transition $\gamma(\mathbf{q})$ behaves as
$1/|\mathbf{q}|$ for $\mathbf{q} \to 0$, reflecting the order parameter conservation
in a metallic ferromagnet. For a generic antiferromagnetic transition
(as well as the pair-breaking superconducting transition),
$\gamma(\mathbf{q}) = \gamma_0 = \textrm{const}$ for $\mathbf{q} \to 0$.\footnote{The
Hertz-Millis theory does not apply to all quantum phase transitions in metals,
and at least two scenarios leading to its breakdown have been identified.
(i) The coupling of the order parameter to other (generic) soft modes present in the system can invalidate
the entire expansion in powers of the order parameter \cite{BelitzKirkpatrickVojta05}.
(ii) Additional degrees of freedom other than
the order parameter fluctuations become critical at the transition point. This mechanism seems
to play a role at magnetic transitions in some heavy-fermion compounds \cite{LRVW07,GegenwartSiSteglich08}.}

To find the LGW theory for the case of a disordered metal, one can repeat the derivation
starting from a model of electrons in the presence a random potential or some other kind
of disorder. Hertz \cite{Hertz76} found that the structure of the action remains
essentially unchanged.\footnote{In a ferromagnet, $\gamma(\mathbf{q})$ now behaves as $1/\mathbf{q}^2$
because the electron motion is diffusive rather than ballistic.} However,
the distance from criticality becomes a random function of spatial position, $r \to r +\delta r(\mathbf{x})$.
This means, the action contains random-mass disorder.

It is instructive to compare the Hertz-Millis action
to the LGW theory of the  dissipative random-transverse-field Ising model considered
in Sec.\ \ref{subsec:smeared_quantum}. This LGW theory, which can be obtained by integrating out
the Ohmic oscillator baths in the partition function, turns out to be
identical to the Hertz-Millis action with a scalar order parameter and
$\gamma(\mathbf{q})=\textrm{const}$. The nonanalytic frequency dependence
$|\omega_n|$ is a reflection of the Ohmic character of the dissipative baths.

In following, we discuss the properties of quantum phase transitions occurring
in the Hertz-Millis LGW theory with random mass disorder. We first focus on the case
$\gamma(\mathbf{q})=\textrm{const}$, valid for antiferromagnetic transitions
(and pair-breaking superconducting transitions). The more complicated ferromagnetic case,
$\gamma(\mathbf{q}) \sim 1/\mathbf{q}^2$ will be addressed later.

\subsubsection*{Ising symmetry: smeared quantum phase transition}

\begin{figure}
\includegraphics[width=6cm,clip=]{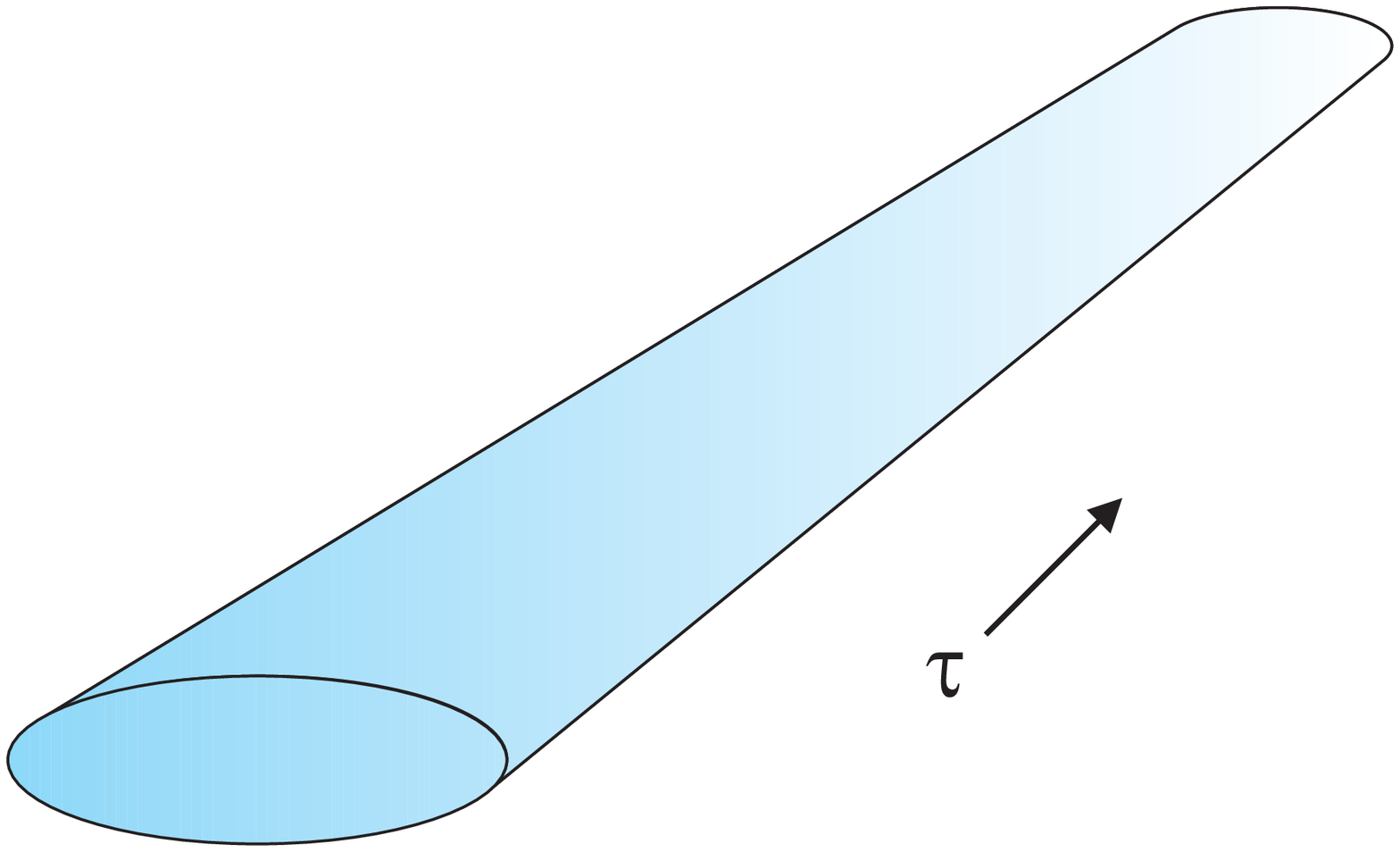}~ \includegraphics[width=7.5cm,clip=]{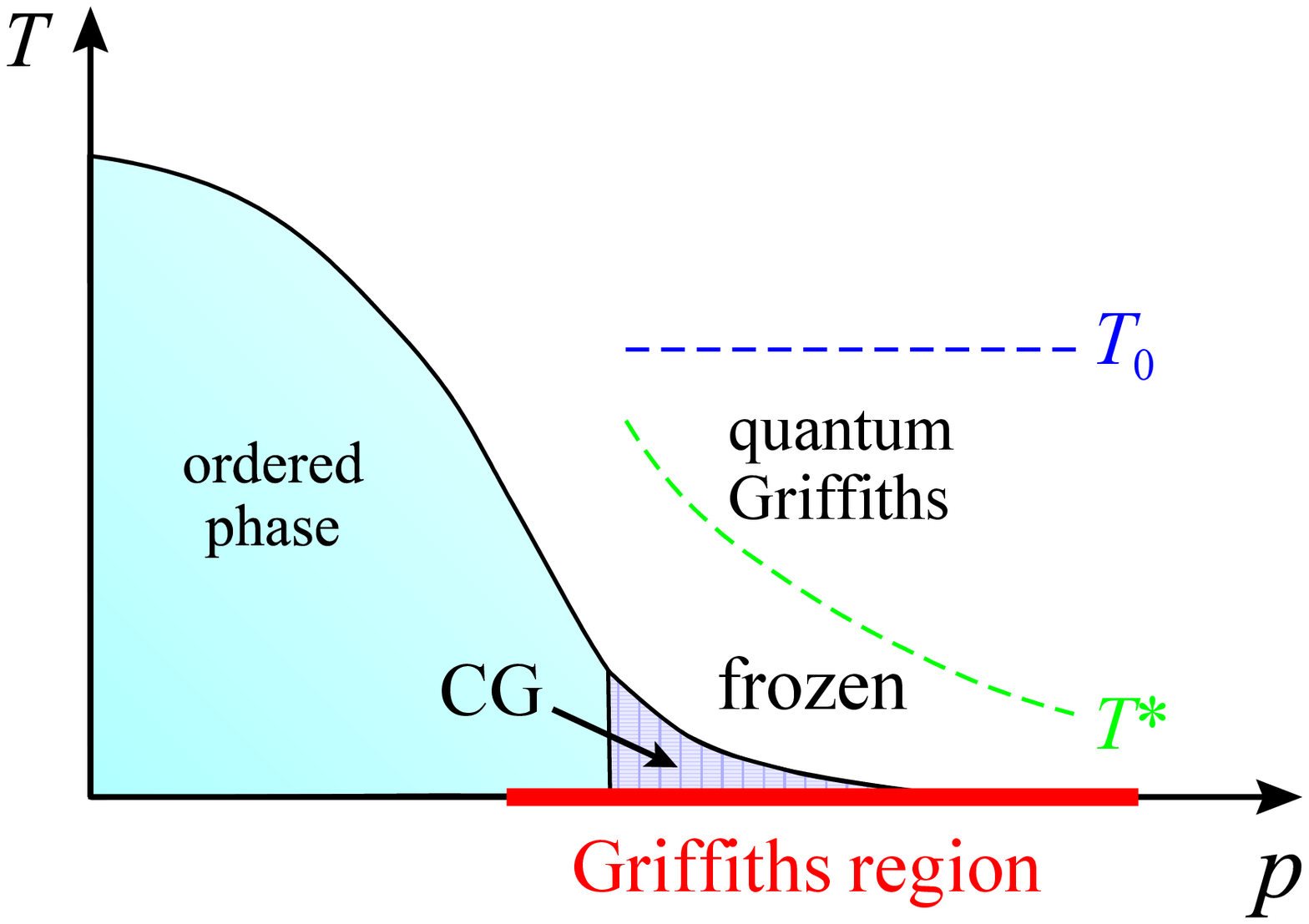}
\caption{Left: A rare region in the LGW theory (\ref{eq:Hertz_action},\ref{eq:bare_Gamma})
  is finite in space but infinitely long in the imaginary time direction. Right: Schematic finite-temperature
  phase diagram near the smeared quantum phase transition. $p$ stands for the quantum
  tuning parameter, and CG denotes the cluster glass expected due
  to the RKKY interaction between the rare regions in a realistic metal. $T^*$ marks the crossover between
  fluctuating and frozen rare regions, and the behavior above the microscopic cutoff $T_0$
  is nonuniversal.}
\label{fig:1dRR}
\end{figure}

The character of the transition depends crucially on the order-parameter symmetry.
For Ising symmetry, each locally ordered rare region can undergo the quantum phase transition
independently of the bulk system. This follows from the equivalence pointed out above
of the Hertz-Millis theory and the dissipative
transverse-field Ising model. The same conclusion can also be drawn directly from
the LGW theory (\ref{eq:Hertz_action},\ref{eq:bare_Gamma}). After quantum-to-classical
mapping, a rare region is equivalent to a quasi one-dimensional rod with finite size
in the space dimensions, but infinitely long in the imaginary time direction
(see left panel of Fig.\ \ref{fig:1dRR}).
As the nonanalytic $|\omega_n|$ frequency dependence corresponds to a long-range
$|\tau_1 - \tau_2|^{-2}$ interaction in the imaginary time direction, each individual
rare region is equivalent to a one-dimensional Ising model with a $1/\tau^2$
interaction. This model in known to have a phase transition \cite{Thouless69,Cardy81}
implying that each rod can order independently. This result was also reached
by Millis, Morr, and Schmalian via an instanton analysis of the Hertz-Millis action
\cite{MillisMorrSchmalian01,MillisMorrSchmalian02}.
The conclusion of all these considerations is that the global (zero-temperature) quantum phase transition
is smeared \cite{Vojta03a}. Observables can be calculated, for instance, by adapting
the optimal fluctuation theory of Sec.\ \ref{subsec:smeared_classical}.

As all experiments are performed at finite temperatures, it is important to ask: What
happens to the smeared quantum phase transition as the temperature is raised?
In the tail of the smeared transition, the locally ordered rare regions are far apart and
thus only weakly coupled. Their relative alignment is therefore destroyed at a very
low temperature. This leads to an exponential dependence of the critical temperature on the
quantum tuning parameter, as shown in the right panel of Fig.\ \ref{fig:1dRR}.
Above the critical temperature, the rare regions act as independent classical moments.

The behavior at higher temperatures depends on the strength of the damping, i.e.,
on the strength (prefactor) of the leading $|\omega_n|$ frequency dependence compared to the
regular (but subleading) $\omega_n^2$ term that is always present in the LGW expansion.
If the damping is weak, it becomes important only below a crossover temperature $T^*$.
Above this temperature, the quantum dynamics of the rare regions is effectively undamped,
leading to the quantum Griffiths singularities of Sec.\ \ref{subsec:quantum_Griffiths}
in a wide temperature window \cite{CastroNetoCastillaJones98,CastroNetoJones00}. In contrast, if the damping is strong,
this window becomes very narrow or completely unobservable. The actual strength of the
damping in realistic metallic magnets has been discussed controversially in the literature.

As is the case for smeared classical transitions, the properties of smeared quantum
phase transitions are nonuniversal. They depend on the details of the disorder distributions
as well as on how the transition is tuned. The experimentally important case of
tuning by chemical composition  is addressed in Refs.\ \cite{HrahshehNozadzeVojta11,SNHV12}.

\subsubsection*{Continuous symmetry: quantum Griffiths singularities}

In the case of continuous $O(N)$ order parameter symmetry (which includes
XY and Heisenberg symmetry), the behavior of locally ordered rare regions
is qualitatively different. After the quantum-to-classical mapping, each rare region
now corresponds to a quasi one-dimensional $O(N)$ model with $1/\tau^2$ interactions.
These models are known \emph{not} to have a phase transition, but they are exactly
at their lower critical dimension $d_c^-$ \cite{Joyce69,Dyson69,Bruno01}.
For this reason, an isolated rare region cannot independently undergo  the quantum
phase transition. However, its characteristic energy depends exponentially on its volume.
According to the classification of rare region effects discussed in Sec.\
\ref {subsec:Griffiths_classification}, the problem is thus in class B.

This prediction was confirmed by an explicit analysis of the rare region effects in
the LGW theory (\ref{eq:Hertz_action}) with $O(N)$ order parameter
symmetry \cite{VojtaSchmalian05}. This work established power-law quantum Griffiths
singularities very similar to that of the transverse-field Ising model
discussed in Sec.\ \ref{subsec:quantum_Griffiths}.\footnote{Griffiths
singularities appear not just in the thermodynamics but also in  transport
properties \cite{NozadzeVojta11}.}
 Later, Hoyos {\it et al.}
\cite{HoyosKotabageVojta07,VojtaKotabageHoyos09} applied a strong-disorder
renormalization group to this problem. Somewhat surprisingly, they found an infinite-randomness quantum critical
point in the same universality class as the random-transverse field Ising model.

This result points to an unusual kind of \emph{super-universality} that is not
fully understood. Normally, critical points
are in the same universality class (i.e., they share the same exponent values) if the symmetries
of the problems are the same. However, the random transverse-field Ising model
has discrete Ising order parameter symmetry and undamped dynamics while the disordered
$O(N)$ Hertz-Millis theory has a continuous symmetry order parameter and Ohmic damping.
Nonetheless, the critical points are in the same universality class.

We emphasize that all these studies assume that the interactions between different
rare regions are weak. Dobrosavljevic and Miranda \cite{DobrosavljevicMiranda05}
studied the effects of the long-range RKKY interaction between magnetic moments.
It is not contained in the Hertz-Millis theory but does exist in realistic metals.
They found that this interaction leads to a spin-glass like freezing of the
rare regions at the lowest temperatures (well below the energy scale of the single
rare-region effects) even in the continuous-symmetry case.

\subsection{Experiments}
\label{sec:experiments}

Clear-cut experimental verifications of the strong-disorder effects discussed in these lectures
were lacking for a long time. In the last five years, however, a number of promising
experiments have appeared.

Guo et al.\ \cite{GYMBHCHD07,GYMBHCHD10a,GYMBHCHD10b} studied the ferromagnetic quantum phase transition occurring in
the magnetic semiconductor  Fe$_{1-x}$Co$_x$S$_2$ at a concentration $x_c$ of about $0.007\pm 0.002$.
Close to the critical concentration, the authors found fluctuating magnetic moments with sizes significantly larger
than the expected spin-1/2 moments  of individual Co atoms in an FeS$_2$ host. This suggests that locally
ordered magnetic clusters are forming. The unusual transport, magnetic, and thermodynamic
properties in this concentration range can be described in terms of the quantum Griffiths power laws listed in
section \ref{subsec:quantum_Griffiths}.

The $f$-electron Kondo lattice system CePd$_{1-x}$Rh$_x$ undergoes a ferromagnetic quantum phase
transition as a function of the rhodium concentration $x$ \cite{SWKCGG07}. CePd is a ferromagnet with a critical
temperature of $T_c=6.6$ K while CeRh has a non-magnetic ground state. The ferromagnetic phase develops
a pronounced tail from about $x=0.7$ to 0.9. Above this tail (at temperatures above $T_c$), the behavior of the
magnetic susceptibility, the specific  heat, and other observables is characterized by nonuniversal
power laws in $T$ whose exponents vary systematically with $x$ \cite{Westerkampetal09}. Moreover, at the lowest temperatures, indications of a ``cluster glass'' are found. All these observations can be attributed
to the scenario outlined in Sec.\ \ref{sec:LGW_metals} (see Fig.\ \ref{fig:1dRR}).

\begin{figure}
\includegraphics[width=10cm]{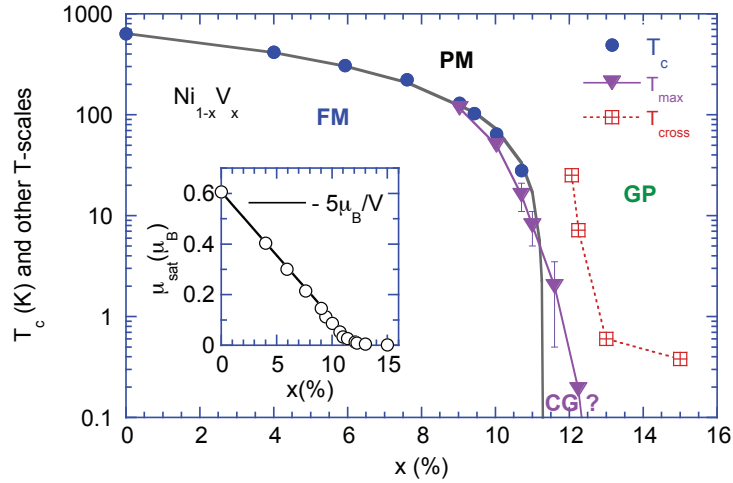}
\caption{Temperature-concentration phase diagram of Ni$_{1-x}$V$_x$
showing ferromagnetic (FM), paramagnetic (PM), quantum Griffiths (GP), and cluster glass
(CG) phases. Inset: saturation magnetization $\mu_{sat}$ versus $x$ (after Ref.\ \cite{UbaidKassisVojtaSchroeder10}).}
\label{fig:NiV_pd}
\end{figure}

\begin{figure}
\includegraphics[width=10cm]{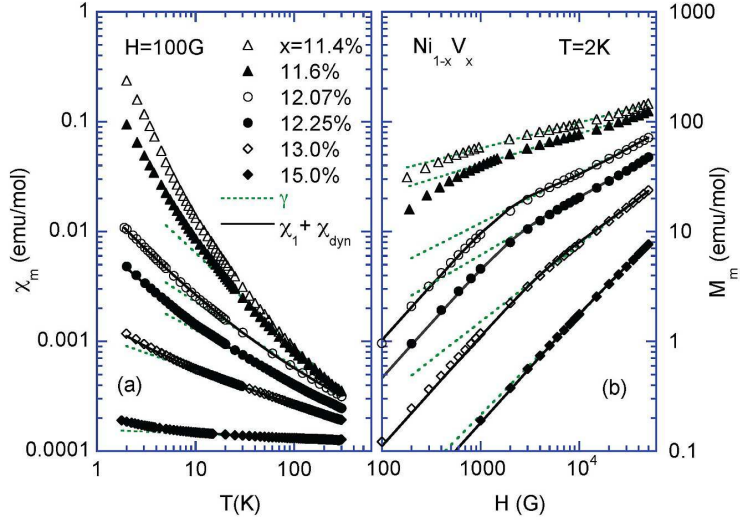}
\caption{(a) Low-field susceptibility $\chi_m$ versus temperature $T$ for several Ni$_{1-x}$V$_x$ samples
with concentrations $x=11 - 15\%$ and (b) low-temperature magnetization $M_m$ vs magnetic field
$H$. Dotted lines indicate power-law fits for $T>10$ K and $H>3000$ G in (a) and
(b), respectively  (after Ref.\ \cite{UbaidKassisVojtaSchroeder10}).}
\label{fig:NiV_chi_M}
\end{figure}

One of the most convincing verifications of quantum Griffiths behavior in a metallic
Heisenberg magnet can be found in the
transition metal alloy  Ni$_{1-x}$V$_x$.
Nickel is a ferromagnetic metal with a high Curie temperature of about $T_c=630$ K.
Vanadium substitution quickly suppresses the ferromagnetic order, leading to the phase diagram shown
in  Fig.\ \ref{fig:NiV_pd}.
Ubaid-Kassis {\it et al.\ }\cite{UbaidKassisVojtaSchroeder10,SchroederUbaidKassisVojta11}
performed magnetization and a.c.\ susceptibility measurements of several samples located
just on the paramagnetic side of the quantum phase transition (samples with
vanadium concentrations $x$ between the critical concentration of about 11\% and 15\%).
The results are summarized in Fig.\ \ref{fig:NiV_chi_M}.
The figure shows that the data (for $T>10$ K and $H>3000$ G) can be well described by
nonuniversal power laws, as predicted in Sec.\
\ref{subsec:quantum_Griffiths} for quantum Griffiths singularities. The Griffiths exponent $\lambda=d/z'$
can be determined from fits of the data to (\ref{eq:chi(T)_Griffiths}) and (\ref{eq:m(H)_Griffiths}).
As expected in the quantum Griffiths scenario, the values extracted from the susceptibility
and from the magnetization-field curve agree. Moreover, $\lambda$ vanishes (and $z'$ diverges)
at the critical point $x_c \approx 11.4\%$, as predicted. We note in passing that
the measurements show indications of a cluster glass at concentrations close to $x_c$
and very low temperatures (below about 1 K), in agreement with the scenario of Ref.\ \cite{DobrosavljevicMiranda05}.

Our final example is the ferromagnetic quantum phase transition that occurs in
Sr$_{1-x}$Ca$_x$RuO$_3$ as a function of composition $x$. The group of K\'ezsmarki
studied this transition using a highly sensitive magneto-optical technique \cite{Demkoetal12}.
The resulting phase diagram, which was already presented in Fig.\ \ref{fig:QPT_examples},
shows a pronounced tail of the ferromagnetic phase boundary. The magnetization-concentration
curve features a similar tail. Both tails can be fitted well with the theory of composition-tuned
smeared quantum phase transitions developed in Refs.\  \cite{HrahshehNozadzeVojta11,SNHV12}.

The attentive reader may have noticed that all the above examples involve \emph{ferromagnetic}
quantum phase transitions in metals while the theoretical analysis of Sec.\
\ref{sec:LGW_metals} focused on the antiferromagnetic transition. This is \emph{not} an accident.
Somewhat ironically, all good experimental examples of quantum Griffiths physics appear to be
found at ferromagnetic transitions
while the theory of rare region effects in metals is better developed for antiferromagnetic transitions.
The reason that quantum Griffiths singularities in ferromagnetic metals remained unexplored by theory
for a long time
is the nonlocal character $|\omega|/\mathbf{q}^2$ of the damping term which complicates
the theoretical analysis. This problem was solved very recently in Ref.\ \cite{NozadzeVojta12}.
The resulting rare region density of states is not a pure power law but rather takes the
form $\rho(\epsilon) \sim (1/\epsilon) \, \exp[-\tilde \lambda \ln^{3/5}(1/\epsilon)]$.
The functional forms of observables are changed accordingly. It turns out that these
ferromagnetic Griffiths singularities actually lead to improved fits \cite{NozadzeVojta12} of the Ni$_{1-x}$V$_x$
data shown in Fig.\ \ref{fig:NiV_chi_M}.

\section{Conclusions and outlook}
\label{sec:conclusions}

These lectures notes have given an introduction into the effects of random disorder on
phase transitions in quantum systems and into the exotic Griffiths phases occurring in
the vicinity of such transitions.

We have seen that zero-temperature quantum phase transitions generically display stronger
disorder effects than thermal (classical) phase transitions. The reason is that quenched disorder
is perfectly correlated in the imaginary time direction which needs to be taken into account
at a quantum phase transition. At zero temperature, the defects are effectively infinite in
this direction, enhancing their effects.

This implies that similarly strong disorder effects should occur at thermal transitions in
the presence of extended defects. This has indeed been observed not just in the famous
McCoy-Wu model \cite{McCoyWu68,McCoyWu68a,McCoyWu69} or in the layered classical
Ising model \cite{SknepnekVojta04} discussed in Sec.\ \ref{subsec:smeared_classical},
but also in randomly layered Heisenberg \cite{MohanNarayananVojta10,HrahshehBarghathiVojta11}
and XY \cite{MGNTV10} models. The latter problem is particularly interesting because it
applies to the superfluid phase transition of ultracold bosons \cite{PekkerRefaelDemler10}.

Two different classifications of disorder effects have emerged from our discussion.
First, one can classify critical points according to the
behavior of the average disorder strength in the limit of large length scales \cite{MMHF00}.
Three cases can be distinguished.
(i) If the Harris criterion is fulfilled, the disorder strength goes to zero, and the critical behavior
is identical to that of the clean system. The other two cases occur if the Harris criterion is violated.
(ii) If the effective disorder strength remains nonzero and finite in the large-length scale
limit, the clean critical behavior is unstable. The disordered critical point still features
conventional power-law scaling, but with exponents that differ from the clean ones.
(iii) If the effective disorder strength diverges, the resulting exotic
infinite-randomness critical point features activated (exponential) dynamical scaling
rather than the usual power-law scaling. Smeared phase transitions do not fit very well
into this classification scheme because qualitatively new physics (namely, the freezing of the
locally ordered rare regions) happens at a \emph{finite} length scale, destroying the
critical point.

The second classification \cite{VojtaSchmalian05,Vojta06} focuses on the properties of the rare regions and was discussed in Sec.\
\ref{subsec:Griffiths_classification}. Depending on the
effective dimensionality $d_{RR}$ of the rare regions, the following classes can be distinguished.
(A) The rare regions are below the lower critical dimension $d_c^-$ of the problem. This leads to (exponentially) weak
Griffiths singularities and critical points with conventional scaling.
(B) If the rare regions are right at the lower critical dimension (but still cannot order independently)
the Griffiths singularities are of power-law type and the critical point is an exotic
infinite-randomness critical point.
(C) If the rare regions are above the lower critical dimension, they can undergo the phase transition
independently of the bulk system. The global transition is thus smeared.
These two classifications are not independent of each other, as can be seen in Table \ref{table:class}.
Together, they form the basis of our current understanding of quantum phase transitions in the presence of random-mass
disorder.

The scenarios developed in these lectures apply to quantum phase transitions that can be described
by Landau-Ginzburg-Wilson order parameter field theories with real actions (such that the
quantum-to-classical mapping is valid). The effects of disorder on other, unconventional quantum
phase transitions is less well understood. This includes transitions in Fermi liquids not described
by the Hertz-Millis theory, Kondo lattice systems, quantum magnets in which the Berry phases are
important, or other systems with topological excitations. Some such transitions have already been
studied and revealed novel, exotic types of phase transitions. For example, Fernandes and Schmalian
studied the quantum phase transition in a diluted Josephson junction array \cite{FernandesSchmalian11}.
They found that a topological Berry phase in this problem renders the critical exponents complex,
a striking deviation from the Landau paradigm.

A systematic investigation of disorder effects at unconventional (non-Landau) transitions remains
a task for the future.


\begin{theacknowledgments}
The author is grateful for the hospitality of the organizers of the XVIIth Training Course in the
Physics of Strongly Correlated Systems, Dr. Avella and Prof. Mancini, as well
as the local staff in Vietri sul Mare. It was a great pleasure to lecture at the course.
Parts of the work reported here were supported by the National Science Foundation under Grant Nos. DMR-0339147, DMR-0906566,
and DMR-1205803, as well as by Research Corporation.

\end{theacknowledgments}

\bibliographystyle{aipproc}   

\bibliography{../00Bibtex/rareregions}

\end{document}